\documentclass[11pt]{book}
\usepackage{Wiley-AuthoringTemplate}
\usepackage[sectionbib,authoryear]{natbib}% for name-date citation comment the below line
\usepackage[margin=1.0in]{geometry}

\makeindex

\begin{document}

\newcommand\apj{Astrophysical Journal}
\newcommand\apjl{Astrophysical Journal Letters}
\newcommand\solphys{Solar Physics}
\newcommand\grl{Geophysical Research Letters}
\newcommand\mnras{Monthly Notices of the Royal Astronomical Society}
\newcommand\aap{Astronomy \& Astrophysics}

\frontmatter
%%%%%%%%%%%%%%%%%%%%%%%%%%%%%%%%%%%%%%%%%%%%%%%%%%%%%%%%%%%%%%%%
%% Title Pages
%% Wiley will provide title and copyright page, but you can make
%% your own titlepages if you'd like anyway
%% Setting up title pages, type in the appropriate names here:

\booktitle{Sun and Solar Wind}

%\subtitle{Efficient Multirate Loss Models}

\AuAff{Nour E. Raouafi\\ Applied Physics Laboratory}

\AuAff{Angelos Vourlidas\\ Applied Physics Laboratory}

%% \\ will start a new line.
%% You may add \affil{} for affiliation, ie,
%\authors{Robert M. Groves\\
%\affil{Universitat de les Illes Balears}
%Floyd J. Fowler, Jr.\\
%\affil{University of New Mexico}
%}

%% Print Half Title and Title Page:
%\halftitlepage
%\titlepage

%%%%%%%%%%%%%%%%%%%%%%%%%%%%%%%%%%%%%%%%%%%%%%%%%%%%%%%%%%%%%%%%
%% Copyright Page

%\begin{copyrightpage}{year}
%Title, etc
%\end{copyrightpage}

% Note, you must use \ to start indented lines, ie,
% 
% \begin{copyrightpage}{2004}
% Survey Methodology / Robert M. Groves . . . [et al.].
% \       p. cm.---(Wiley series in survey methodology)
% \    ``Wiley-Interscience."
% \    Includes bibliographical references and index.
% \    ISBN 0-471-48348-6 (pbk.)
% \    1. Surveys---Methodology.  2. Social 
% \  sciences---Research---Statistical methods.  I. Groves, Robert M.  II. %
% Series.\\

% HA31.2.S873 2004
% 001.4'33---dc22                                             2004044064
% \end{copyrightpage}

%%%%%%%%%%%%%%%%%%%%%%%%%%%%%%%%%%%%%%%%%%%%%%%%%%%%%%%%%%%%%%%%
%% Only Dedication (optional) 

%\dedication{To my parents}

\tableofcontents

%\listoffigures %optional
%\listoftables  %optional

%% or Contributor Page for edited books
%% before \tableofcontents

%%%%%%%%%%%%%%%%%%%%%%%%%%%%%%%%%%%%%%%%%%%%%%%%%%%%%%%%%%%%%%%%
%  Contributors Page for Edited Book
%%%%%%%%%%%%%%%%%%%%%%%%%%%%%%%%%%%%%%%%%%%%%%%%%%%%%%%%%%%%%%%%

% If your book has chapters written by different authors,
% you'll need a Contributors page.

% Use \begin{contributors}...\end{contributors} and
% then enter each author with the \name{} command, followed
% by the affiliation information.

 \begin{contributors}

 \name{Mausumi Dikpati,} High Altitude Observatory, Boulder, CO, USA

 \name{Rachel Howe,} School of Physics and Astronomy, University of Birmingham, UK

 \name{Mark G. Linton,} Space Science Division, Naval Research Laboratory, Washington, DC, USA

 \end{contributors}

%%%%%%%%%%%%%%%%%%%%%%%%%%%%%%%%%%%%%%%%%%%%%%%%%%%%%%%%%%%%%%%%
% Optional Foreword:

%\begin{foreword}
%\lipsum[1-2]
%\end{foreword}

%%%%%%%%%%%%%%%%%%%%%%%%%%%%%%%%%%%%%%%%%%%%%%%%%%%%%%%%%%%%%%%%
% Optional Preface:

%\begin{preface}
%\lipsum[1-1]
%\prefaceauthor{}
%\where{place\\
% date}
%\end{preface}

% ie,
% \begin{preface}
% This is an example preface.
% \prefaceauthor{R. K. Watts}
% \where{Durham, North Carolina\\
% September, 2004}

%%%%%%%%%%%%%%%%%%%%%%%%%%%%%%%%%%%%%%%%%%%%%%%%%%%%%%%%%%%%%%%%
% Optional Acknowledgments:

%\acknowledgments
%\lipsum[1-2]
%\authorinitials{I. R. S.} 

%%%%%%%%%%%%%%%%%%%%%%%%%%%%%%%%
%% Glossary Type of Environment:

% \begin{glossary}
% \term{<term>}{<description>}
% \end{glossary}

%%%%%%%%%%%%%%%%%%%%%%%%%%%%%%%%
%\begin{acronyms}
%\acro{ASTA}{Arrivals See Time Averages}
%\end{acronyms}

\setcounter{page}{1}
\setcounter{chapter}{1}

%\begin{introduction}

\mainmatter
%\include{ch01}

%%%Begin Solar Interior chapter
%
\chapter{Solar Interior \\ By Mark Linton, Mausumi Dikpati, and Rachel Howe \\ In book ``Sun and Solar Wind," edited by Nour E. Raouafi and Angelos Vourlidas}

\section{Abstract}

This chapter introduces the reader to the solar interior, in particular the convection zone.
The first section explores the solar cycle and the dynamo models which have been studied
to explain this cycle. The second section explores helioseismology obserations and analysis
of the solar interior, and reviews the fundmental knowledge that has been gained from
these studies. The third and final section reviews observations and theory of magnetic
fields emerging from the convection zone into the solar corona. 
%--------------------------------------------Enter here Authors and affiliations
%\author{M. Dikpati
%\affil{High Altitude Observatory, NCAR, 3280 Center Green Dr.,
%Boulder, CO 80301, USA}}

\section{Solar Dynamo (Mausumi Dikpati)\label{section:dynamo}}

\subsection{Solar dynamo models: overall context}

The Sun's magnetic field is very complex, with fields observed in the photosphere 
showing spatial variations over a wide range of scales, from the smallest resolvable 
scales ($\sim 100$km) to the size of the Sun itself ($\sim 10^5-10^6$ km). The smallest 
scales vary on time scales of minutes, while the global scales can take months to 
years to change much. The Table I presents a few examples of the hierarchy 
of magnetic field structures and evolutionary patterns.

\begin{tabular}{lll}
Hierarchy in & Hierarchy in & Hierarchy in \\
length scales & flux and/or & temporal variations \\
& field strength & \\
\hline
&& \\
Network fields & Diffuse fields & small-scale mixed polarity \\
$\sim 100$ km size & outside sunspots & turbulent fields \\
& a few tens of Gauss & random features \\
&& \\
Sunspots, ephemeral & Plage & butterfly diagram, \\
regions, plage & a few hundreds of & polar reversal \\
$\sim 10000-30000$ km & Gauss & cyclic features \\
diameter & & \\
&& \\
Large-scale, & spots, active regions, & active longitudes\\
unipolar regions & network fields & persistent features \\
$\sim 100,000$ km extent & a few thousand & \\
& Gauss & \\
\hline
&& \\
\end{tabular}

The generation, maintenance and evolution of the solar magnetic fields,
in random as well as in systematic fashions, are most likely due to a
`dynamo' operating inside the Sun. A dynamo is a process by which the
magnetic field in an electrically conducting fluid is maintained
against Ohmic dissipation. In astrophysical objects, there can always
be a dynamo whenever the plasma consists of seed magnetic fields and 
flow fields. The Sun, consisting of plasma, can have its own 
dynamo. But to this day, there is no unified, single solar dynamo 
model that can explain all solar magnetic features.

While in a pure hydrodynamic (HD) situation, there is evidence that the 
large-scale motions develop from the small-scale ones through inverse 
cascading \citep{Miesch2000}, the situation is more complicated in the 
magnetohydrodynamic (MHD) case. 
There exists observational evidence of both the cascading and inverse 
cascading, for example, the break-up of large active regions is a 
cascading phenomena and their coalescence to form the large-scale, diffuse 
magnetic fields \citep{Leighton1964} is an inverse cascade. 

However, the organization of the small-scale magnetic fields into the 
large-scale patterns, which cyclically evolve with an $\sim$11-year 
periodicity has not yet been established theoretically. Most likely two 
different types of dynamos, the small-scale type and the large-scale
type, are operating in the Sun coexisting together -- each type is
responsible for producing certain aspects of the solar magnetic
features. Small-scale turbulent dynamos are primarily responsible for
producing the generation and evolution of small-scale magnetic fields 
in short-term and in a more random fashion, whereas the generation and
systematic evolution (in a cyclic fashion with 11-year periodicity) of 
large-scale magnetic fields is due to global dynamos.

\subsection{Small-scale turbulent dynamos and their observational signatures}

Numerical simulations of local dynamo action in a 3D box, by \citet{Emonet2001} 
\citep[see also][]{Meneguzzi1989, Cattaneo1999, Emonet2001}, have indicated that 
a turbulent flow interacting with a seed 
magnetic field, which can be either a random field or an imposed uniform 
vertical field, can produce dynamo action and generate the small-scale 
magnetic fields. Depending on the flow fields, the generated magnetic fields 
can correspond to the intra-network fields (in the case of granular scale flows), 
and the ephemeral regions, network fields (in the case of supergranular scale
flows).

A few among many observational signatures of the small-scale surface dynamo 
action in the Sun, are: magnetic flux evolution in plage -- observed in high 
resolution G-band \citep{Berger1996}, the presence of weak magnetic field 
of $\sim 1$ Gauss over 1 ${\rm arc} \, {\rm s}^2$ in the quiet Sun -- observed 
from the Zeeman splitting in high resolution infrared lines \citep{Lin1999}, 
generation of mixed-polarity network fields in the quiet Sun -- observed 
in the Solar and Heliospheric Observatory / Michelson Doppler Imager \citep[SOHO/MDI:][]{1995SoPh..162..129S} magnetogram images \citep{Schrijver1997}, 
the existence
of unsigned, weak ($5-15$ Gauss) fields in the photosphere, chromospheric
spicules -- observed using Hanle depolarization effect \citep{Stenflo1999, TrujilloBueno2005}, 
and the existence of the small-scale, turbulent magnetic 
structures of sizes below the mean free path of a photon -- derived from the 
observed asymmetry in Stokes V profile \citep{SanchezAlmeida2000}.

The continuous churning of the small-scale magnetic fields in the granular 
lanes outside the sunspot can be compared with the simulation output of 
\citet{Emonet2001} (see their figure 1).

\subsection{Global dynamo models and solar cycle}

\begin{figure}
\begin{center}
\includegraphics[clip,width=0.6\textwidth]{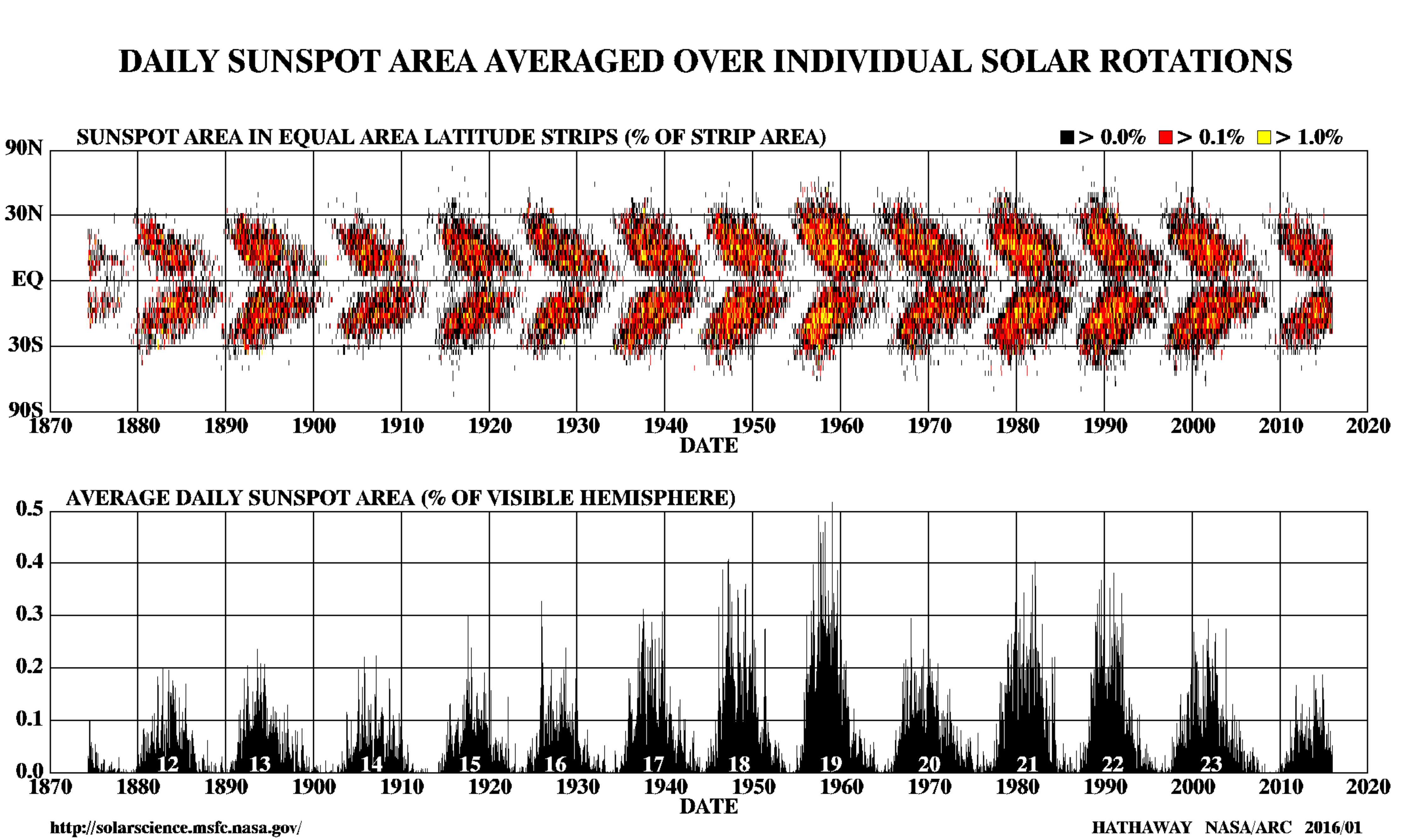}
\end{center}
\caption{Upper:Time-latitude diagram of spot area; Lower:total spot area
as function of time. Credit: Hathaway/NASA/ARC \citep{Hathaway2015}.
    \label{fig:dynamo-1}}
\end{figure}

The existence of a 'solar cycle' has been known since the middle of the 19th
century. Panels a and b of Figure \ref{fig:dynamo-1} show the classical representations of 
this cycle. Figure \ref{fig:dynamo-1}a is the so-called `butterfly diagram' which 
is a time latitude plot of the
fractional area covered by sunspots, with North and South hemispheres averaged
together. The pattern is rather regular, with each new cycle starting in the
neighborhood of $30^{\circ}$ and ending close to the equator. We can see that 
the cycle period averages about 11 years, but there is a variance of more than 
one year, shorter and longer. As seen in Figure \ref{fig:dynamo-1}b, solar cycle amplitudes as 
measured by sunspot area vary by a factor of 2-3 from cycle to cycle, not 
randomly, but with an envelope of cycle amplitudes that varies over longer 
time scales than a single cycle. Among other manifestations of solar cycle are 
the polar field reversal every 11 year, evolution of `active longitudes', 
variation in global coronal structure. 

There is a general consensus that the solar cycle is produced by a global MHD
dynamo operating in the solar convection zone and the solar 
tachocline just below. General dynamo theory began in the 1930's, motivated 
mainly by the existence of the Earth's magnetic field, thought to be generated 
by a dynamo operating in the liquid interior. Dynamo models with specific 
application to the Sun first appeared in \citet{Parker1955b}. Solar dynamo models 
have advanced greatly since then, but there are still many challenges to overcome 
in the quest to build a dynamo model that simulates the most prominent features 
of a solar cycle, and allows prediction of amplitude and other properties of 
future cycles. 

We describe in this review the current state of global solar dynamo modeling, 
including various classes of models, important mathematical tools used to build 
them, what has been accomplished to date and what outstanding challenges remain
to be answered to make further progress. There are several excellent, recent 
reviews, such as \citet{Charbonneau2010, Charbonneau2014} for solar cycle models, and 
\citet{Hathaway2015} and \citet{Petrie2015} for observations, which the readers are encouraged to 
consult. This review is primarily from a perspective of what needs to be 
done to go forward, in order to motivate the younger generation in this field 
of research.

Many stars have dynamo-maintained magnetic fields, some similar to the Sun,
others quite different, some with magnetic cycles, some not. Solving the solar
dynamo problem has obvious value for understanding stellar dynamos generally;
conversely, successfully modelling the properties of stellar dynamos will have
great value for understanding the solar cycle problem.

Ever since the first solar dynamo model developed by \citet{Parker1955b}, solar dynamo 
research has proceeded following two different approaches in parallel 
\citep[see][for a detailed review]{Charbonneau2010}. The first one uses the mean field 
formalism \citep{Steenbeck1966}, which solves the magnetic 
induction equation to obtain the solution for the large-scale, mean magnetic 
field, by approximating the small scale motions and magnetic fields. The 
equatorially antisymmetric helical part of the correlation between small scale 
motions and magnetic fields gives rise to the so-called "$\alpha$-effect", and 
the mirror-symmetric nonhelical part is the turbulent diffusivity ($\eta_T$). 
These models are computationally fast and useful for simulating solar activity 
features, but one of the limitations is that they rely on the parametric form of
$\alpha$-effect processes and turbulent magnetic diffusivity, and hence rely on 
capturing the physics of these processes. Many solar dynamo models have been built 
over the past 70 years using the mean field formalism, primarily with the aim of 
reproducing the sunspot butterfly diagram, the most well-known characteristic of 
the solar cycle.

In the second approach, full 3D magnetic induction equations are solved
along with the flow and energy equations including thermodynamics under the
the anelastic approximation \citep{Gilman1983, Glatzmaier1985, Brun2004}. 
These models have recently advanced to the level of sophistication 
that they produce cycles of solar-like cycle length \citep{Racine2011, Fan2014}, 
as well as intricate details, such as, wreathe-like formations of 
toroidal fields \citep{Brown2010, Nelson2013}. These models also have 
their limitations. Even though these models solve the full set of MHD equations 
in 3D with the aim of resolving all processes relevant for dynamo action, the 
contributions from subgrid scale processes are parameterized. Furthermore, it 
has been a challenge to reach the right parameter regime in terms of Reynolds
and Rossby numbers.
Currently this class of models does not produce solar like differential rotation if 
solar type rotation and luminosity are used \citep[see, e.g.][]{Fan2014, Gastine2014}, 
indicating a too high Rossby number, i.e. too fast convective flows.

\subsubsection{Overview of past accomplishments}

Successful modelling of the global fluid flow and magnetic fields in a planetary 
atmosphere or ocean, or the convection zone of a rotating star, has generally begun 
with relatively simple models that make no attempt to model all the small-scale 
processes of the whole system, but rather focus on those physical processes thought 
to be important for that system. For example, models for instability of global flows 
in the Earth's atmosphere began to be developed in the 1940's \citep{Charney1947, Eady1949}.
It took another twenty years to develop models that had enough physical processes 
and enough realism and spatial resolution to plausibly model the atmospheric 
circulation \citep{Trenberth1992}. Since then, such models have become vastly more 
detailed and realistic, simulating and predicting ever greater detail of the global 
circulation. 

Dynamo models for solar cycles are developing in an analogous way, but are at
a much earlier stage. Parker's model contained very relevant physical
processes (in particular, shearing of magnetic field lines by differential
rotation and twisting and lifting of field lines by helical turbulence), but
certainly was not realistic enough to simulate the primary features of a particular
solar cycle. But now such models do exist, and they can and have been calibrated
to major features of solar cycles. Despite the complexity and realism of current 
global models for the general circulation of the Earth's atmosphere, such models 
still must parameterize important physical processes that are too small in spatial 
scale to be calculated explicitly everywhere on the globe. The situation is the same 
for modelling solar cycles with a dynamo model. Currently calibrated solar cycle 
simulation models are virtually always in two dimensions, latitude and radius, and 
can simulate only axisymmetric solar cycle features -- which, fortunately, describe
a large part of what constitutes a cycle. In these models, all non-axisymmetric
processes are parameterized with axisymmetric representations. As we discuss
in later sections, the most successful of these models in simulating solar
cycles have been the so-called Babcock-Leighton flux-transport (BLFT) dynamos.

So-called `full 3D' solar dynamo models exist \citep{Gilman1983, Fan2014, Lawson2015}, 
which do include departures from axisymmetry; these models 
have recently produced cyclic dynamo solutions, but they have yet to be calibrated 
for the Sun. Furthermore, they too do not spatially resolve energetically important 
scales of motion and magnetic fields in the Sun, and so still rely heavily on 
averages of and parameterizations of unresolved processes. This limitation on 
spatial resolution will not be overcome for many generations of computing power 
growth, and hence, despite their higher spatial resolution, including variations 
in longitude, the `full 3D' dynamos will be continuing to average over processes 
occurring on smaller spatial scales for the foreseeable future. 

\subsubsection{New challenges \label{section:new_challenges}}

The previous section described in broad terms where solar cycle modeling using
dynamo models currently stand. Here we outline what the major challenges are to
making meaningful progress toward realistic simulation of actual solar cycles.

The most obvious, and perhaps most important, challenge is to move from 2D
to 3D solar cycle simulation models, in order to correctly simulate global
but longitude-dependent solar cycle features such as magnetic flux emergence,
active longitudes, and solar sector structure, which has so much influence in
organizing and modulating the solar wind and interplanetary magnetic field
that transmits the effects of solar activity to the Earth, particularly to
its upper atmosphere. Since the non-axisymmetric solar cycle features to be 
simulated are themselves global in scale, it is not clear the full 3D convective
dynamo simulation approach, which would spatially resolve a whole spectrum of 
convection, is either warranted or practical. Perhaps it may be more productive 
to generalize the BLFT dynamo models to include the most important global 3D 
effects \citep[see, e.g.][]{Miesch2014}. 

There is incomplete information available about key components of solar dynamo 
models, particularly meridional circulation (see \S \ref{section:hsm-meridional}). 
Observations of meridional circulation 
are good near the surface, but much more uncertain for deeper layers. Helioseismic 
methods are generating some possible profiles, but they show large variations for 
different analysis techniques and time periods. It will be necessary to develop, 
in parallel with the 3D dynamo model itself, better theories for meridional 
circulation, which take account of what we do know about flow in the convection 
zone, such as the differential rotation there. This knowledge can be used to 
constrain meridional circulation profiles that are possible.  

Modern data assimilation (DA) techniques, in the early stages of being included
in solar dynamo models, will provide a powerful tool for inferring the form and
amplitude of meridional circulation with depth and latitude 
\citep{Dikpati2014, Dikpati2016a, Dikpati2016b, Jouve2015, Hung2015}, using 
so-called observing 
system simulation experiments (OSSE's). Even more important, DA will provide the 
computational framework for optimizing the use of observations of solar velocities 
and magnetic fields to initialize and generate the best possible simulations of 
actual solar cycles, including global 3D effects. More advanced use of such 
methods has been enormously beneficial to improving simulations and predictions 
of global atmospheric flows. Even though it is true that solar data comes much more 
from photosphere and above than in the solar interior, it is far more important 
to use the available solar data for assimilation by implementing modern DA 
methods for estimating spatio-temporal profiles of unknown ingredients for the 
regions where we lack observational data. In this context we recall the statement 
of Kalnay -- the use of the model forecast is essential in achieving four-dimensional 
data assimilation. The model transports information from data-rich to data-poor 
regions, and it provides a complete estimation of the four-dimensional state of 
the atmosphere. Including the data-rich Northern hemisphere in the assimilation 
scheme makes much bigger improvements in the forecasts in the Southern hemisphere 
\citep{Kalnay2003}.

The kinematic system can be the place to start, because we do not need a model 
to simulate differential rotation, since it is nearly constant, but crucial 3D 
dynamical processes, such as the role of global MHD tachocline and convection
zone dynamics can be included as part of input of solar flow fields. 

\subsection{Solar cycle observations to be modelled by a cyclic dynamo \label{section:dynamo-solar_cycle}}

\begin{figure}
\begin{center}
\includegraphics[clip,width=0.6\textwidth]{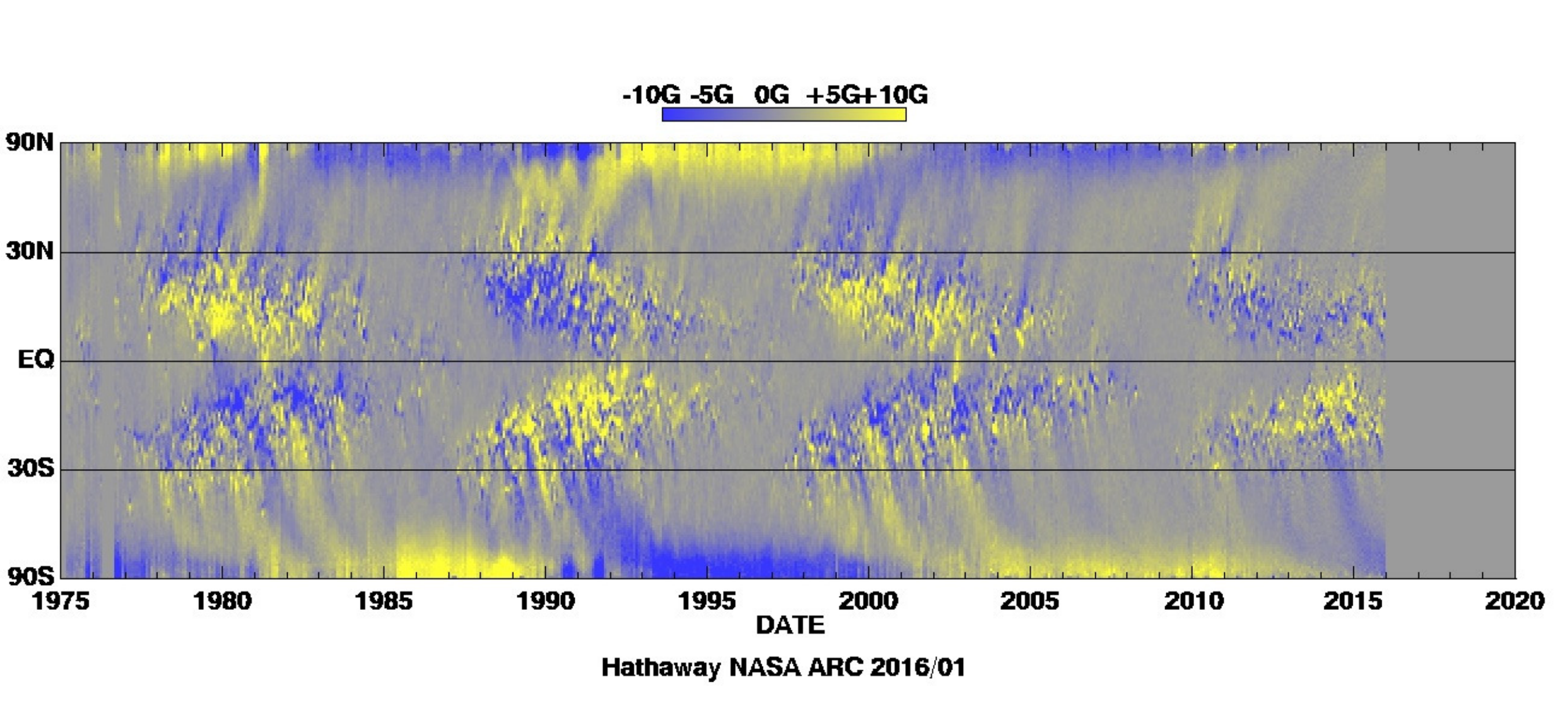} 
\end{center}
\caption{Magnetic butterfly diagram, constructed from longitudinally averaged
radial magnetic fields obtained from Kitt Peak and SOHO measurements. Blue
and yellow shading denote opposite polarities. Note the clear presence of
polar field reversals and migration of flux to the poles (the nearly vertical 
streaks). Credit: Hathaway NASA/ARC). \label{fig:dynamo-2}}
\end{figure}

Figure \ref{fig:dynamo-1} characterizes only sunspot patterns and statistics. But there
is a very important relationship between sunspot fields and other surface
magnetic fields, so called `poloidal' fields, especially those near the poles.
Figure \ref{fig:dynamo-2} shows this relationship. Blue and yellow shading 
represent opposite
polarity fields. We see the low latitude butterfly diagram, but we also see
rather rapid migration of poloidal fields toward the poles (the steeply slanted
streaks). One sign of field predominates in this migration, opposite in North
and South hemispheres, and opposite in each succeeding cycle. We can clearly
see that this migration reverses the sign of the polar field near the maximum
of the sunspot cycle, The polar field then holds its sign in each hemisphere
until the next maximum. 

\begin{figure}
\begin{center}
\includegraphics[clip,width=0.6\textwidth]{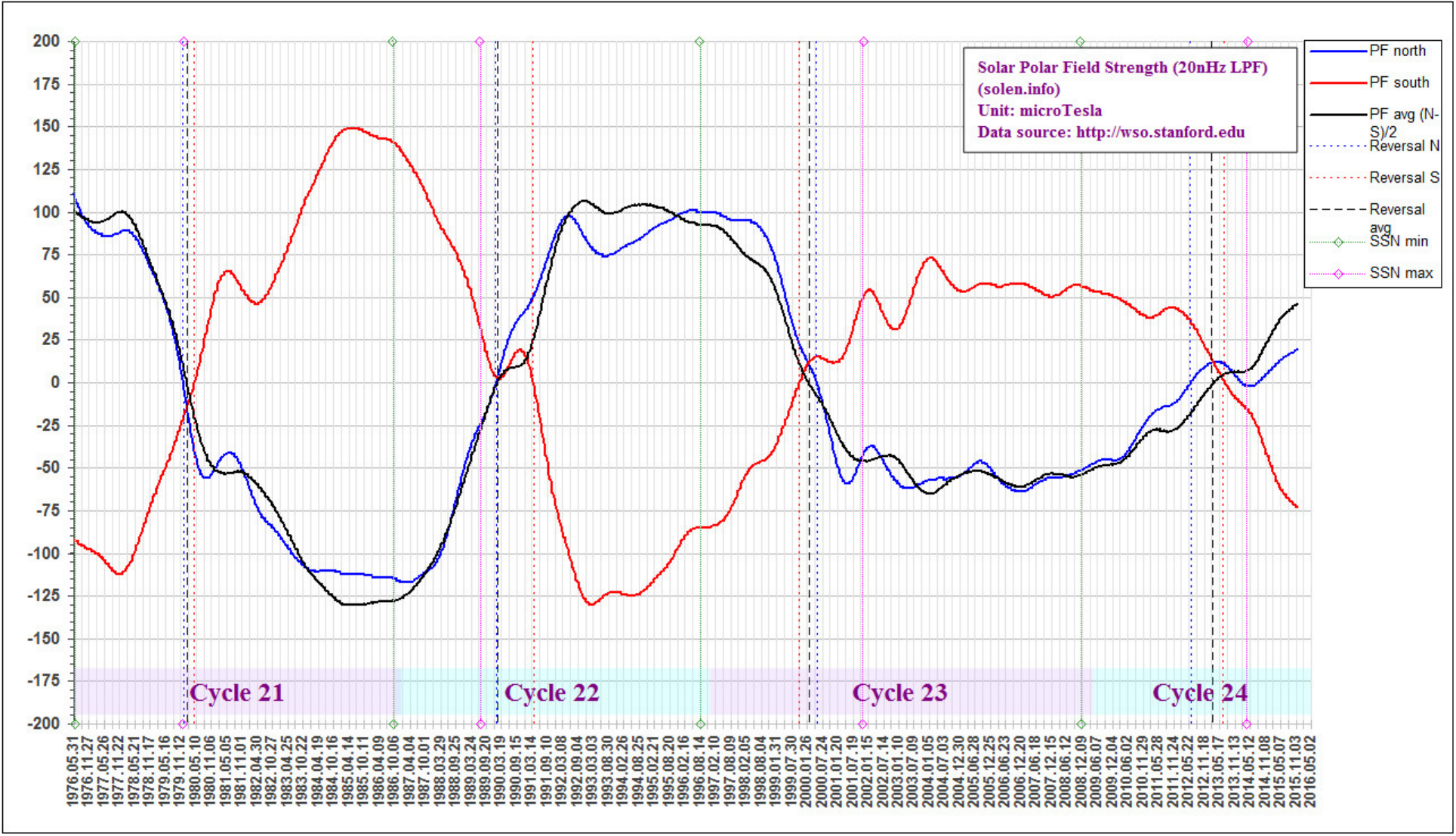} 
\end{center}
\caption{Polar field amplitudes for cycles 21-24 from the Wilcox Solar
Observatory (WSO). Shown are North and South poles separately and together. 
Credit: Hoeksema and WSO website. \label{fig:dynamo-3}}
\end{figure}

Figure \ref{fig:dynamo-3} shows a trace of the polar field amplitude for cycles 21 
to the present \citep[see also][]{Petrie2015}. 
We see that the strength of this polar field itself 
varies from cycle to cycle, and for the past few cycles has shown a downward 
trend. The transition from one polarity to the other takes place over a fairly 
small fraction of a cycle, followed by a longer period over which the field is 
often (but not always) nearly constant. There are differences in timing of 
polar field reversal between North and South, and occasionally there can be 
multiple reversals in field -- but always an odd number, so that the basic 
pattern of field reversal is sustained. It is clear from Figure \ref{fig:dynamo-2} that a 
mechanism for producing different polar field amplitudes in succeeding cycles 
is variations in the amount of surface magnetic flux that migrates to the 
poles in each cycle. If the new cycle is weaker than the previous one, then 
there is less new flux available to reverse the polar field, so when it does 
reverse the new polar field peak is likely to be weaker than the previous one 
(and vice versa).

\begin{figure}
\begin{center}
\includegraphics[clip,width=0.6\textwidth]{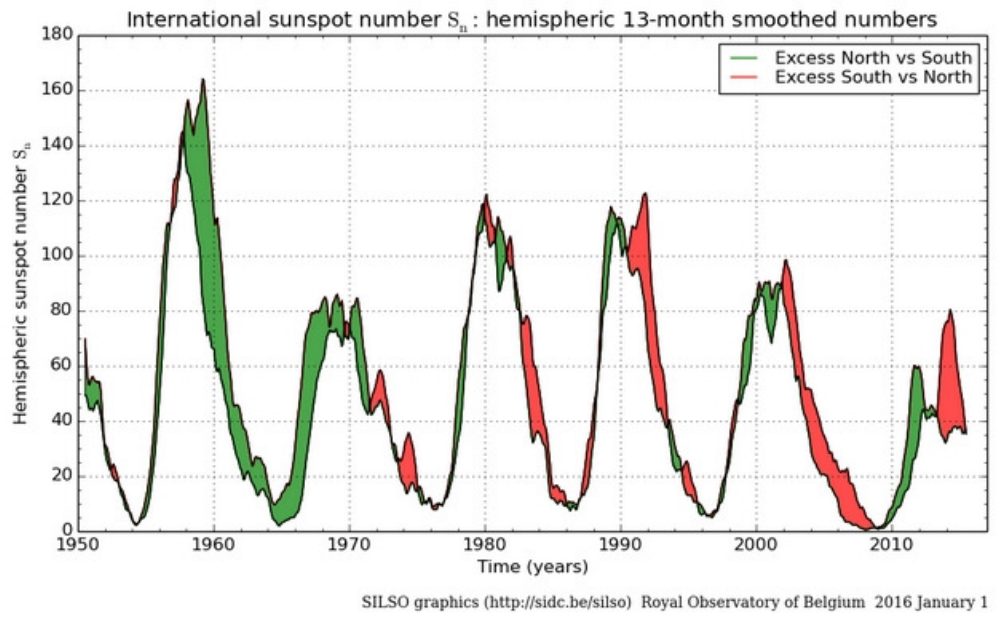} 
\end{center}
 \caption{Excess hemispheric sunspot number, North (green) over South (red), 
 showing that differences in phase between hemispheres can persist for multiple 
 cycles. Source: Royal Belgian Observatory SILSO website. \label{fig:dynamo-4}}
\end{figure}

Some properties of sunspot cycles, such as amplitude and phase, can vary 
substantially between North and South hemispheres. If one correlates cycle
peak amplitude between North and South, the correlation coefficient is about 
0.5 for 3 month running averages, rising to over 0.8 for running averages
exceeding 1.5 years (20 solar rotations). Thus, on short 
time scales the amount of activity can be very different between hemispheres. 
But even over a whole cycle there are substantial differences. Figure \ref{fig:dynamo-4} 
shows hemispheric sunspot number, with the excess of one hemisphere over the other 
shaded in green or red, depending on which hemisphere has more activity. We 
see that for several successive years one hemisphere has consistently more 
flux than the other.

\begin{figure}
\begin{center}
\includegraphics[clip,width=0.6\textwidth]{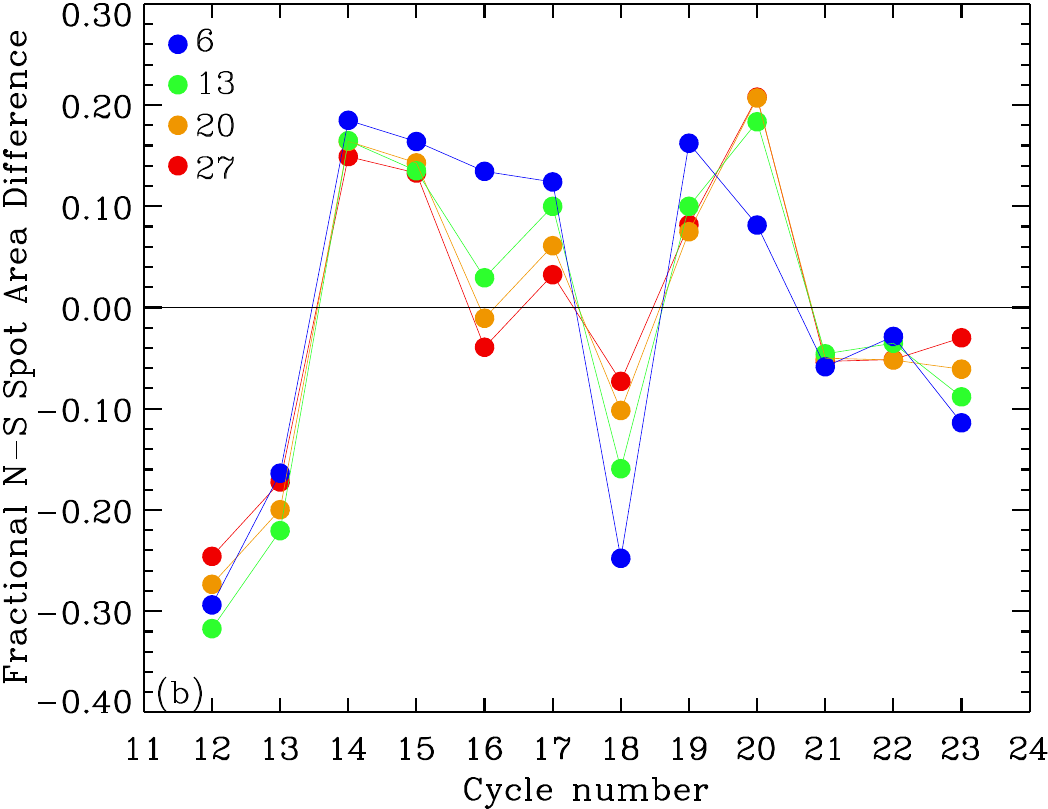} 
\end{center}
\caption{ Fractional difference between spot area in North and South, adapted 
from \citet{Dikpati2007}. Shows that the difference can be up to $30\%$ of 
the average between hemispheres, and persist for multiple cycles. Color key 
with numbers denotes the length of the running average used, in solar rotations, 
showing that the basic result is evident for all averaging intervals. \label{fig:dynamo-5}}
\end{figure}

The same sign of the difference often persists for more than one cycle.
Figure \ref{fig:dynamo-5} shows the fractional difference in sunspot area between hemispheres
for whole sunspot cycles, starting with cycle 12. We see that one hemisphere
can be consistently stronger than the other for more than one cycle, but 
followed by an abrupt switch to the opposite hemisphere. It appears that each
hemisphere has some memory of the strength of the previous cycle in that 
hemisphere, but there must be enough connection between hemispheres to cause
a reversal in amplitude difference.

\begin{figure}
\begin{center}
\includegraphics[clip,width=0.8\textwidth]{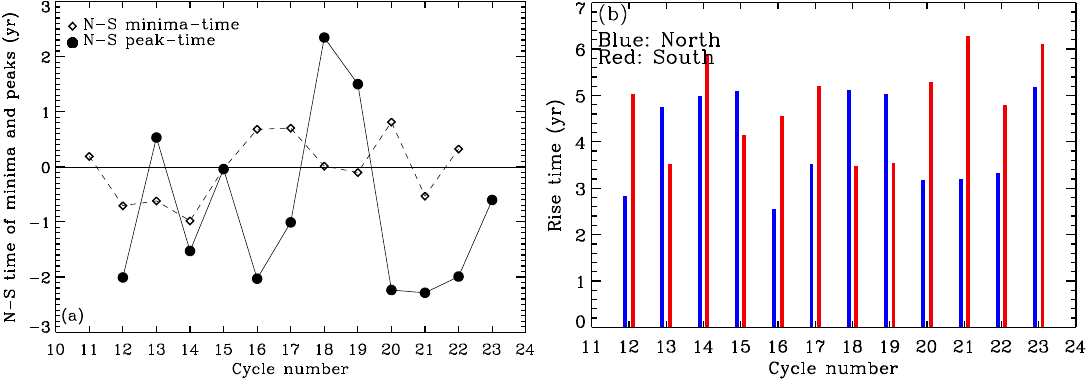} 
\end{center}
\caption{Difference between North and South in timing of peaks and minima, from
\citet{Dikpati2007}. Frame (a) shows that the difference in timing of cycle peaks
is usually more than one year, often more than two years. By contrast the timing
of minima differ by less than one year. The difference in these timing differences
comes from the differences in rise time of a cycle between North and South,
which is virtually always greater than one year, and sometimes more than two
years. One hemisphere always jumps ahead of the other at the beginning of a new
cycle. \label{fig:dynamo-6}}
\end{figure}

The timing of the peaks also can differ between hemispheres, by up to more 
than 2 years. Figure \ref{fig:dynamo-6}a shows this clearly. In fact, it can be argued that
the North and South hemispheres avoid having synchronized activity
peaks. By contrast, the timing of minimum in the two 
hemispheres varies by no more than a year, and usually less. So some
interaction between hemispheres tends to re-synchronize the cycles by the end 
of a cycle. This has to occur during the declining phase, because, from Figure
\ref{fig:dynamo-6}b, the ascending phase is always much shorter in one hemisphere than the 
other, commonly by substantially more than a year. 

Which hemisphere ascends faster typically persists for more than one cycle. 
These statistics also indicate that there is a relatively weak but not 
insignificant interaction between the hemispheres that keeps the timing of 
cycle minimum fairly well in phase, but with each hemisphere pursuing its own
pace within the cycle. Thus the `shape' of a sunspot cycle can be rather 
different between North and South. Correctly simulating differences between 
hemispheres is a significant challenge for dynamo models; succeeding in 
simulating them would be a strong indication of the validity of the dynamo 
model used.

\begin{figure}
\begin{center}
\includegraphics[clip,width=0.6\textwidth]{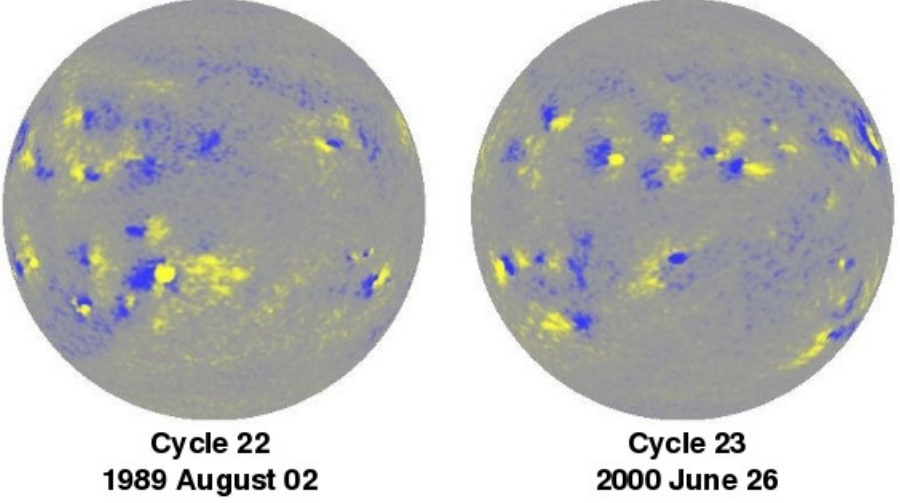} 
\end{center}
\caption{Typical synoptic magnetograms from cycles 22 and 23 (credit: 
Hathaway NASA/ARC). Hale's polarity laws and Joy's laws are very evident.
\label{fig:dynamo-7}}
\end{figure}

The properties of sunspot cycles described above are all axisymmetric, but
we know that the emergence of solar activity in a cycle is fundamentally
longitude-dependent, and this longitude dependence must be taken into account
in solar dynamo models. Figure \ref{fig:dynamo-7} shows typical examples of solar magnetic
fields on the visible disk near the maximum of cycles 22 and 23. Again,
blue and yellow shading denote opposite field polarities. In Figure \ref{fig:dynamo-7} we see 
clearly Hale's polarity law, which says the leading and following parts of
each active region have opposite polarities in North and South hemispheres,
which polarities reverse from one cycle to the next. We also see that all
active regions by eye are tilted with respect to latitude circles, such that
the follower polarities are closer to the poles, on average, than are the 
leader polarities. This tilt, called Joy's law, is crucial for determining 
the sign of the net magnetic flux that migrates to the poles to reverse the 
polar fields seen in Figure \ref{fig:dynamo-3}. All dynamo models must take this tilt 
into account, either by including it in parameterizations of surface poloidal 
flux emergence (2D models) or actually calculating it, in a way that is 
calibrated to observed tilts. 

The average amount of this tilt is now well known from observations of active
regions \citep[see, e.g., figure 1 of][]{Wang2015}. The angle is greatest at the 
start of the cycle, declining nearly linearly to the end. There are some 
differences between cycles, but in the averages these are relatively small. 
Different data sets give similar results. Particularly when coupled with 
statistics for the latitude centroid of sunspot occurrence, the 
combination of tilt angle, centroid and cycle amplitude can give us good 
estimates of emerged surface poloidal flux available for migration to the 
poles to reverse the polar fields.

There are other measures of longitude dependence in surface magnetic fields
that are important for dynamo models to simulate. These involve particularly
the tendency for new magnetic flux to emerge in the same longitude band for
many months, the so-called active longitudes. These persistent features are
known to lead to global longitude dependent magnetic structures in the corona
and interplanetary medium, the so-called magnetic sectors. 

Finally, it is known that solar cycles can seem to virtually disappear for
several decades, an extreme variation in the envelope of cycle amplitudes.
The so-called Maunder minimum is the most recent solar manifestation of this
phenomenon, for which there were very few spots seen on the Sun from about 
1645-1715. There is some evidence that the cycle continued, but at a very 
low amplitude. Simulating these multi-cycle minima is a particularly difficult 
challenge for dynamo models, particularly from actual solar data.

\subsection{Recent models and results}

\begin{figure}
\begin{center}
\includegraphics[clip,width=0.9\textwidth]{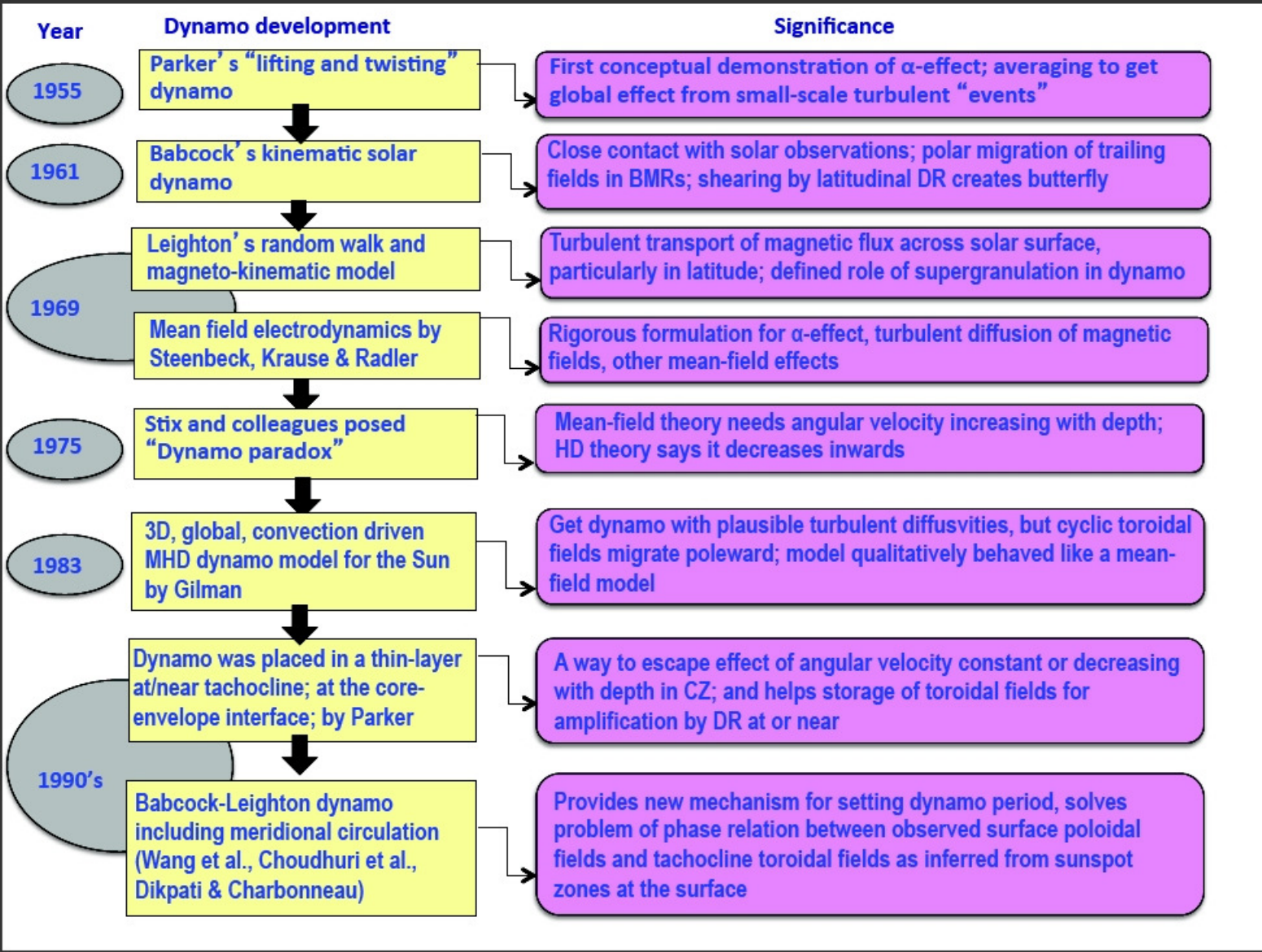} 
\end{center}
\caption{Flow-chart of the history of development of solar cycle dynamo models
through the 1990's, from \citet{Parker1955b}. \label{fig:dynamo-8}}
\end{figure}

\subsubsection{Brief history of solar dynamo model developments}

A brief history of developments in solar dynamo theory is diagrammed in Figure
\ref{fig:dynamo-8}. 
The modern era of solar dynamos really began with Parker's `lifting and twisting' 
dynamo, which showed how a combination of helical flow and differential rotation 
could produce a `dynamo wave' that propagated in latitude with time, with the 
right choice of helical flow and differential rotation, could roughly simulate 
the `butterfly diagram' seen in Figure \ref{fig:dynamo-1}. This model was followed 
by more heuristic 
models of \citet{Babcock1961} and \citet{Leighton1964, Leighton1969} 
that used observed patterns 
and migration of fields to create magnetic cycles. The rise of `mean-field' 
dynamo theory from the German school \citep{Radler2007} gave solar dynamo theory 
much more rigor and connection to MHD turbulence concepts. 

Mean-field dynamos gave plausible cycles and butterfly diagrams provided the 
angular velocity increased inwards in the bulk of the convection zone, but this 
picture was overturned by the discoveries of helioseismology (see 
\S \ref{section:helioseismology}). And the first full 
3D convectively driven solar dynamo of \citet{Gilman1983} gave plausible differential 
rotation but an `anti-solar' butterfly diagram in which toroidal fields migrated 
toward the poles. These developments led to shifting the focus of the dynamo from 
the bulk of the convection zone to its base. The problem of reversed butterfly 
diagrams was overcome by the demonstration of a new mechanism for toroidal field
migration, provided by meridional circulation 
\citep{Wang1991, Durney1995, Choudhuri1995, DikpatiCharbonneau1999}. Thus was 
born the so-called Babcock-Leighton flux transport dynamos, which have proved so 
far to be the most successful dynamos for simulating solar cycles 
\citep{Nandy2001, DikpatiGilman2001, Dikpati2004, Guerrero2004, MunozJaramillo2009, Hotta2010, Belucz2015}. Extensive references on BLFT dynamos can be found in 
\citet{Charbonneau2010}. These are 
2D (latitude-radius) models, but efforts have begun to generalize them to 3D.

\begin{figure}
\begin{center}
\includegraphics[clip,width=0.8\textwidth]{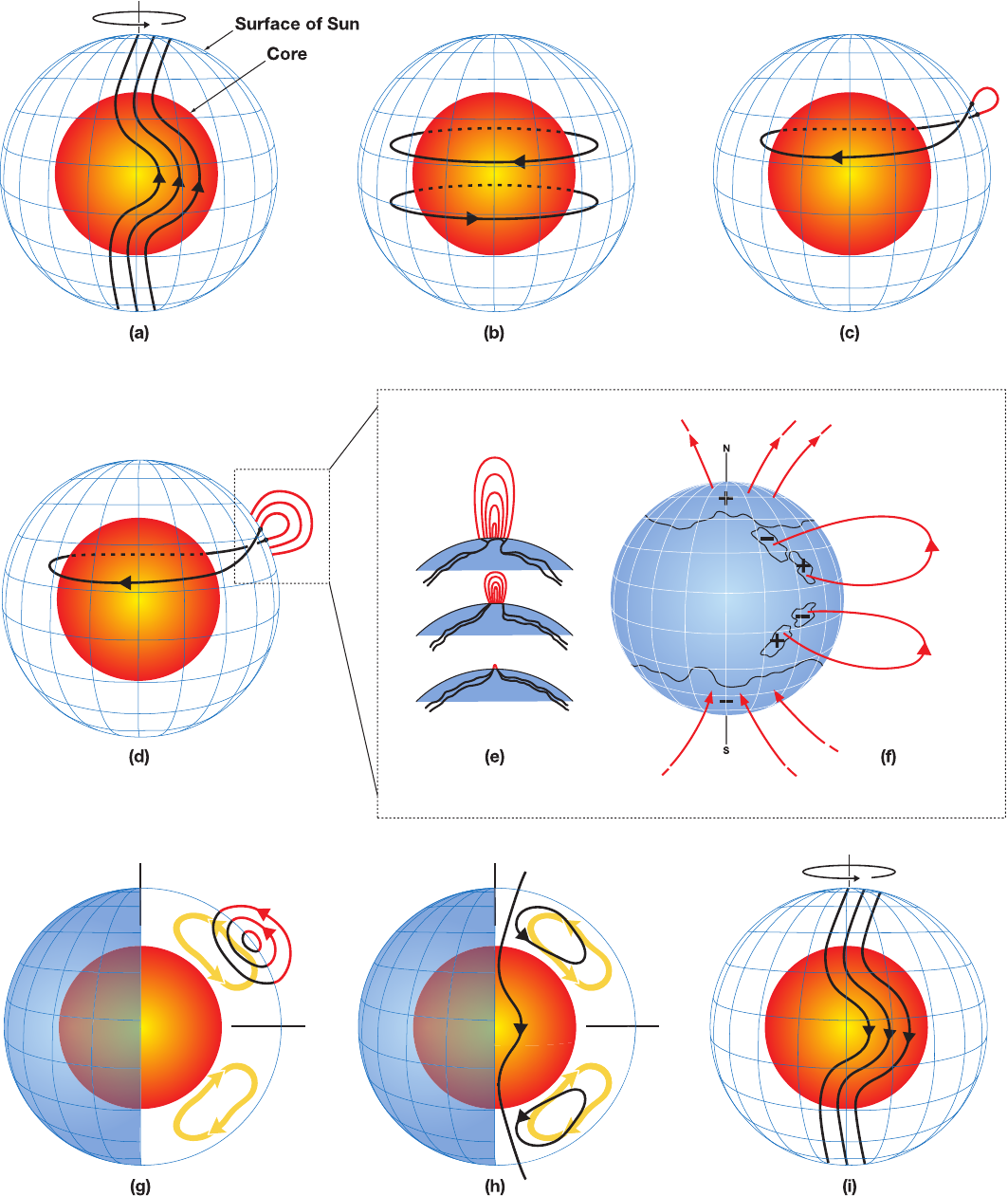} 
\end{center}
\caption{Schematic sequence of induction and transport processes contained in
Babcock-Leighton flux-transport dynamo models, adapted from \citet{DikpatiGilman2006}. \label{fig:dynamo-9}}
\end{figure}

Figure \ref{fig:dynamo-9} provides a schematic diagram that shows how this dynamo works, from
\citet{DikpatiGilman2006}. It begins with the shearing of an antisymmetric
poloidal field by the latitudinal differential rotation (frames a,b) to produce a 
toroidal field near the base of the convection zone where the turbulent 
magnetic diffusivity is relatively low. This is followed (frames c,d) by
toroidal flux loops rising to the surface while twisting, creating new
poloidal loops there (frames e,f), which are tilted with respect to a latitude
circle. These then contribute to a new axisymmetric poloidal field (frame g)
that is carried toward the poles by meridional circulation and then down into 
the convection zone to its base (frame h), which has the opposite sign from the
original poloidal field (frame a), from which a new toroidal field, also of
opposite sign, is generated, leading to a repeat of the whole process, with all
magnetic field signs reversed.

In actual dynamo simulations with the flux-transport dynamo equations, if
they are done with a full spherical shell with no symmetry conditions imposed
at the equator, there is no guarantee that the model will select the correct
field symmetry for the Sun. In fact, various calculations showed a tendency
to pick the opposite, or `quadrupole' symmetry. \citet{DikpatiGilman2001} 
looked into this question in detail, and found that even if the simulation
was started with the correct solar symmetry, if only the surface 
Babcock-Leighton mechanism for forming surface poloidal flux from emerging toroidal 
fields (Figure \ref{fig:dynamo-9}d-f) was present, within a few hundred years of simulation 
the symmetry switched over to quadrupole type. In this situation, a butterfly 
diagram remains plausible, but Hale's polarity law is violated. \citet{DikpatiGilman2001} 
showed that if there was a second mechanism for generating poloidal 
from toroidal field, such as the lifting and twisting ($\alpha$-effect) due to 
global hydrodynamic instability of differential rotation in the tachocline, the 
original dipole type symmetry was retained. The effect of the bottom 
$\alpha$-effect preserves the original symmetry to provide a booster to 
the poloidal field brought down from surface polar regions, which is then swept
toward the equator before it can dissipate, getting in synch with the other 
hemisphere to create poloidal field lines that cross the equator rather than 
closing there within each hemisphere. The equator-ward flowing meridional 
circulation is a key factor, since it brings stronger poloidal field from 
high latitudes to merge near the equator with its opposite-hemisphere
counterpart. Without the bottom $\alpha$-effect, there is little poloidal field 
available to merge with its opposite-hemisphere counter part through the equator
to form a dipole structure, which can produce antisymmetric toroidal field 
after being sheared by the differential rotation.

Symmetry selection processes have been studied by many others later on 
\citep{Bonanno2002, Chatterjee2004, Hotta2010, Belucz2015}, 
and proposed and tested additional possible mechanisms to ensure the correct
symmetry is selected.

\subsubsection{Calibrated and benchmarked models}

\begin{figure}
\begin{center}
\includegraphics[clip,width=0.6\textwidth]{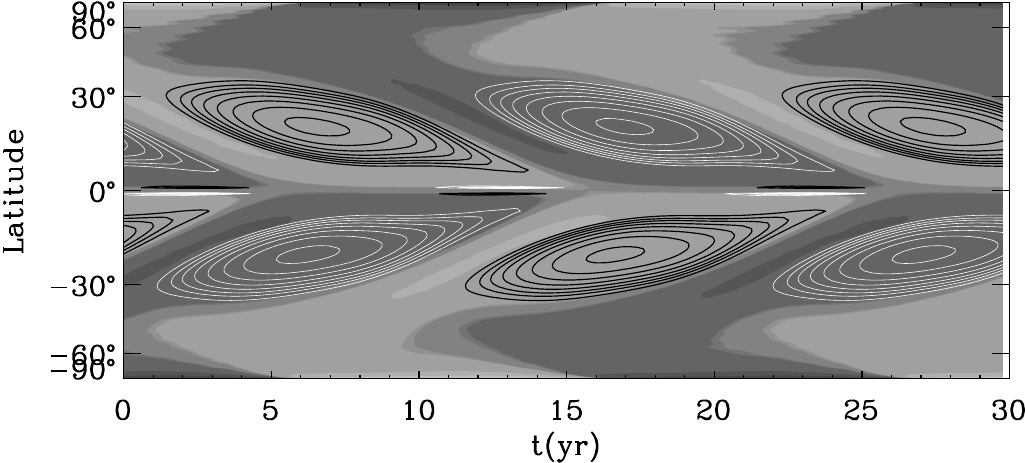} 
\end{center}
\caption{Calibrated butterfly diagram from a Babcock-Leighton flux-transport dynamo. 
Shading is surface poloidal field (light and dark
shading opposite polarities); contours are for toroidal field amplitude at the bottom
of the dynamo domain. This diagram closely resembles the observed diagram shown in
Figure \ref{fig:dynamo-2}. The narrow black/white streaks at the equator near the end 
of each cycle
are the weak remains of the toroidal field from the previous cycle, which are slow 
to cancel across the equator because of the low turbulent magnetic diffusivity at 
the bottom. \label{fig:dynamo-10}}
\end{figure}

It makes sense to calibrate a dynamo model to solar cycle properties only if
the model at least gives the correct dominant symmetry of magnetic fields
about the equator. Figure \ref{fig:dynamo-10} shows a calibrated butterfly diagram, in which
the spot-producing toroidal fields, taken from the tachocline, are shown as 
contours, the surface poloidal field as light and dark shading. We see that 
there is a strong resemblance between Figure \ref{fig:dynamo-10} and 
Figure \ref{fig:dynamo-2}. The shape of 
both field patterns is about right, the polar fields reverse near the maximum 
in toroidal fields, and reach maximum in between sunspot cycles. Eight versions 
of this class of model, using different numerical algorithms, have been 
benchmarked against each other for certain parameter choices in common, and 
have been found to agree very closely \citep[see][]{Jouve2008}.

\subsubsection{Solar cycle simulation and prediction}

The quality of the calibration suggested that the flux-transport dynamo model
was ready to use to attempt the dynamo-based simulations of particular
solar cycles, and even the first dynamo-based prediction of a solar cycle.
Dikpati and colleagues \citep{DikpatiGilman2006, Dikpati2006} 
used the same calibrated flux-transport dynamo model that generated 
Figure \ref{fig:dynamo-10} to 
simulate the peaks of cycles 12-23, and predict the peak amplitude of the then 
future cycle, cycle 24. For cycles 12-23, they achieved a correlation between 
the observed and simulated peaks in excess of 0.94 for all turbulent magnetic 
diffusivities chosen, up to $3 \times 10^{11} {\rm cm}^2\,{\rm s}^{-1}$. To do 
these calculations, data `nudging' was used to drive the model with the observed 
time history of surface poloidal flux. Thus, magnetic surface magnetic fields 
from previous cycles led to the amplitude of the next cycle. When North and South 
hemispheres were simulated separately \citep{Dikpati2007}, the model produced 
the larger observed differences between hemispheres. The correlation between 
simulated and observed peaks was somewhat lower, but still in excess of 0.8. 
Simulations of this length were possible because for the turbulent magnetic 
diffusivities chosen, the `memory' of the model was about two sunspot cycles. 
In the diffusivity range chosen, the dynamo was definitely in the 
advection-dominated regime. However, the duration of the model's memory may 
not be settled yet, as \citet{Yeates2008} produced only one sunspot cycle.

Unfortunately, the observed peak in cycle 24 has turned out to be substantially 
lower than predicted by Dikpati and colleagues. Also using a flux-transport dynamo
model, but higher turbulent magnetic diffusivities and input of surface poloidal 
fields only near the poles and only near solar minimum, \citet{Choudhuri2007} 
predicted a much lower peak for cycle 24, actually lower than observed.  
Their model is in the diffusion-dominated regime, with a much shorter
memory. 

There is currently no consensus as to what the turbulent magnetic diffusivity
is in the convection zone, so it is hard to choose among models that use
different values. But there is a problem with high diffusivity models as 
true self-contained dynamo models, in that it can be shown by scale analysis
that for diffusivities in excess of $10^{12} {\rm cm}^2 \, {\rm s}^{-1}$, such 
as suggested by mixing length applied to the Sun, and as used by \citet{Choudhuri2007}, 
it would take a very unrealistically large surface poloidal source 
or $\alpha$-effect to sustain a dynamo at all. 

One way to see this is by calculating the critical dynamo number $P$ for the 
model, which must be exceeded for a dynamo to be sustained, following, for 
example, \citet{Stix1976}. The original argument is for $\alpha$-$\omega$ dynamo 
models, but it holds equally as well for flux-transport dynamos. 
$P=\alpha_0 \Delta\omega \lambda^3 r_{\odot}^3/\eta_T^2 $,
in which $\alpha_0$ is the $\alpha$-effect, $\Delta\omega$ a measure of the 
difference in rotation across the shell, $\lambda$ the shell thickness in 
fractions of its radius, $r_{\odot}$ the solar radius, and $\eta_T$ the 
turbulent magnetic diffusivity. It is well established that $P$ should 
exceed $3 \times 10^3$ for sustained dynamo action. For a mixing-length amplitude 
diffusivity $\sim 3\times10^{12} {\rm cm}^2 \, {\rm s}^{-1}$, this requires 
$\alpha_0\sim 1.5\times10^3\, {\rm m}{\rm s}^{-1}$, which is completely unrealistic 
for any scale of solar convection anywhere in the convection zone, or for any \
surface poloidal source. By contrast, $\eta_T\sim 10^{11} {\rm cm}^2 \, {\rm s}^{-1}$ 
would require $\alpha_0 \sim 1.5 \,{\rm m}{\rm s}^{-1}$, a much more plausible 
value. In fact, this is what was argued by \citet{Choudhuri1995} in their 
very first Babcock-Leighton flux-transport dynamo model, and therefore a 
value of $3\,{\rm m}{\rm s}^{-1}$ was carefully considered. However,
Choudhuri and colleagues developed the so-called high-diffusivity dynamo models 
by choosing a high diffusivity to poloidal fields, but a low diffusivity for 
toroidal field, along with the use of an unrealistically high $\alpha$-effect, 
to obtain a sustained dynamo solution. It remains a confusion whether a dynamo 
model with high and low diffusivities applied to respectively the poloidal and 
toroidal fields can be called a diffusion-dominated model or a dynamo with 
diffusion-dominated poloidal-field and advection-dominated toroidal-field.  

If one uses the unrealistically high surface poloidal source, with high 
diffusivity, the calibration is negated because the resulting dynamo 
period is much too short. The \citet{Choudhuri2007} prediction model gets 
around these problems because it is being forced at the top using 
magnetic data from previous cycles. But the Sun does not have that freedom; 
it must be a self-contained calibrated dynamo to begin with.

\begin{figure}
\begin{center}
\includegraphics[clip,width=0.6\textwidth]{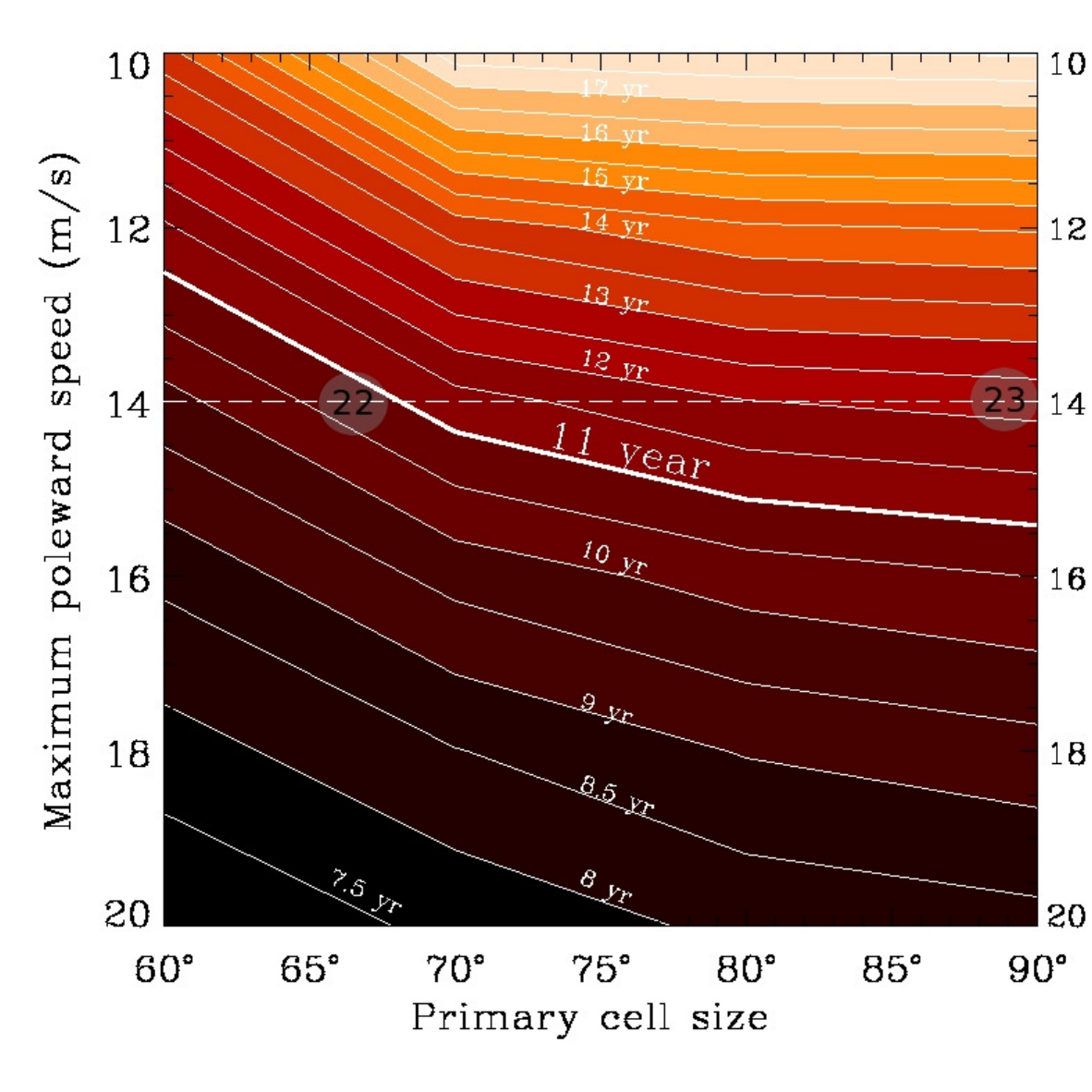} 
\end{center}
\caption{Relationship between latitude of poleward boundary of primary 
meridional circulation and period of magnetic cycles produced by 
Babcock-Leighton flux-transport dynamo. Adapted from \citet{Dikpati2010}.
\label{fig:dynamo-11}}
\end{figure}

In any case, it is more important to analyze reasons why the advection-dominated
flux-transport dynamo-based prediction of cycle 24 amplitude have not succeeded. 
These have to do with the simplifying assumptions that went into the early 
models. In particular, the early prediction models used a steady single celled 
meridional circulation, but we now know that the circulation amplitude and 
profile both vary with time within a solar cycle. Even allowing just for a 
change in the high latitude boundary of the primary meridional circulation 
flow toward the poles is enough to explain the 2-year longer duration for 
cycle 23 compared to cycles 21 and 22 \citep{Dikpati2010}. In the earlier 
cycles this cell reached to about $65^{\circ}$ latitude, while in cycle 23 most 
of the time it reached all the way to the poles. The longer conveyor belt in 
cycle 23 led to a longer cycle, because it took longer for the surface poloidal 
fields in active latitudes to be transported to the poles and down to the 
bottom of the convection zone. Figure \ref{fig:dynamo-11} shows a plot of the relationship 
between cycle period, peak circulation speed, and the latitude of the 
poleward boundary of the circulation.

\begin{figure}
\begin{center}
\includegraphics[clip,width=0.8\textwidth]{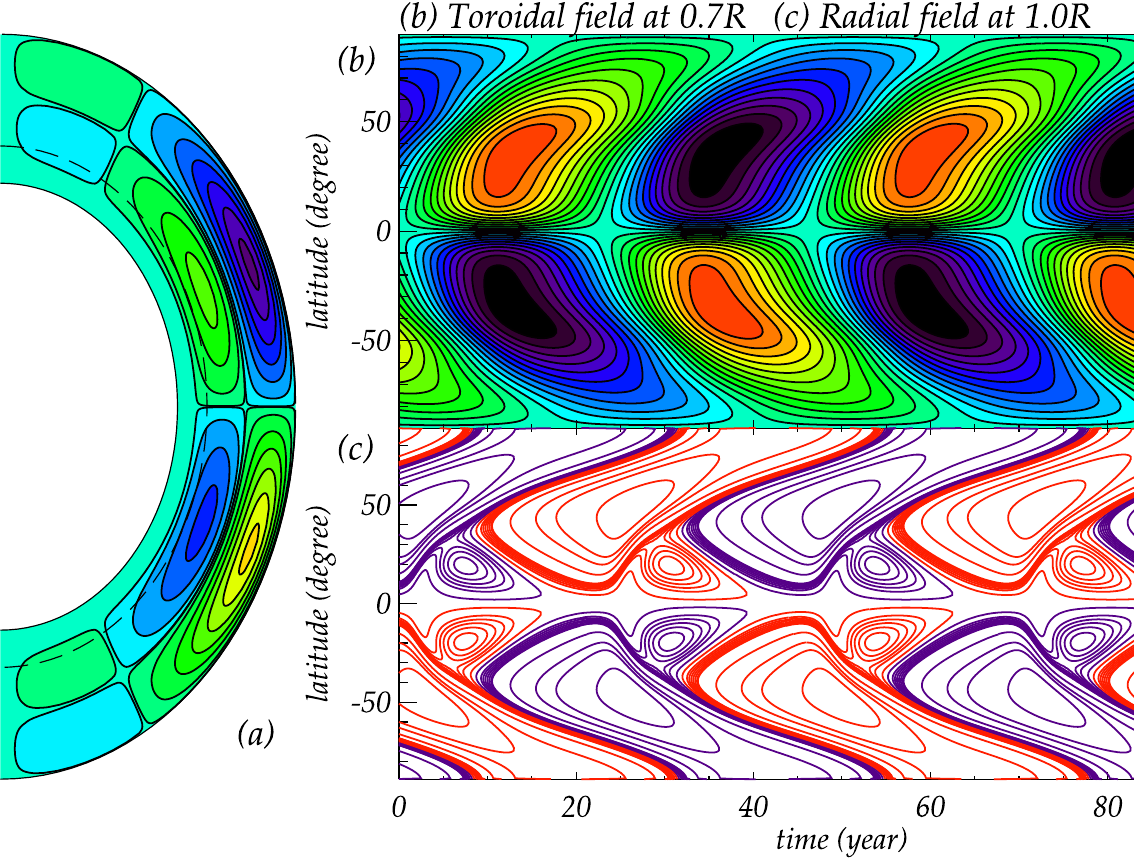} 
\end{center}
\caption{Butterfly diagrams (frames b,c) from flux-transport dynamo containing two 
primary  meridional circulation cells stacked in depth (frame a). Adapted from \citet{Belucz2015}. 
It is evident that two cells in depth leads to a non-solar (reversed) butterfly diagram, 
particularly for the toroidal (sunspot) fields.
This is because the toroidal field at the bottom where it is strongest is being carried
toward the poles rather than the equator. \label{fig:dynamo-12}}
\end{figure}

Recently there has developed evidence of a possible second, reversed meridional
circulation cell below the primary cell that flows poleward, carrying the
surface poloidal flux to the poles to reverse the polar field. Although 
there exists some debates about the existence of the weak reverse cell 
\citep{Kholikov2014}, if such a two-celled pattern persists in the Sun, then the calibration 
of flux transport dynamo models to the Sun is completely destroyed. An example 
is given in Figure \ref{fig:dynamo-12}, from \citet{Belucz2015}. 
Because in this case the bottom 
circulation is toward the poles rather than the equator, as it would be with a 
single circulation cell with depth, the butterfly diagram constructed from the 
bottom toroidal fields is completely reversed from the real Sun. If this is 
what is happening in the Sun, then the Sun must not be a Babcock-Leighton 
flux transport dynamo at all, and there must be another paradigm shift in 
solar dynamo theory. But observations of the second cell with depth are quite 
uncertain, and so it is premature to conclude such a shift is needed.

Beyond changes in meridional circulation, several other factors not previously
accounted for in solar dynamo models may also contribute to the lack of success
in solar cycle prediction models. Asymmetries between North and South 
hemispheres are substantial and virtually always present; the early models 
treated a single hemisphere. There is still uncertainty about the profile of 
turbulent magnetic diffusivity with depth. The data nudging scheme used in 
early prediction models did not make use of much of the available solar 
observations, which could change the predictions substantially. As discussed 
in section \ref{section:new_challenges}, modern data assimilation methods need to 
be incorporated into 
the models to use all available data. Finally, a substantial part of observed
magnetic fields and patterns of flux emergence are longitude-dependent, and all 
prediction models so far are axisymmetric. For example, creating active region 
tilts is strongly longitude-dependent as well as highly variable in time.
Active longitudes, while varying on longer time scales are also obviously
functions of longitude.

\subsubsection{Beginnings of application of data assimilation}

Application of data assimilation (DA) methods to the solar dynamo problem is in its very
early stages. As a precursor, \citet{DikpatiAnderson2012} have found from numerical
experiments that the `response time' of a Babcock-Leighton flux-transport dynamo
model to a change in meridional circulation amplitude is about six months. This is a
key time scale for designing an optimum DA scheme, since the time interval for
updating data needs to be long enough that the model has time to start responding 
to the change. In a more recent study, \citet{Dikpati2014} have 
carried out OSSE's that show it
is possible to reconstruct time changes in meridional circulation from the magnetic
fields generated by the dynamo, using a relatively limited number of `observations',
synthetically generated from the model. This kind of reconstruction will be 
necessary \citep[see also][]{Hung2015} because there do not now exist consistent 
observations of meridional circulation below the solar surface, including near 
the bottom of the convection zone and in the tachocline, where the circulation 
plays a key role in moving the dynamo-generated toroidal field.

\subsubsection{Flux emergence and active longitudes}

\begin{figure}
\begin{center}
\includegraphics[clip,width=0.95\textwidth]{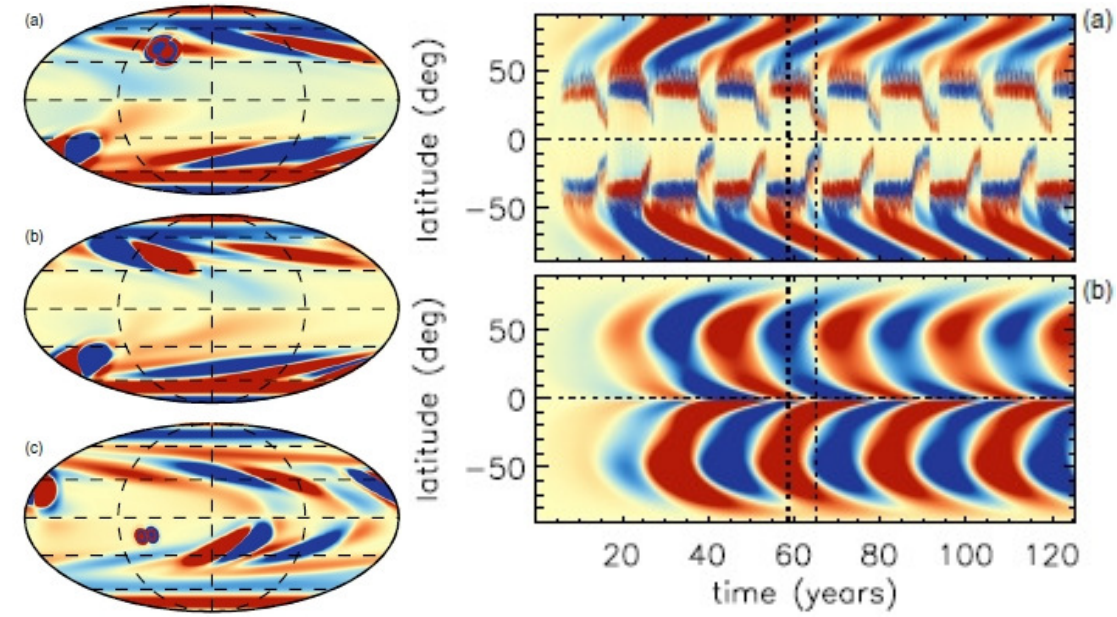} 
\end{center}
\caption{ Early results from  a 3D Babcock-Leighton flux transport dynamo model
(BASH model), adapted from \citet{Miesch2014}. Frames (a)-(c) on the left show typical
evolution of the surface radial field originating from newly emerged flux. Frames
(a) and (b) on the right show, respectively, time-latitude butterfly diagrams of 
the surface poloidal 
and bottom toroidal fields for a 120 year model run. \label{fig:dynamo-13}}
\end{figure}

Work has begun on incorporating longitude dependent flux emergence into flux
transport dynamos. For example, using a `spotmaker' recipe that generates 
statistically appropriate active region tilts, \citet{Miesch2014}
have created a 3D Babcock-Leighton type dynamo which generates a time sequence
of longitude-dependent domains of surface flux similar to those in the original 
Leighton model, as seen in Figure \ref{fig:dynamo-13}a. Averaging in longitude and 
time over this
sequence creates a surface poloidal flux source for the 2D Babcock-Leighton 
flux-transport dynamo. Solutions from this dynamo contain plausible butterfly
diagrams for surface radial field and toroidal field near the base, seen 
respectively in Figures 13a,b. 

The longitude locations where new magnetic flux emerges appear not to be random;
certain locations are repeated sources for new flux, which can persist for many 
months, even years. \citet{DikpatiGilman2005} proposed a theory for these `active
longitudes' in terms of global hydrodynamic (HD) and MHD instabilities occurring in the solar
tachocline. Because the upper boundary of the tachocline can deform, instability
of the combined latitudinal differential rotation and toroidal field there creates
`bulges' of the tachocline into the convection zone above that could be favored 
locations from which toroidal flux can rise through the convection zone to the
photosphere according to the physics of flux emergence (see \$ \ref{section:FE}). 
This kind of effect must be incorporated into global 
flux-transport dynamo models for generalizing them to 3D.

Recently \citet{McIntosh2015} have shown that many manifestations of solar 
activity show global periodicities with periods in the range 6-18 months. This 
range of periods may also be due to global dynamics and MHD of the solar 
tachocline. \citet{Dikpati2018a} showed, using a nonlinear MHD shallow water model of 
unstable latitudinal differential rotation in the tachocline, that there is a 
well defined nonlinear oscillation between energy in the differential rotation and 
in the low longitudinal wave number `Rossby waves' that gain energy from it. When 
these waves extract enough energy from the differential rotation, it ceases to be 
unstable, and subsequently extracts energy from the disturbances, growing until 
it becomes unstable again. In this way, the differential rotation and perturbation 
energies oscillate out of phase with each other, with the total energy of the 
system remaining virtually constant. Similar oscillations should still occur 
when toroidal fields are added; we would expect that the maximum amount of new 
flux would emerge at the surface when the perturbations and therefore the bulges 
have maximum amplitude.

\subsection{Outstanding challenges to be explored}

The overriding priority for the 21st century solar cycle dynamo models is to 
build a 3D calibrated dynamo. 3D is essential for simulating the 
substantial departures from axisymmetry seen in typical synoptic magnetograms 
produced starting in the 1950's, and now available many times per solar rotation 
from both the Heliospheric and Magnetic Imager \citep[HMI:][]{2012SoPh..275..207S} onboard the Solar Dynamics Observatory (SDO) and from ground-based observing systems such as the National Solar Observatory's Synoptic Optical Long-term Investigations of the Sun (SOLIS) facility and the Mount Wilson Observatory (MWO). 
Calibration, including for longitude-dependent features, is essential for 
determining that the model simulates typical solar field patterns through a 
solar cycle and that the model can be used to simulate past cycles as well as applied 
to forecasting future cycles.

In principle, there are at least two approaches that could be taken. One is to 
use available full 3D convectively driven MHD dynamo models; the other is to 
generalize an already calibrated 2D Babcock-Leighton flux transport model to 
include longitude-dependent fields and flows. The latter approach is advantageous
over the former, because no current full 3D dynamo model has been successfully 
calibrated for the Sun. 

What form should this generalization take? It should be strongly influenced by 
solar observations. For example, the typical solar synoptic magnetogram shows 
photospheric magnetic fields have truly global patterns, with high amplitude 
in low longitudinal wave-numbers $m$. Only a few low $m$'s can capture much of 
the important departures from the axisymmetric or $m=0$ part of the field. In 
addition, as we go above the photosphere into the corona and beyond, the lowest 
$m$'s dominate even more in the magnetic structure. Solar `sectors' generally 
are well represented by just $m=1,2$ modes. For these reasons, the model that 
is built should allow for picking just the lowest wave-numbers, and allow 
experimentation to determine how few are needed. Therefore, fast-Fourier 
transform representation of interactions between modes should be avoided in the 
formulation.

It is straightforward to expand the kinematic induction equation to include 
longitude dependence of this type. In the linear case, one simply gets separate 
equations for the amplitude of each longitudinal wave-number, including the 
original 2D $m=0$ mode. Linear solar dynamo models of this type were first 
studied in the early 1970's by \citet{Stix1971}, but have not been pursued that much since. 
In these models, the flow fields included were still axisymmetric, as was the 
parameterized helical turbulence included. For future models, we have strong 
motivation, and theoretical support, for going far beyond this limited 3D model,
to include velocity fields that are themselves dependent on longitude. The 
theoretical support is provided by the substantial theory of global HD and MHD 
in the solar tachocline that has been developed over the past 20 years
\citep{Gilman1997, DikpatiGilman1999, DikpatiGilman2001, Cally2001, Garaud2001, CallyDikpati2003, Cally2003, Dikpati2012, Dikpati2017, Dikpati2018a, Dikpati2018b}. 
This theory also favors the existence of global helical velocity
and magnetic field disturbances in the the tachocline, for which the same low 
longitudinal wave-numbers $m=1,2$ are the most unstable, and therefore the most 
likely to dominate in the tachocline and layers above. The only exception to that 
is if the toroidal field present there is confined to a narrow ($< 10^{\circ}$) 
band of latitudes. In this case, somewhat higher $m$'s are also excited.

In concentric spherical shells, starting from the tachocline, the model would 
include 3D helical flow from global tachocline instabilities, from which bulges 
into the convection zone above would be generated. In the bulk of the convection 
zone there would be differential rotation, meridional circulation, global helical 
flows and buoyantly rising magnetic flux (see e.g. \S \ref{section:FE}). Near the 
outer boundary there would be Babcock-Leighton surface poloidal field source, 
generated from newly emerged flux from active longitude locations. 

Unstable modes in the tachocline can have at least three important effects on the 
dynamo. First, they can `imprint' longitude dependent patterns on 
the bottom of the convection zone, particularly in magnetic field, which can be 
transmitted to the photosphere where they can be seen. Second, they can provide 
limited longitudinal bands where upward bulges in the tachocline can appear, which, 
when they coincide with the presence of strong toroidal fields, can provide
favored sites for toroidal flux tubes that become buoyant and rise to the 
photosphere. Third, they also are a powerful source of kinetic helicity for 
the dynamo itself, supplementing the Babcock-Leighton type surface poloidal 
source in creating, amplifying and evolving the dynamo's poloidal field. All 
three of these effects could be important in the solar dynamo, so it is 
essential that their effects be included in the model.

It is feasible to incorporate all three of these effects into a global 3D 
flux-transport dynamo model. The imprinting of magnetic and velocity patterns 
at the bottom of the dynamo domain follows directly from including 
non-axisymmetric flow and fields from solutions to the linear instability 
equations, with assigned amplitudes. Amplitudes for such modes can be calculated 
directly from nonlinear MHD models for global tachocline instabilities \citep{Dikpati2018a}. 

This 3D class of model will make much heavier use of solar observations, 
particularly at the surface; incorporating all this data into the model in a 
sensible way will require sophisticated data assimilation algorithms, most likely 
of the sequential type, to update the model integrations. The choice of update 
interval will be constrained by the time resolution of the observed data as well 
as the 'response' time of the dynamo model to changes in an input, such as 
meridional circulation. Experience so far with simpler 2D models with data 
assimilation point to an update interval of 15-30 days, which is consistent with 
the time cadence of the solar observations to be used.

The initial 3D models should be kinematic; they should solve just the induction 
equation. In later versions it should be possible, and desirable, to couple the 
induction equation to equations that calculate the global tachocline instability 
modes. The toroidal field produced by the dynamo as a function of time can be 
incorporated into linear and nonlinear models of tachocline instability to 
calculate how the unstable modes evolve. Then these can be included in 
computational updates of the induction equation, that include all changes in the 
dynamo inputs.

While mean field dynamo models have been quite successful in simulating 
important solar cycle magnetic features, they still rely on assumptions and 
parametric representations of certain physical processes they can not 
resolve spatially. Full 3D convective dynamo simulations for the Sun, 
starting with \citet{Gilman1983}, and continuing to this day by Toomre and 
colleagues with the various Anelastic Spherical Harmonic (ASH) codes
\citep{Browning2006, Augustson2015}, 
as well as by \citet{Racine2011} and \citet{Fan2014}, have produced
cycles, but yet to be calibrated with solar cycle features. In addition, 
there is a third major track of model developments that avoid the 
complications of spherical geometry and work in boxes 
\citep{Tobias2008, Tobias2011, Tobias2013, Cattaneo2014, Rempel2014, Pongkitiwanichakul2016, Nigro2017}, that focus in greater 
detail on important processes contributing to dynamo action, including rotation, 
shear, convection, and various types of turbulence, including magneto-rotational 
turbulence. These studies allow development of greater physical insight into the 
interplay of various important physics, which may lead to better estimates of various 
parameters used in mean field models, as well as better understanding of 
full 3D spherical shell MHD dynamo simulations. All three of these tracks are 
likely to contribute to the `ultimate', or physically most realistic, solar 
cycle dynamo model.

Once a 3D global solar dynamo as described above is produced, which calibrates 
well to the Sun, then it will be possible to extend such a model into the corona 
to calculate dynamo-produced coronal structures with longitude dependence that 
can be compared with observed coronal structures. Then it should be possible to 
relate these structures to features such as coronal holes, polar structures, 
sectors, and high speed solar wind streams (see Chapters 3-7). Finally, if the model 
calibrates well to the Sun, it can be tested for its ability to predict both 
longitude-averaged and longitude dependent features of future solar cycles, 
using real solar data.

In closing we judge that enormous progress has been made in modelling the solar 
dynamo, including simulating important features of solar cycles. But even now 
there is much to do before there is a consensus that the solar cycle problem has 
been `solved.' It remains a very exciting challenge, one for which powerful tools 
available now should help us reach that goal.

\section{Helioseismology (Rachel Howe) \label{section:helioseismology}}

%\subsection{This is Second Level Heading}
%
%\subsubsection{This is Third Level Heading}
%
%\paragraph{This is Fourth Level Heading}
%
%\subparagraph{This is Fifth Level Heading}

\subsection{Basic Principles}
A wealth of information comes from the layers of the Sun's atmosphere, but how can we study what is happening below the opaque photosphere? Helioseismology is one of the few ways in which observers can probe the Sun's interior, using sound waves that travel through it.

\begin {figure}
  \begin{center}
    \includegraphics[width=0.45\linewidth]{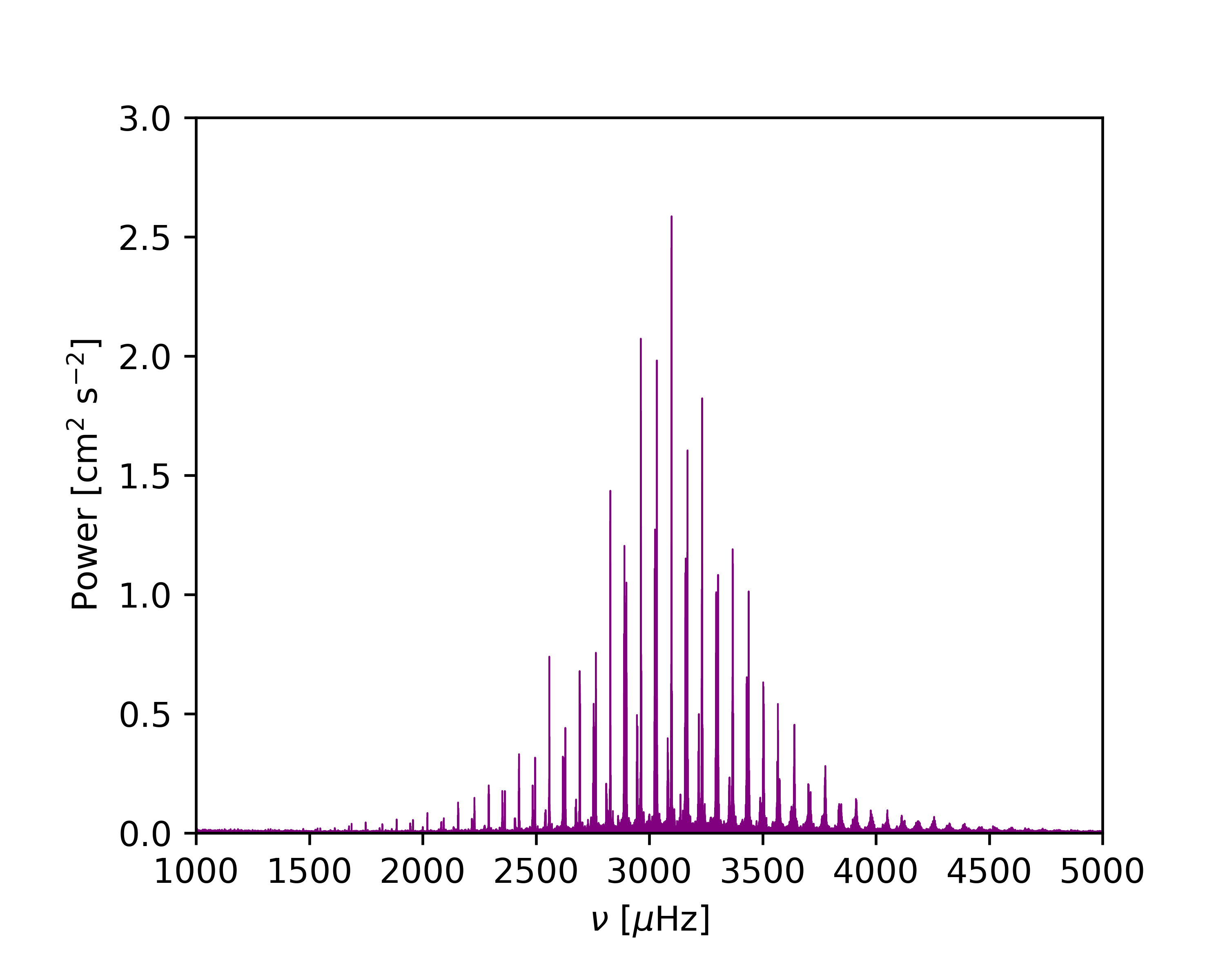}
    \includegraphics[width=0.54\linewidth]{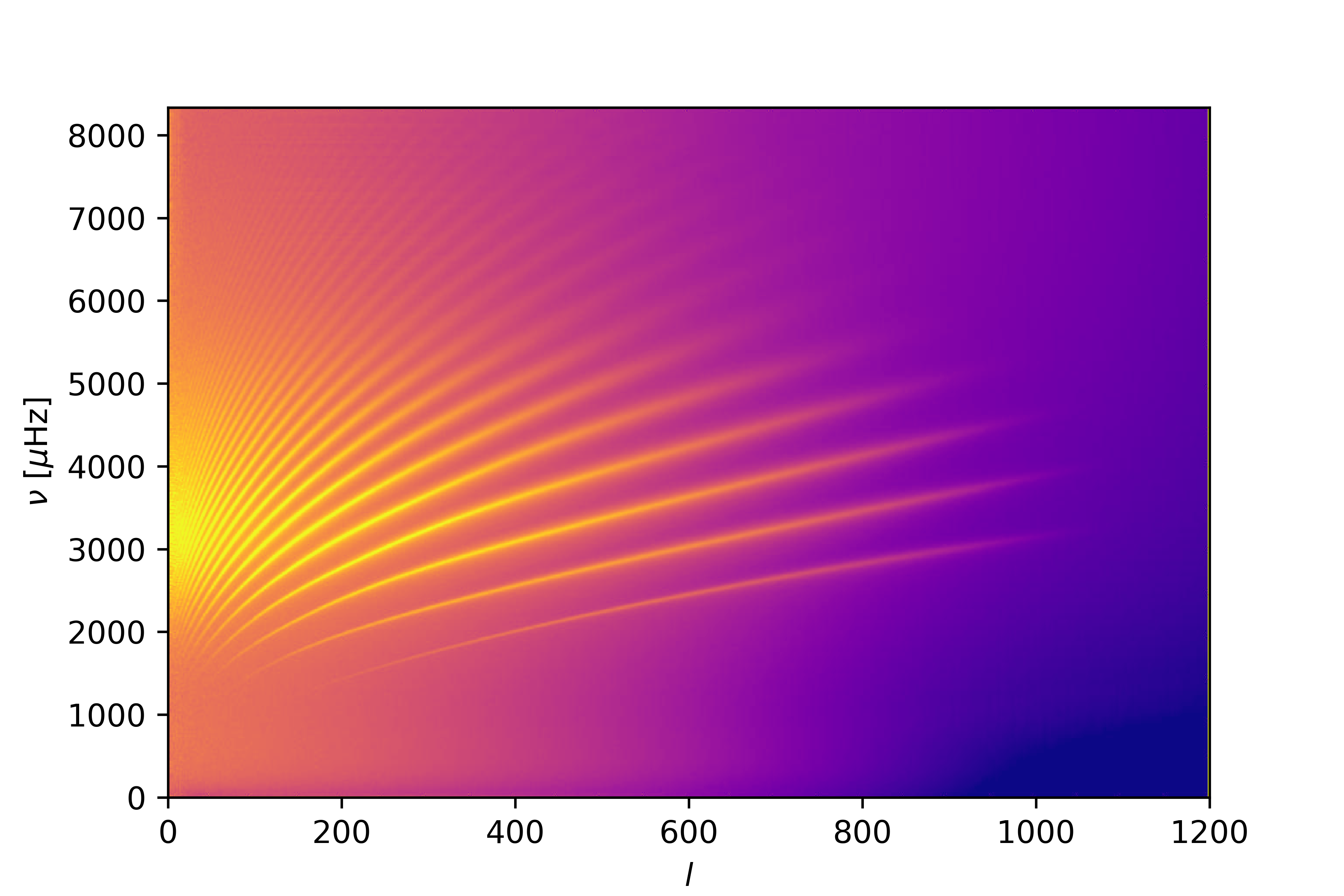}
    \end{center}
  \caption{Left: Sun-as-a-Star power spectrum from the BiSON network. Right: Rotation-corrected, $m$-averaged power spectrum in the $l-\nu$ plane, from one day of GONG observations. Each ridge corresponds to a single radial order $n$. Original image courtesy GONG/NSO/AURA.}
  \label{fig:rhfig7}
  \end{figure}

The surface of the Sun oscillates with a period -- or rather, many periods -- of around five minutes, corresponding to a frequency of about 3\,mHz. \citet{1975A+A....44..371D} showed that the power in ``five-minute oscillations'' of the solar surface, when plotted in the wave-number--frequency plane, revealed a structure that identified them as nonradial acoustic eigenmodes of the solar interior (see Figure~\ref{fig:rhfig7} for a more modern example). The cavity in which these ``$p$-modes'' (so-called because their restoring force is pressure) propagate is bounded by the solar surface and by the depth at which inward-propagating waves are refracted back towards the surface. The depth of this inner turning point varies monotonically with $\nu/L$, where $\nu$ is frequency and $L$ is related to the degree $l$ by $L\equiv\sqrt{l(l+1)}$; the radial modes $(l=0)$ penetrate to the core, while modes of $l=100$ penetrate only about $0.05R_{\odot}$ below the surface. Both ``local'' and ''global'' modes of oscillation exist; the global modes are those whose lifetime is longer than the time to make a complete circuit of the Sun, so that they can form a coherent global standing-wave pattern, while local modes -- in general, the modes of shorter wavelength -- are damped out more quickly and are coherent only over a small region of the solar surface. Both are used in helioseismology.

The acoustic modes can be described by spherical harmonics of degree $l$, azimuthal order $m$, and radial order $n$. Modes with $n=0$ are a special case known as $f$-modes; these are surface gravity waves confined to the outer layers of the Sun. In the case of perfect spherical symmetry, modes of the same $l$ and $n$ but different $m$ would be degenerate, but this degeneracy is lifted both by rotation and by the more subtle violation of spherical symmetry caused, for example, by the latitudinal distribution of magnetic activity. This makes it possible to use inversion techniques to infer a two-dimensional profile of the interior rotation, as we will discuss below. 

\subsection{Observations}

The acoustic modes are observed by monitoring velocity -- with Doppler techniques -- or brightness variations at the photosphere, either in integrated light (``Sun as a star'') or using imaging. Sun-as-a-star observations are limited to the lowest-degree modes, mostly those of $l \leq 3$. Early results came from relatively short series of observations from a variety of single instruments -- see, for example, the summary given by \citet{Howe2009}, of which perhaps the best known is that operated at the Big Bear Solar Observatory from 1986\,--\,1990 by Libbrecht and co-workers \citep[e.g.][]{1990Natur.345..779L}. For ground-based observations it is preferable to use a global network of observing sites, to reduce the systematic errors introduced by interruptions. The Birmingham Solar-Oscillations Network
\citep[BiSON:][]{hale16}
has been operating as a six-site network for low-degree helioseismic observations since 1992, with sparser observations stretching back into the late 1970s; Figure~\ref{fig:rhfig7}(left) shows a sample BiSON spectrum. For medium- and (later) high-degree observations (Figure~\ref{fig:rhfig7}(right)), the Global Oscillations Network Group
\citep[GONG:][]{1996Sci...272.1284H}
has been in operation since 1995. Another strategy for obtaining near-continuous data is to observe from space. MDI \citep{1995SoPh..162..129S} observed medium-degree oscillations from SOHO almost continuously from 1996\,--\,2011, with a few months of observations at high degree each year. MDI was then superseded by the HMI \citep{2012SoPh..275..207S} instrument onboard SDO, which has been in use since 2010.

\subsection{Inversions}

The principle on which helioseismic inversions are based is that each observation $d_i$ (for example a frequency, rotational splitting, or rotational splitting coefficient) represents a spatially weighted average of the quantity of interest over the solar interior, with an added error term $\epsilon_i$.

\subsubsection{Rotation Inversions}
\begin{figure}
  \includegraphics[width=\linewidth]{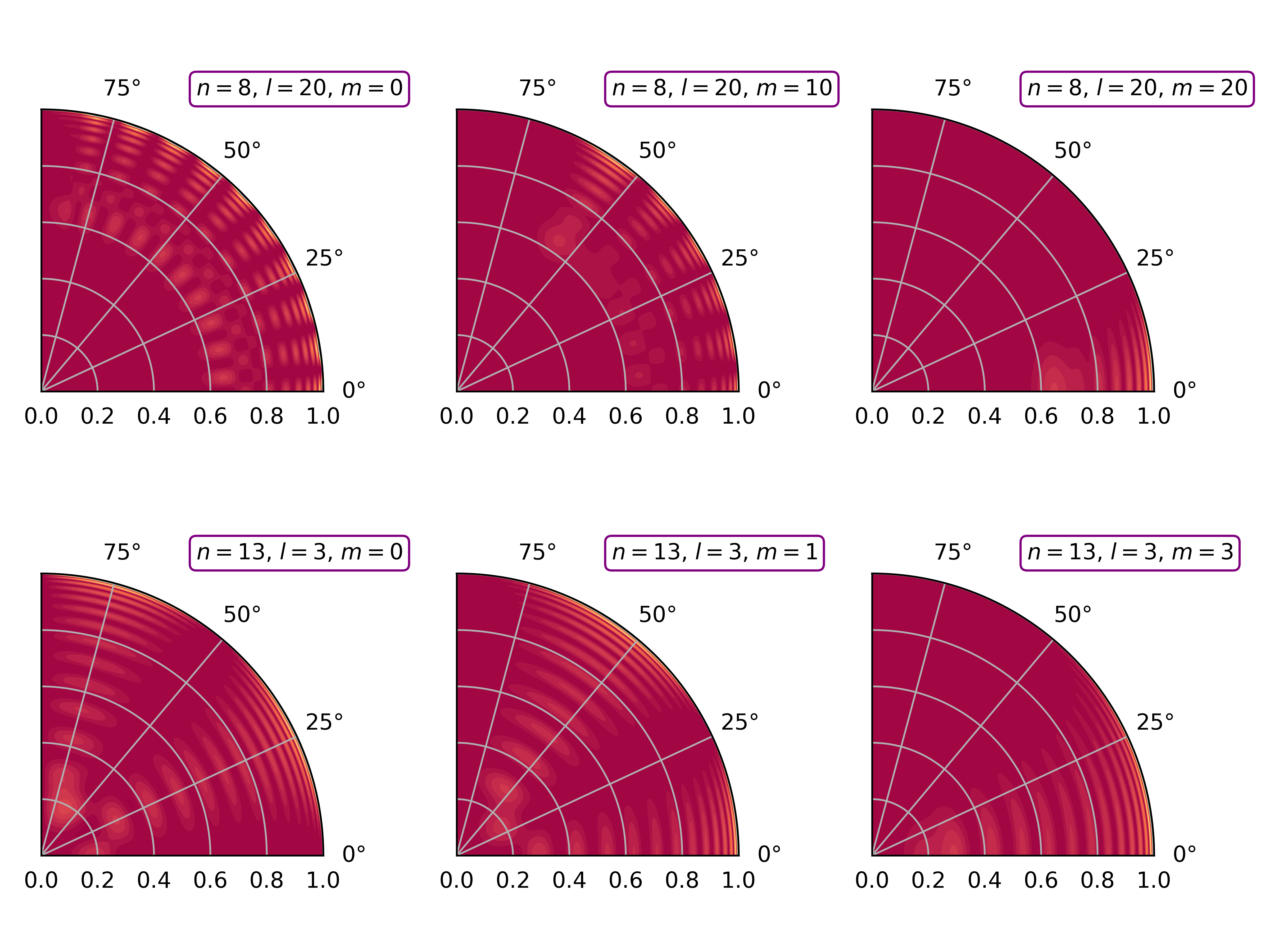}
  \caption{Sample rotation kernels based on Equation~\ref{eq:rheq1b}}
  \label{fig:rhfig4}
  \end{figure}

In the case of a rotation inversion this can be written in the form
\begin {equation}\label{eq:eqrh1}
d_i=\int_0^{R_\odot}\int_0^\pi{K_i(r,\theta)\Omega(r,\theta) dr d\theta} +\epsilon_i,
\end{equation}
where $R_{\odot}$ is the solar radius, $r$ and $\theta$ are respectively the distance from the center of the Sun in units of the solar radius and the colatitude, and $K$ is the kernel, a spatial weighting function that depends on the solar model being used  \citep{1977ApJ...217..151H,1980A+A....89..207C}.
 \citet{1994ApJ...433..389S} write
the 2d-rotation kernel of an $(n,l,m)$ mode as  
\begin{equation*}
K_{nlm}(r,\theta)= \frac{m}{I_{nl}}\left\{{{\xi_{nl}(r)\left[{\xi_{nl}(r)-\frac{2} {L}\eta_{nl}(r)}\right]{P_l^m(x)^2}}}\right.
\end{equation*}
\begin{equation}\label{eq:rheq1b}
+\frac{\eta_{nl}(r)^2}{L^2}\left[\left(\frac{dP_l^m}{dx}\right)^2(1-x^2)
-  \left .{2P_l^m\frac{dP_l^m}{dx}x} + \frac{m^2}{1-x^2}P_l^m(x)^2\right]\right\}\rho(r)r\sin{\theta},
\end{equation}
with
\begin{equation}
I_{nl}=\int_0^{R_\odot}{[\xi_{nl}(r)^2+\eta_{nl}(r)^2]\rho(r)r^2 dr},
\end{equation}
$x=\cos{\theta}$, and $L^2=l(l+1)$, where the radial and horizontal displacements  are
$\xi_{nl}$ and  $L^{-1}\eta_{nl}$ and $\rho(r)$ is the density. $P_m^l$ is an associated Legendre function. Figure~\ref{fig:rhfig4} shows some example kernels.

The inferred rotation rate  $\bar{\Omega}(r_0, \theta_0)$ at the location $(r_0,\theta_0)$ can be written as 
\begin{equation}\label{eq:eqrh2}
 \bar{\Omega}(r_0, \theta_0)=\sum_{i=1}^M{c_i(r_0,\theta_0)d_i},
\end{equation}
where the $c_i$ are weighting coefficients to be determined by the inversion.
We can substitute Equation~(\ref{eq:eqrh1}) into the right hand side of Equation~\ref{eq:eqrh2} to obtain
\begin{equation}\label{eq:eqrh3}
\bar{\Omega}(r_0,\theta_0) = \int_0^{R_\odot}\int_0^\pi{{\cal K}(r_0,\theta_0;
r,\theta)\Omega(r,\theta) dr d\theta} +\epsilon_i,
\end{equation}
where
\begin{equation}\label{eq:eq4}
  {\cal K}(r_0,\theta_0;r,\theta) \equiv \sum_{i=1}^M{c_i(r_0,\theta_0) K_i(r,\theta)}
\end{equation}
is called the averaging kernel for location $(r_0,\theta_0)$. The averaging kernels do not depend on $d_i$, but they are affected by the uncertainties in $d_i$, as these are used to weight the data in the inversion calculation.

The formal uncertainty on the input profile is given by 
\begin{equation}
\sigma^2[\Omega(r_0,\theta_0)]=\sum_i[c_i(r_0,\theta_0)\sigma_i]^2.
\label{eq:erreq}
\end{equation}
where $\sigma_i$ are the uncertainties on the input data, provided that these are uncorrelated and normally distributed.

There are several possible approaches to solving the basic inversion problem. One commonly used method is  ``regularized least squares'' (RLS), which uses, essentially, a least-squares fit with a penalty term to impose smoothness to find the rotation profile that best fits the input data \citep{1994ApJ...433..389S}. Another approach is ``subtractive optimally localized averaging'' (SOLA) \citep{1968GeoJ...16..169B,1970RSPTA.266..123B,1992A+A...262L..33P,1994A+A...281..231P}, where the quantity to be minimized is the difference between the actual averaging kernels and a target kernel. For examples of other methods, see \citet{1998ApJ...505..390S}. The localization is never perfect, and there is always a trade-off between error and resolution; as only modes of lower degree penetrate into the deep interior, the resolution that can be achieved by the inversions is poorer at greater depth. For more details, see \citet{Howe2009}.

It is important to remember that these global inversion techniques always average the northern and southern hemispheres; to study the two separately requires local helioseismology or other more mathematically sophisticated techniques.

\subsubsection{Structure Inversions}

\begin{figure}
  \begin{center}
    \includegraphics[width=0.7\linewidth]{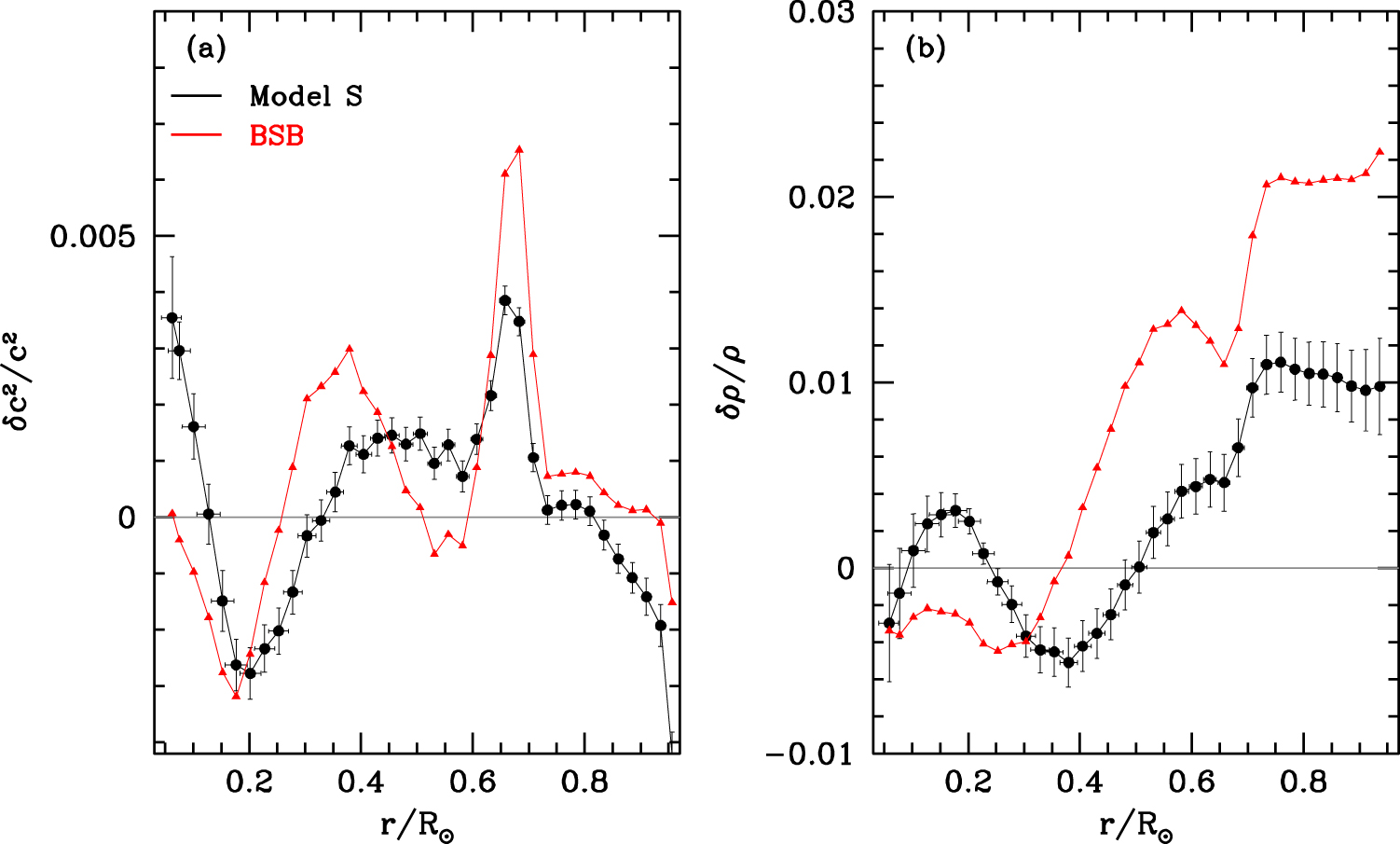}
    \end{center}
    \caption{The relative differences of the squared sound speed between the Sun and two other solar models, Model S
and model BSB. (b) The relative density differences between the Sun and the two models. Reproduced from Figure 3 of \citet{2015ApJ...808..123S}, {\textit {The Astrophysical Journal}}, 808, 123, by permission of the AAS.}
\label{fig:rhfig2}
  \end{figure}

When helioseismologists talk about ``structure'' they generally mean the density and sound speed profiles of the solar interior, or quantities related to these.
The problem of inverting for such quantities is quite complex; for a review, see \citet{Basu2016}. In practice, structure inversions are carried out using not the ``raw'' mode frequencies but the difference between the observed frequencies and those predicted by a model. It is usually necessary to remove a frequency-dependent ``surface term'' that arises because the near-surface layers are not well modeled \citep[see, for example][ and references therein]{2014A&A...568A.123B}.
Once this term has been taken into account, the remaining differences between observation and model are very small -- less than one part in a thousand, as seen for example in Figure~\ref{fig:rhfig2}.
``Model S'' of \citet{1996Sci...272.1286C} is a particularly good match to the data and is therefore popular with helioseismologists; unfortunately, since the model was developed there have been changes to the calculation of element abundances in the Sun that have worsened the agreement between models and data \citep{2005ApJ...618.1049B}, and the discrepancy has proved difficult to reconcile. \citet{2008PhR...457..217B} and \citet{2009ApJ...699.1403B} have suggested that the problem may lie with the revised element abundances.

\subsection{Solar-Cycle Variations}

%\subsubsection{Early Findings}

It has been known since quite early in the history of helioseismology that the frequencies -- and other properties -- of the $p$-modes vary with the solar cycle. \cite{1985Natur.318..449W} looked at the mode frequencies in Sun-as-a-Star observations from the Active Cavity Radiation Monitor instrument onboard the Solar Maximum Mission and found that they were significantly lower at solar minimum than at solar maximum. The result was confirmed by \cite{1989A&A...224..253P} and \cite{1990Natur.345..322E} using ground-based instruments that were to become part of the BiSON network. Around the same time, \cite{1990Natur.345..779L} found variations in medium-degree observations from the Big Bear Solar Observatory over the years 1986\,--\,1990. The medium-degree results showed that not only did the frequencies increase with increasing activity, but that higher-frequency modes experienced greater changes than those at lower frequency, following the variation of the so-called ``mode inertia''. This is a quantity defined by \cite{1986ASIC..169...23C} as being equal to 
\begin{equation}
{E_{nl}}={\frac{\int_0^R{[\xi_r^2(R)+l(l+1)\xi^2_t(R)]\rho r^2 {\mathrm d}r}}{4\pi M[\xi_r^2(R)+l(l+1)\xi^2_t(R)]}},
\end{equation}
with $\xi_r$ and $\xi_t$ being the radial and tangential components of the displacement, $M$ the total mass of the Sun, and $R$ its photospheric radius. The fact that the frequency changes scale in this way shows that they are concentrated in the near-surface layers; higher-degree modes with shallower lower turning points, or higher-frequency modes with shallower upper turning points, are most strongly affected.

%\subsubsection{Localization}

\begin{figure}
  \includegraphics[width=\linewidth]{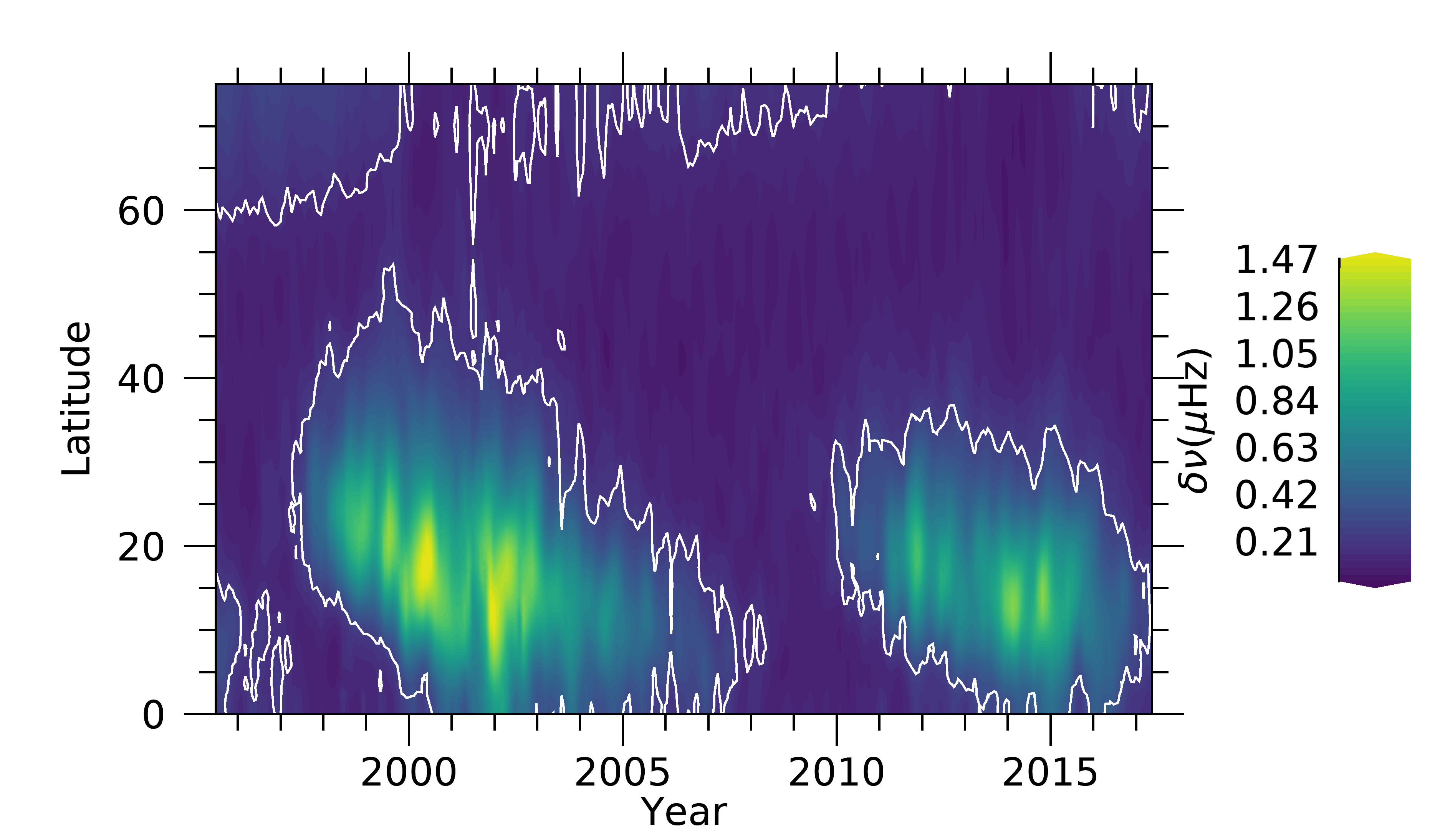}
  \caption{Latitudinal distribution of frequency shifts from GONG, averaged over modes with $40 \leq l \leq 80$ and $9 \leq n \leq 11$, based on the analysis of \citet{2002ApJ...580.1172H}. The white contour represents the 5\,G level of the north-south averaged unsigned longitudinal magnetic field strength from National Solar Observatory synoptic maps.}
  \label{fig:rhfig6}
  \end{figure}

The variation of mode frequencies with activity is not only temporal; it also reflects the geometry of surface activity. In medium-degree modes, the even-order rotational splitting coefficients vary with time in a way that correlates with the coefficients of a Legendre polynomial expansion of the latitudinal distribution of surface activity \citep[e.g.][and references therein]{2001MNRAS.327.1029A,2002ApJ...580.1172H}. \citet{2002ApJ...580.1172H} used a one-dimensional inversion technique to map the latitudinal distribution of the $m$-dependent $p$-mode frequency shifts from GONG observations and showed that they mapped closely to the (north--south symmetrized) magnetic butterfly diagram, as shown in Figure~\ref{fig:rhfig6}

For local modes \cite{2000SoPh..192..363H} showed that the mode frequencies in ``ring-diagram analysis'' -- which uses the 3-dimensional power spectrum of the oscillations in a small patch of the solar surface \citep{1988ApJ...333..996H} -- increased in patches corresponding to active regions, and \citet{2001ApJ...563..410R} also found increased ring-diagram frequencies in active regions, with larger shifts for higher frequencies.

%\subsubsection{Variations in Interior Structure}
While the activity-related changes in mode frequencies are relatively easy to detect, their effect seems to be largely confined to the outer couple of per cent of the solar radius, and in inversions it is dominated by the surface term \citep[e.g][]{2001MNRAS.327.1029A}. It is much more challenging to find signatures of changing magnetic activity deeper in the interior. \citet{2002ApJ...580..574E} placed an upper limit of $3\times 10^{-5}$ on any fractional change in the interior sound speed. The subtle changes found at the base of the convection zone by \citet{2008ApJ...686.1349B} are just within this limit.

\subsection{Rotation}

\begin{figure}
  \begin{center}
  \includegraphics[width=0.8\linewidth]{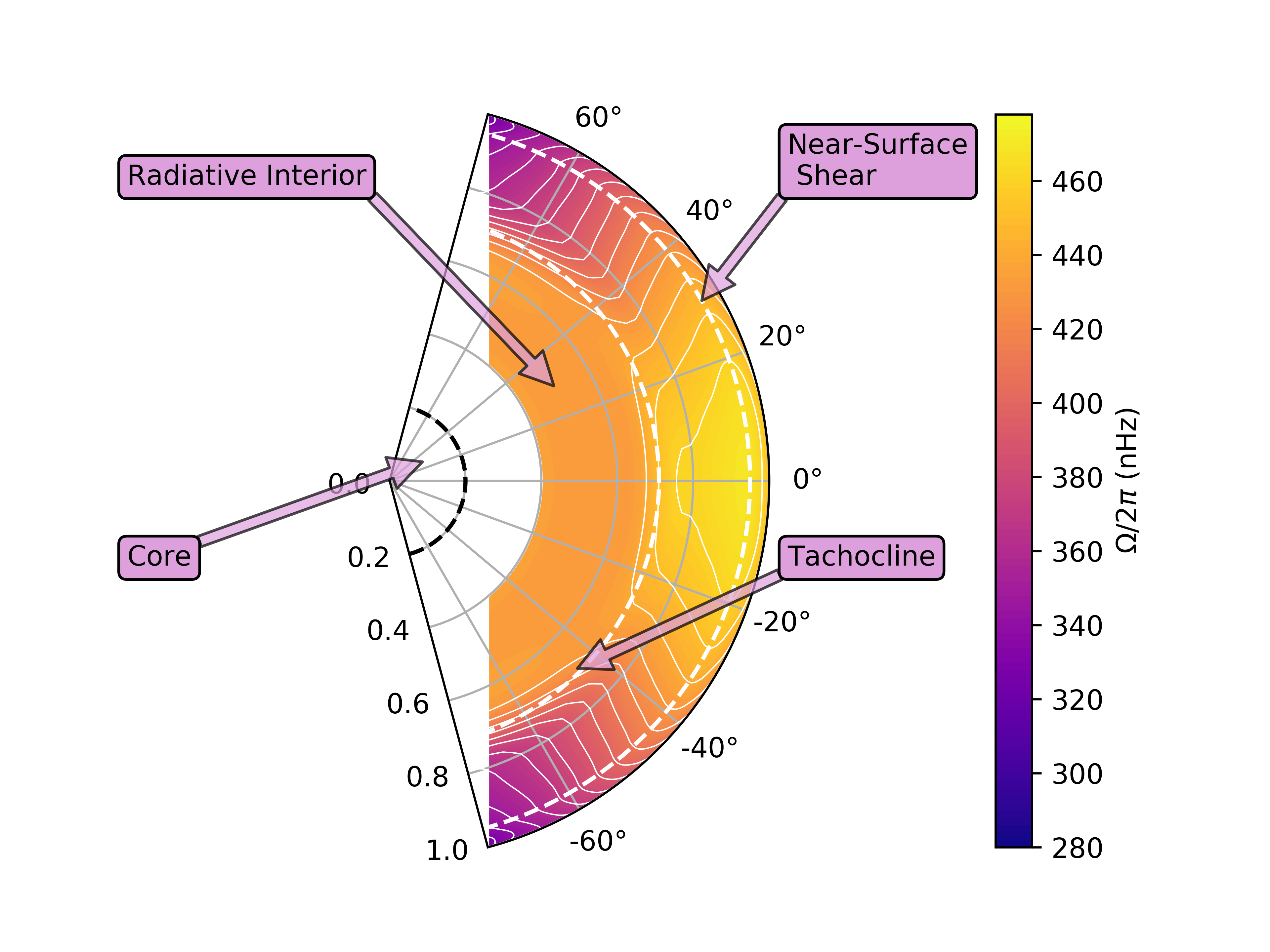}
  \end{center}
  \caption{Solar interior rotation profile, based on the average of 2dRLS inversions of 45 72-day periods of HMI data from 2010\,--2018. The white contours are at intervals of 10\,nHz. The dashed arcs represent the boundary of the core at $0.2\,R_{\odot}$, the base of the convection zone at $0.71\,R_{\odot}$, and the base of the near-surface shear layer at $0.95\,R_{\odot}$.}
  \label{fig:rhfig3}
  \end{figure}

The interior rotation of the Sun can be divided into five regimes, working from the surface inwards: the near-surface shear layer, the bulk of the convection zone, the tachocline, the radiative interior, and the core (see Figure~\ref{fig:rhfig3}). The maximum rotation rate at the equator corresponds to a speed of about 330 ms$^{-1}$. We consider each of these regimes in turn.

\subsubsection{Near-Surface Shear}
The rotation rate increases inwards in about the outer five per cent of the solar radius. An early helioseismic detection of an inward-increasing rotation gradient was made by \citet{1979A&A....72..177D} using high-degree observations that probed about the upper 20~Mm of the convection zone. There were also hints that such a shear layer might exist in  non-helioseismic observations, with a higher rotation rate seen in direct Doppler measurements at the photosphere than from tracking surface features \citep[e.g.][]{1989ApJ...343..526B}, which might be explained by the idea that features such as sunspots are anchored in a faster-rotating layer below the surface. The shear layer was clearly seen in the early global rotation inversion results from GONG \citep{1996Sci...272.1300T} and MDI \citep{1997SoPh..170...43K}; in both cases the shear apparently reversed at high latitudes. 

Because the shear layer is not fully resolved using medium-degree $p$-modes alone, it is best studied using $f$-modes, as was done by \citet{2002SoPh..205..211C} with MDI data. They found the rotation rate decreased away from the rotation axis at a rate of about $-400$~nHz/$R_{\odot}$ at low latitudes, but this slope decreased to a small value by $30\deg$ latitude and possibly reversed sign at latitudes above that (but only in the outermost 5~Mm). However, they pointed out that the high-latitude result depended on the handful of modes with $l\geq 250$. They also found that the rotation rate varied almost linearly with depth, which is unexpected, as considerations of conservation of momentum would suggest the rotation rate should vary with the inverse square of distance from the rotation axis.  
\citet{2014A&A...570L..12B}, using HMI data and re-analyzed MDI data, confirmed the low-latitude results of \citet{2002SoPh..205..211C} but found the sign of the shear remained negative up to at least $60\,\deg$ latitude.

\subsubsection{The Bulk of the Convection Zone}

Like the surface rotation, the solar rotation in the bulk of the convection zone is differential -- fastest near the equator and slowest near the poles. Below the near-surface shear layer, the rotation profile is almost -- but not quite -- constant along radial lines. The contours of constant rotation in fact make an angle of about $25\,\deg$ to the rotation axis over a wide range of latitudes \citep{2003ESASP.517..283G,2005ApJ...634.1405H}; this may be due to Coriolis forces acting on the meridional flow.

\subsubsection{The Tachocline}

Below the convection zone lies a relatively thin region of strong radial shear known as the tachocline, marking the transition from differential rotation to the more or less rigid rotation of the radiative interior. This region is not fully resolved in inversions, meaning that the transition appears wider and smoother in inversion results than it is in reality, so forward calculations or other strategies are needed to estimate its thickness. Table 2 of \citet{Howe2009} summarizes the work of various authors to parametrize the shape of the tachocline at equatorial latitudes. The consensus is that the tachocline covers a few per cent of the solar radius and is centered slightly below the base of the convection zone ($0.713\,R_{\odot}$) as determined from structure inversions \citep{1997MNRAS.287..189B}. \citet{2001MNRAS.324..498B} also found that the tachocline is prolate -- shallower at higher latitudes -- a result more recently confirmed by \citet{2011ApJ...735L..45A}.

\subsubsection{The Radiative Interior}

Below the tachocline, the latitudinal and depth resolution of the inversion results is relatively poor, but this region appears to rotate rigidly; \citet{2002ApJ...573..857E} found a rate of approximately 435\,nHz with no significant differential rotation between 0.4 and 0.7$R_{\odot}$.

\subsubsection{Core Rotation}

Only the radial ($l=0$) modes penetrate to the very center of the Sun, and these carry no information about the rotation. The $l=1$ modes do have their lower turning points within the nuclear-burning core of the Sun, but even these modes spend the majority of their time closer to the surface, so estimating the core rotation is not as simple as measuring the $l=1$ rotational splitting. It is difficult to localize an averaging kernel below about $0.25\,R_{\odot}$ \citep[e.g.][]{1999MNRAS.308..405C}, and to date it has not been possible to put strong constraints on the core rotation rate using $p$-modes. It would be helpful if we could measure the solar $g$-modes (modes where the restoring force is gravity, and which have their greatest amplitude in the interior and a very small amplitude at the surface), but there are no independently confirmed $g$-mode observations for the Sun, though the results of \citet{2007Sci...316.1591G} are intriguing. Using observations from the Global Oscillations at Low Frequencies (GOLF) instrument onboard SOHO, they found a periodic structure in the low-frequency spectrum consistent with $g$-mode peaks favouring a faster rotation rate than that in the rest of the radiative interior. However, this result remains controversial.

\subsubsection{The Torsional Oscillation}

\begin{figure}
  \includegraphics[width=\linewidth]{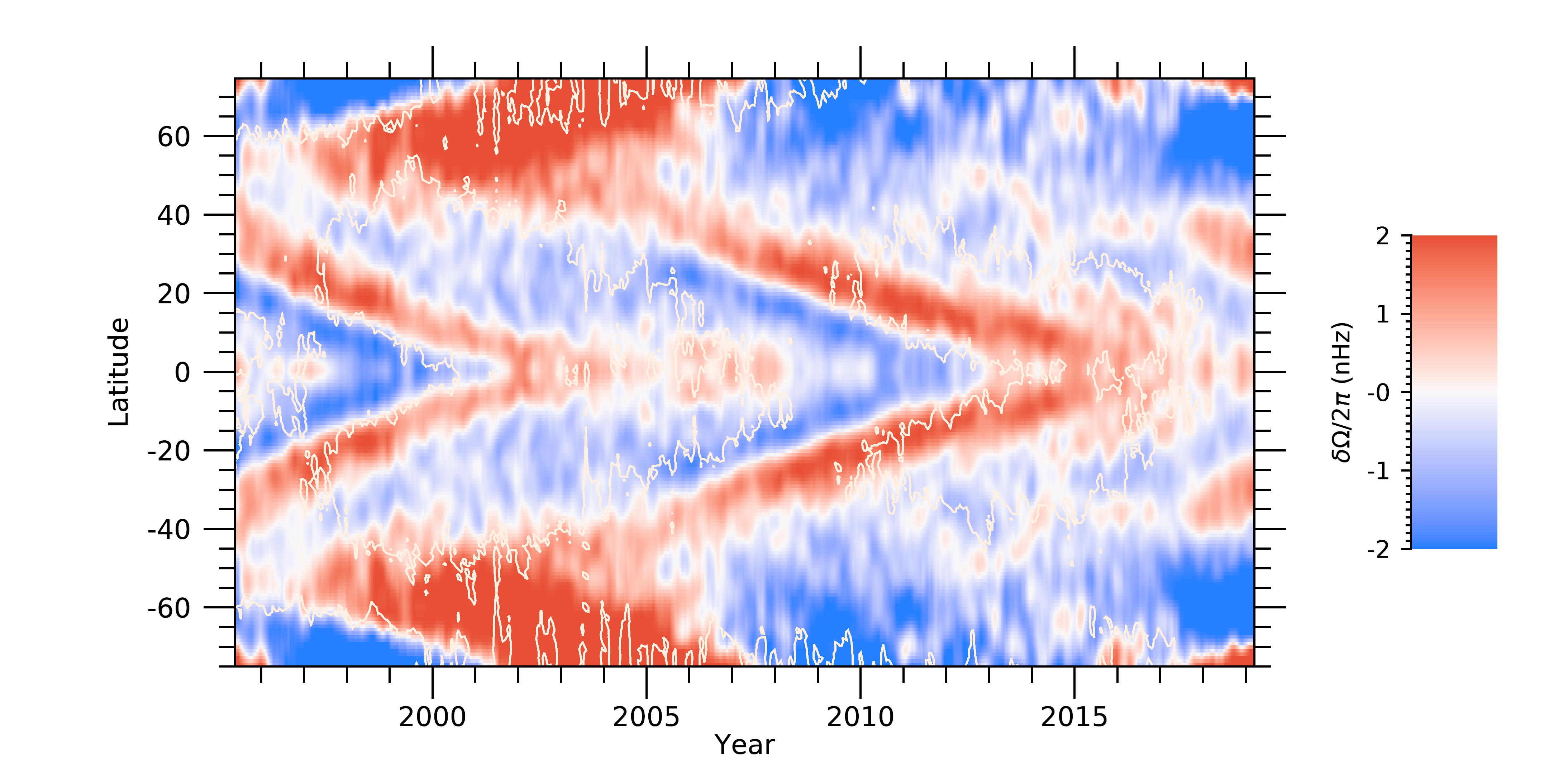}
  \caption{Zonal-flow residuals at $0.99\,R_{\odot}$ from 2dRLS inversions of GONG, MDI and HMI data after subtraction of a temporal mean at each latitude. The white contours mark the 5\,G level of the longitudinal magnetic field strength from synoptic maps produced by the National Solar Observatory. Color map from \citet{2015arXiv150903700K}.}
  \label{fig:rhfig5}
  \end{figure}

The solar rotation at the surface is not constant over time; it shows a pattern of migrating bands of faster and slower rotation (or in other words, zonal flows that are prograde and retrograde in the frame of the overall solar rotation) that is closely associated with the migration of the active latitudes towards the equator over the course of a solar cycle (Figure~\ref{fig:rhfig5}). The variation in the rotation rate at the equator corresponds to a speed change of a few meters per second. This pattern was first identified, and dubbed the ``torsional oscillation'' by \cite{1980ApJ...239L..33H} in direct Doppler observations from the Mount Wilson Observatory. The bands were first detected by helioseismic means by \cite{1997ApJ...482L.207K} using $f$-mode splittings from MDI. As data from GONG and MDI accumulated during the rising phase of Solar Cycle 23, it became possible to detect the migrating flows using $p$-mode data. This revealed that the flows penetrated to at least $0.92\,R_{\odot}$, or about 30\,Mm, below the photosphere \citep{2000ApJ...533L.163H,2000SoPh..192..437T,2000ApJ...541..442A,2002Sci...296..101V}. The helioseismic observations also revealed the presence of a poleward-propagating branch at latitudes above about $45\,\deg$ which appears to penetrate through the whole depth of the convection zone \citep{2001ApJ...559L..67A}. As more data were collected and the signal-to-noise of the observations improved, it became apparent that the pattern may be coherent through the whole of the convection zone, with the signal near the base of the convection zone seeming to lead that at the surface by about two years -- or, to look at it another way, the phase of the pattern is constant along lines of constant rotation rather than lines of constant latitude \citep{2005ApJ...634.1405H}.

Around solar minimum, when the main bands of faster-than average rotation are close to the equator, an acceleration in the near-surface zonal flows at mid-latitudes appears to precede the appearance of new-cycle magnetic activity. During the unusually long minimum following Solar Cycle 23, the new equatorward branch associated with Cycle 24 was visible, but its migration towards the equator was delayed \citep{2009ApJ...701L..87H}. As noted by \cite{2011JPhCS.271a2074H}, the new cycle activity appeared only when the flow belt reached a latitude of about $25\,\deg$, as had been seen in previous cycles.  

During Solar Cycle 24, the poleward-propagating branch of the torsional oscillation pattern appeared much weaker, but this appearance was shown by \cite{2013ApJ...767L..20H} to be partly due to a slow-down in the underlying rotation rate at higher latitudes, which may be related to the weaker polar fields during this cycle \citep{2012ApJ...750L...8R}. In the most recent observations \citep{2018ApJ...862L...5H}, the beginnings of the equatorward branch for Cycle 25 can be clearly seen.

\subsection{Meridional Flow \label{section:hsm-meridional}}

\begin{figure}
  \begin{center}
  \includegraphics[width=0.45\linewidth]{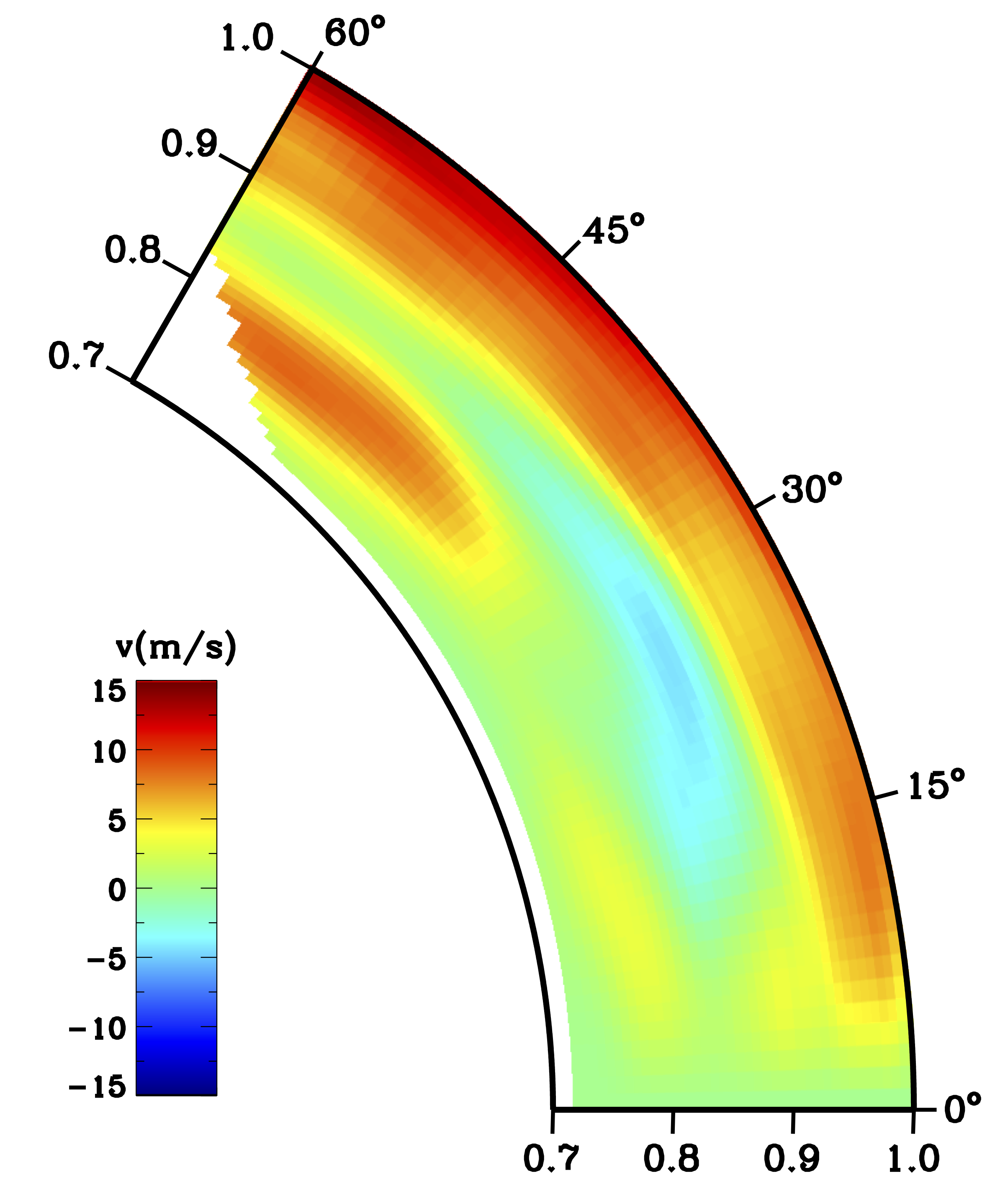}
\end{center}
  \caption{North-south average of meridional flow profile from Figure 8 of \citet{2017ApJ...849..144C}, {\textit{The Astrophysical Journal}} 849, 144, reproduced by permission of the AAS}.
  \label{fig:rhfig1}
\end{figure}

The Sun has a well-known pattern of meridional flow from the equatorial regions towards the poles, as found by \citet{1979SoPh...63....3D} using the Stanford magnetograph and by \citet{1982SoPh...80..361L} in Mount Wilson Doppler observations at the surface. This flow plays an important part in the solar magnetic cycle \citep[e.g.][]{1989Sci...245..712W}{; the flow carries magnetic flux from decaying active regions poleward, until at solar maximum the sign of the polar field reverses. Conventional global helioseismic techniques cannot measure this flow, but it can be accessed by local helioseismic techniques, at least in the outer layers of the convection zone. \citet{1997Natur.390...52G} used ``helioseismic tomography'' -- a variant of the time-distance technique \citep{1993Natur.362..430D}, which measures the travel-time of waves between pairs of points on the solar surface -- to measure the longitudinally-averaged flow in the outer 4\,\% of the solar radius, and found that the speed of the flow was approximately constant over that depth range. Subsequent results \citep[e.g.][]{1999ApJ...510L.153G,2000SoPh..192..335H,2004ApJ...603..776Z} confirmed that the flow in the outer convection zone is poleward up to at least $50\,\deg$ latitude and down to a depth of 30\,Mm. \citet{2004ApJ...603..776Z} also reported a time-varying component to the meridional flow, with flows converging towards the magnetic activity belt above 12\,Mm, and diverging flows below 18\,Mm.  \cite{2014ApJ...789L...7Z} again found flows converging towards the activity belts, and also reported that the meridional-flow speed above latitude 30° was slower when following-polarity flux was being transported and faster when leading-polarity flux was  being transported. Similar solar-cycle variations of the meridional flow have been seen in other work, for example by \cite{2011ApJ...729...80H}, who studied the surface flows by correlation tracking of features in MDI magnetograms and found, in addition to the inflows into the activity belts, that cells of equatorward flow appeared at higher latitudes, in the southern hemisphere during the rising phase of Solar Cycle 23 and in the northern hemisphere during its declining phase. \cite{2011JPhCS.271a2073G} also reported on modulation of the meridional flow by the activity cycle from ring-diagram analysis of GONG data; an interesting feature of this work is that, as with the zonal flows, the changes in the meridional flow pattern start to appear in advance of the new-cycle activity.

The meridional ``return flow'' that must exist (so that mass does not pile up at the poles), or any complex interior structure of the meridional flow,  are much more challenging to observe. Several attempts to probe the meridional flow deep in the convection zone have been made in recent years, particularly since the center-to-limb correction \citep{2012ApJ...749L...5Z} was introduced. Several of these attempts produced somewhat inconsistent results: \citet{2013ApJ...774L..29Z}, using time-distance analysis on two years of HMI data, \citet{2013ApJ...778L..38S} using an analysis of perturbations to global modes caused by the meridional flow, and  \citet{2014ApJ...784..145K} and \citet{2015ApJ...805..133J} -- both using GONG data -- all found some indications of a multi-cell structure, but \citet{2015ApJ...813..114R} did not find strong evidence of a multi-cell flow in their analysis of four years of GONG data. More recently, using seven years of HMI data, \citet{2017ApJ...849..144C} found a two-cell structure in each hemisphere, with an equatorward flow in the middle of the convection zone and poleward flows above and below (Figure~\ref{fig:rhfig1}), in broad agreement with the earlier findings of \citet{2013ApJ...774L..29Z}. On the other hand, \citet{2018ApJ...863...39M} and \citet{2018A&A...619A..99L} reported results favoring a single-cell structure with the return flow close to the base of the convection zone, although the latter work did find an anomalous two-cell structure in the northern hemisphere during Solar Cycle 24. It is probably still too soon to say that the question of the interior meridional flow has been settled.

\section{Flux Emergence (Mark Linton) \label{section:FE}}

\subsection{Observations of Emerging Flux \label{section:FEobservations}}
%Target 10-15 pages per section, arguably 20-30 in this style

\subsubsection{Early Observations of Sunspots}
A sunspot is a darkening of the Sun's surface, corresponding to a reduction in temperature
at the surface from $\sim$ 5700-6500 degrees Kelvin (K) outside sunspots to $\sim$ 3500-4500 K 
in sunspots. 
This darkening is due to the strong concentration in sunspots of magnetic flux. 
The magnetic field, concentrated to several
kiloGauss (kG) in sunspots, inhibits the convective motion of the plasma within that field,
and so prevents the efficient transport of energy by convection through
the convection zone just below the sunspot, where the magnetic field is particularly strong. 
Rather, energy here is transported by radiation, which is less
efficient in convection zone conditions, and so the surface temperature above these
regions is lower. The dark, central regions of these spots are referred to as the
sunspot umbrae. Fully formed sunspots are surrounded by a brighter, striated region
called the penumbra.  The images below, from \citet{Tiwari2015} show the white 
light (continuum intensity) of a well formed sunspot
(Figure \ref{fig:Tiwari2015_white_light}), as well as the temperature, magnetic field
strength, magnetic field inclination from the vertical, and line of sight velocity in 
and around the spot
(Figure \ref{fig:Tiwari2015_four_plots}). 
\begin{figure}[ht]
    \centering
    \includegraphics[width=0.49\textwidth]{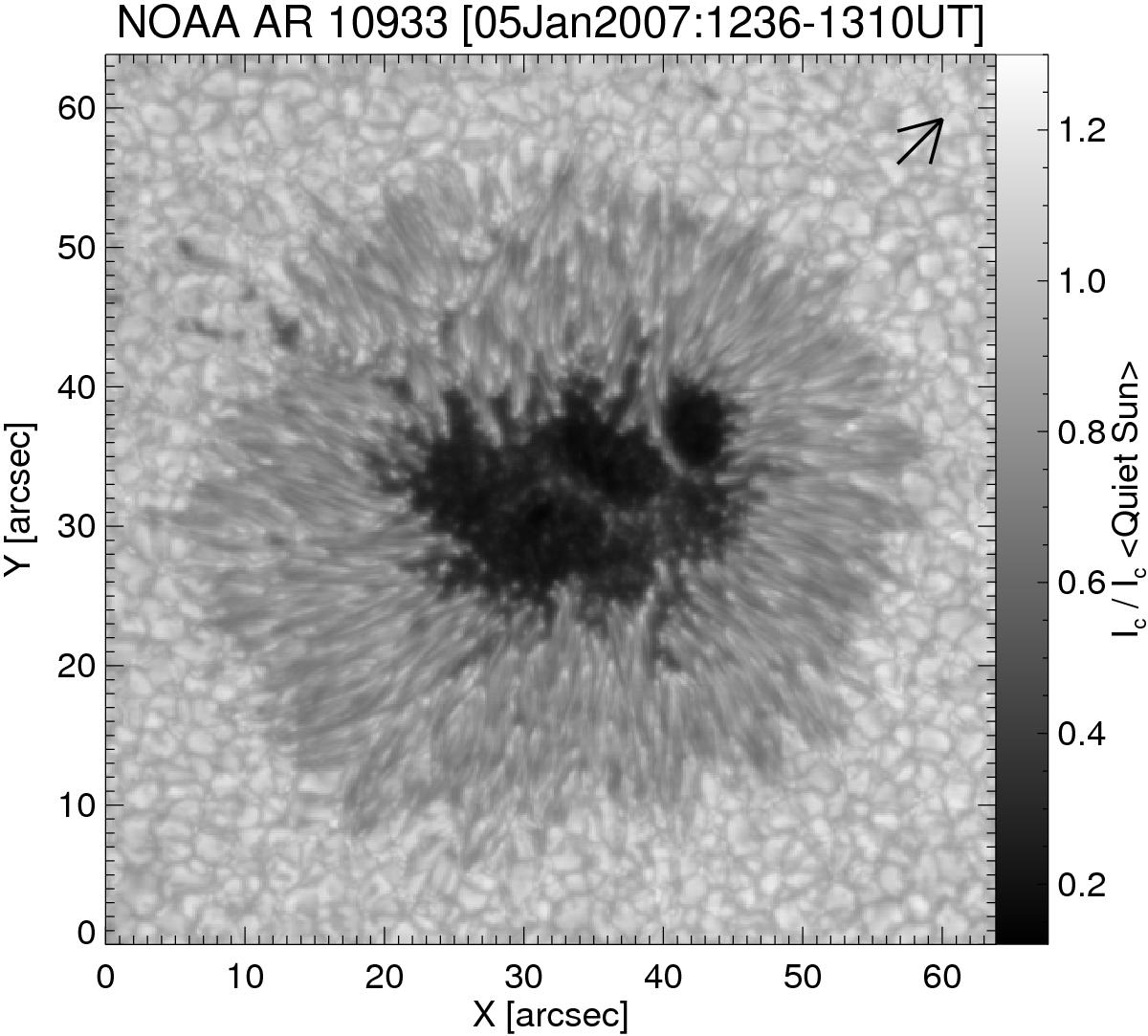}
    \caption{White light observation of a sunspot (National Oceanographic and Atmospheric Administration active region, NOAA AR, 10933), taken by the Solar 
    Optical Telescope (SOT) on board the Hinode satellite. From \citet{Tiwari2015}.
    \label{fig:Tiwari2015_white_light}}
\end{figure}
\begin{figure}[ht]
    \centering
    \includegraphics[width=0.99\textwidth]{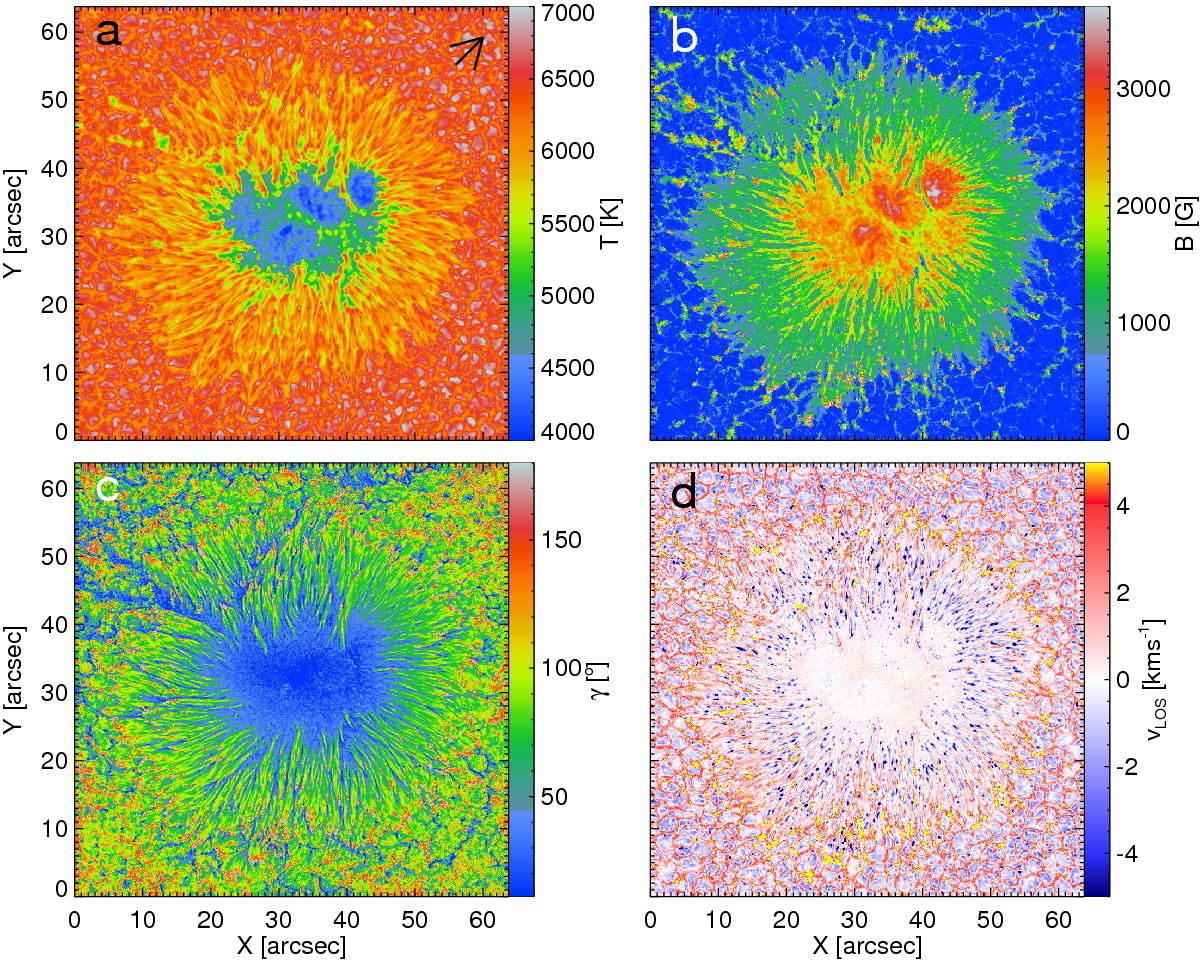}
    \caption{Inferred temperature (a), magnetic field strength (b), inclination of magnetic
    field vector to the solar surface (c), and line of sight velocity (d) for NOAA AR 10933.
    Data taken by Hinode/SOT. From \citet{Tiwari2015}.
    \label{fig:Tiwari2015_four_plots}}
\end{figure}

Sunspots largely appear within $\pm$ 40 degrees in latitude from the solar equator, and last for several days
to a few months. They usually appear as pairs of positive and negative magnetic field
polarity, suggesting that they are caused by the emergence of an $\Omega$-shaped rope of magnetic flux
through the photosphere: where one part of the emerged rope intersects the photosphere,
positive vertical field is measured, and where the other part intersects the photosphere,
negative vertical field is measured. Bright loops of magnetic field are observed
between these two spots in extreme ultraviolet and X-ray observations of the solar atmosphere, 
completing the picture of a rope of magnetic field arching from one sunspot 
to the other.

\citet{Coyne1991} presents an informative review of early sunspot observations.
The first mention of sunspot observations is from the fourth century BC by Teophrastus 
of Athens
\citep{Coyne1991}. Naked eye observations of sunspots were then recorded in China, and to
some extent in Japan and Korea, starting in the second century BC 
\citep{Bray_Loughhead_1964, Yau_Stephenson_1988, Coyne1991}. 
In Europe, telescopic observations of sunspots date 
to around 1611, when a number of astronomers observed and reported them. 
The review by \citet{Coyne1991} of these earliest observations credits 
Galileo Galilei, Johann Goldschmid, Christopher Schneier, Thomas Hariott, and 
Francesco Sizzi with these first observations.  The review by \citet{Morfill1991} credits
\citet{Schwabe1843} with the first reported observation indicating that there is
a sunspot cycle, as discussed in \S \ref{section:dynamo}.  
%[Note, Zirin says \citep{Schwabe1844}]
Within this cycle are several key features of sunspot emergence and evolution. 
See \citet{Zirin1988} for a review of these solar cycle features, 
briefly summarized below.

\subsubsection{Sunspot latitude:} The latitude at which sunspots emerge decreases during the cycle, 
starting with an average latitude of about $\pm$ 30 degrees early in the cycle, and ending near the 
equator \citep{Zirin1988}.  This was first reported by Carrington, and is referred to as 
Sp\"orer's law \citep{Carrington1958} and is discussed in more detail in \S \ref{section:dynamo}. 
The top panel of Figure \ref{fig:dynamo-1} shows this effect:
concentrations of sunspots emerge around $\pm$ 30 degrees latitude at the beginning of each 
cycle with subsequent spots emerging closer to the equator as the spot progresses. 

\subsubsection{Sunspot magnetic field:}
\citet{Hale1912} used the Zeeman effect \citep{Zeeman1897} 
to determine that sunspots have strong magnetic fields.
The Zeeman effect is the process whereby a magnetically sensitive ionized atom in a magnetic field 
will emit linear and circularly polarised light, with the relative characteristics of the different
polarizations depending on the direction and strength of the magnetic field. Panels (b) and (c) 
of Figure \ref{fig:Tiwari2015_four_plots} show the resulting magnetic field strength and 
magnetic field inclination with respect to the vertical for a spot observed by \citet{Tiwari2015}. 
Hale used his Zeeman effect measurements to show that sunspot pairs are aligned so that the 
leading spot (in the sense of solar rotation) is of one magnetic polarity and the following 
spot is of the opposite polarity. Critically, \citet{Hale1919} then found that the polarity 
of the leading spots is same for the the majority of spot pairs in one hemisphere 
(Northern or Southern), and is the opposite for the majority of spot pairs in the other hemisphere. 
This polarity rule then flips with each subsequent solar cycle (see \S \ref{section:dynamo}). 

\subsubsection{Sunspot tilt:} 
An additional observed rule
for sunspot emergence is the tilt of the sunspots relative to the equator. As was pointed out in \S \ref{section:dynamo}, while a sunspot
pair, from leading to following spot is aligned largely parallel to the equator, there
is on average a tilt with respect to the equator, with the leading spot lying closer to the equator.
This tilt varies from about 11 degrees at $\pm$ 30 degrees latitude to about 3 degrees near the equator 
\citep[see][]{Hale1919}, and is referred to as Joy's law \citep{HaleNicholson1925}.
Numerous measurements of this variation of tilt with latitude have been made over the years: 
see \citet{WangSheeley1989}, \citet{Stenflo2012}, and \citet{McClintock2014}, 
as well as the review by \citet{Pevtsov2014SSR}.  
Figure \ref{fig:Tlatova2018} shows one such recent measurement of sunspot tilt versus latitude
by \citet{Tlatova2018}. 

\begin{figure}[ht]
    \centering
    \includegraphics[width=0.49\textwidth]{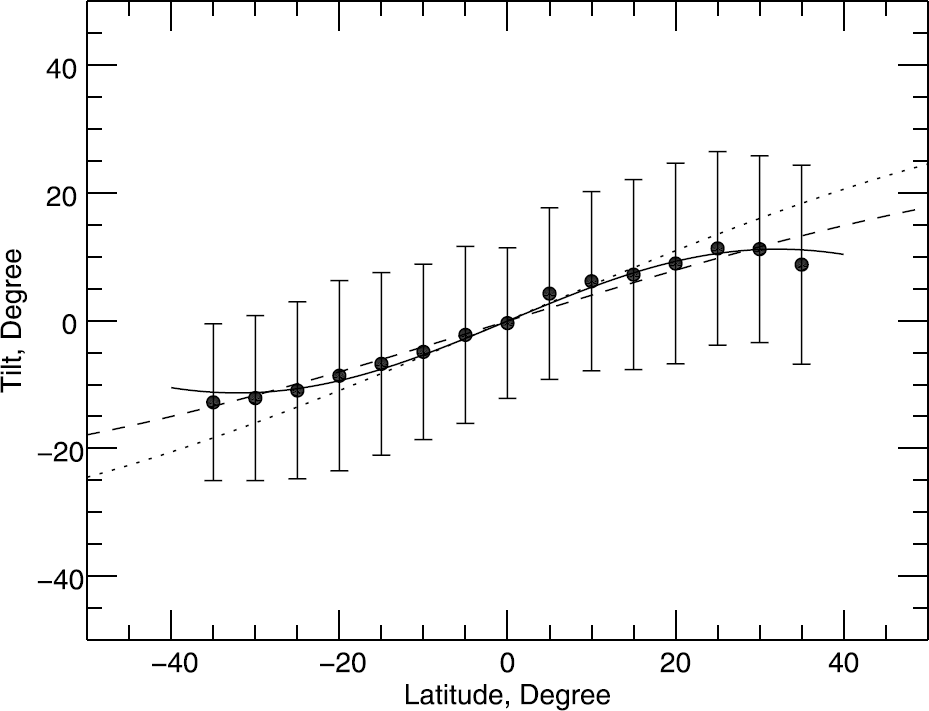}
    \caption{Measured tilt of solar magnetic sunspot pairs, taken from Mount Wilson Observatory 
    measurements from 1917 to 2017. The filled circles show the measured mean tilts, while
    the error bars show the measured standard deviations. The dashed line shows the fit to this data, and the dotted line shows a fit from the earlier study of \citet{Stenflo2012}. This figure is reproduced from \citet{Tlatova2018}.
    \label{fig:Tlatova2018}}
\end{figure}

\begin{figure}[ht]
    \centering
    \includegraphics[width=0.49\textwidth]{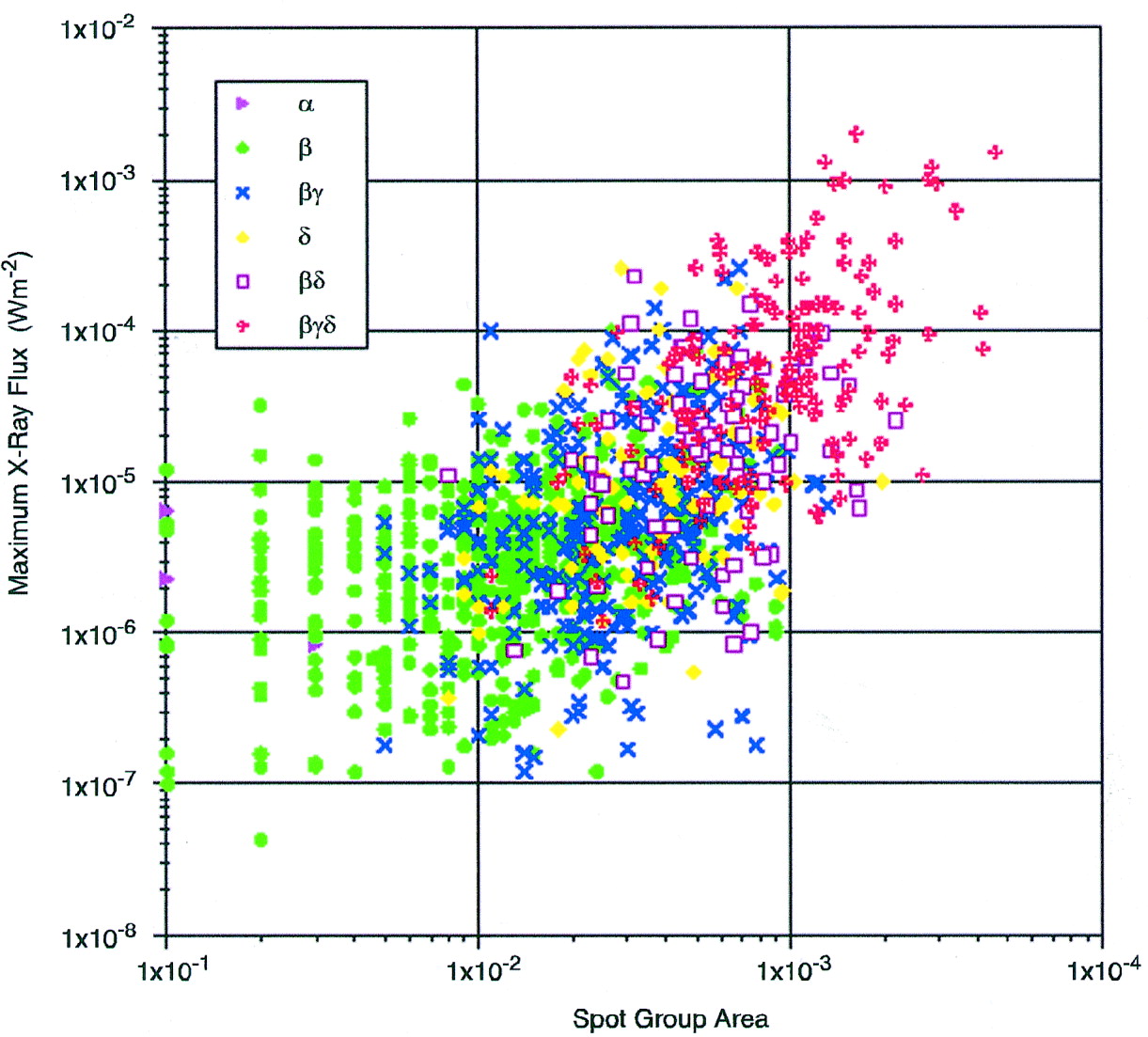}
    \caption{Sunspot group area versus peak X-ray flux for 2789 sunspot regions observed by the
    US Air Force Solar Optical Observing Network and by the Mount Wilson Observatory from 1989-1997. 
    Sunspot type is indicated by symbols (see inset). This shows that the majority of high X-ray
    flux flares originate from sunspot groups with a $\delta$ classification. This figure is
    reproduced from \citet{Sammis2000}.
    \label{fig:Sammis2000}}
\end{figure}

\subsubsection{Sunspot Complexity:}
While sunspots in general appear as a positive and negative polarity pair, or a positive
and negative polarity group of spots, and obey the leading-following (Hale) and tilt (Joy)
laws outlined above, sunspot regions, also referred to as active regions, can sometimes 
form in more complex configurations. \citet{Hale1919}
developed a classification system, often referred to as the Mount Wilson classification, 
to describe these types of spots. Single spots with
no corresponding opposite spot are termed $\alpha$ spots. These spots obviously have
magnetic field which connects to opposite magnetic field at the photosphere - but in
this case that field is weaker and more dispersed, and not strong enough to form a sunspot.
The standard configuration with a pair of opposite polarity spots is termed $\beta$. 
Spots with an "irregular distribution of polarities" \citep{Zirin1988} are termed $\gamma$,
and spots with "umbrae of opposite polarity within a single penumbra" are termed $\delta$
\citep{Kunzel1960}. These different types can overlap and combine together and so
extended classifications, generally consisting of combinations of these $\alpha$, $\beta$, 
$\gamma$, and $\delta$ classifications, have been developed and used over the years.
\citet{Jaeggli2016} analyzed the statistics of active region classifications from 1992 to 2015 
and found that $\sim$ 20\% of active regions are classified as $\alpha$ or $\alpha\gamma$ spots, 
$\sim$ 75\% are classified
as $\beta$ or $\beta\gamma$ spots, and the remaining 5\% are some form of $\delta$-spots.

\subsubsection{$\delta$-spots:} 
While the flaring activity generated above sunspots will be the focus of Chapter 4, 
we focus briefly here on the contribution that $\delta$-spots make to
strong solar flares. The complexity associated with $\delta$-spots has
been shown to indicate a stronger likelihood of eruption \citep{Kunzel1960}. 
These regions often have inverted polarity relative to Hale's law,
and \citet{Tanaka1980} showed that 90\% of such inverted $\delta$-spot regions
are highly active.  
\citet{ZirinL1987} and \citet{Sammis2000} investigated the flare activity of these different
types of spots. Figure \ref{fig:Sammis2000} shows the result from \citet{Sammis2000}.
From this and similar
studies, they found that the more complex spots, in particular those with a $\delta$ classification,
either on its own or combined with another classification, were responsible for the majority of 
large, or so-called X-class, flares. These X-class flares are represented on this Figure by the
symbols above an X-ray flux of $10^{-4} W/m^2$. The knowledge of this increased activity 
has fostered a number of theoretical and modeling studies to explain this, discussed below.

\begin{figure}[ht]
    \centering
    \includegraphics[angle=90,width=0.49\textwidth]{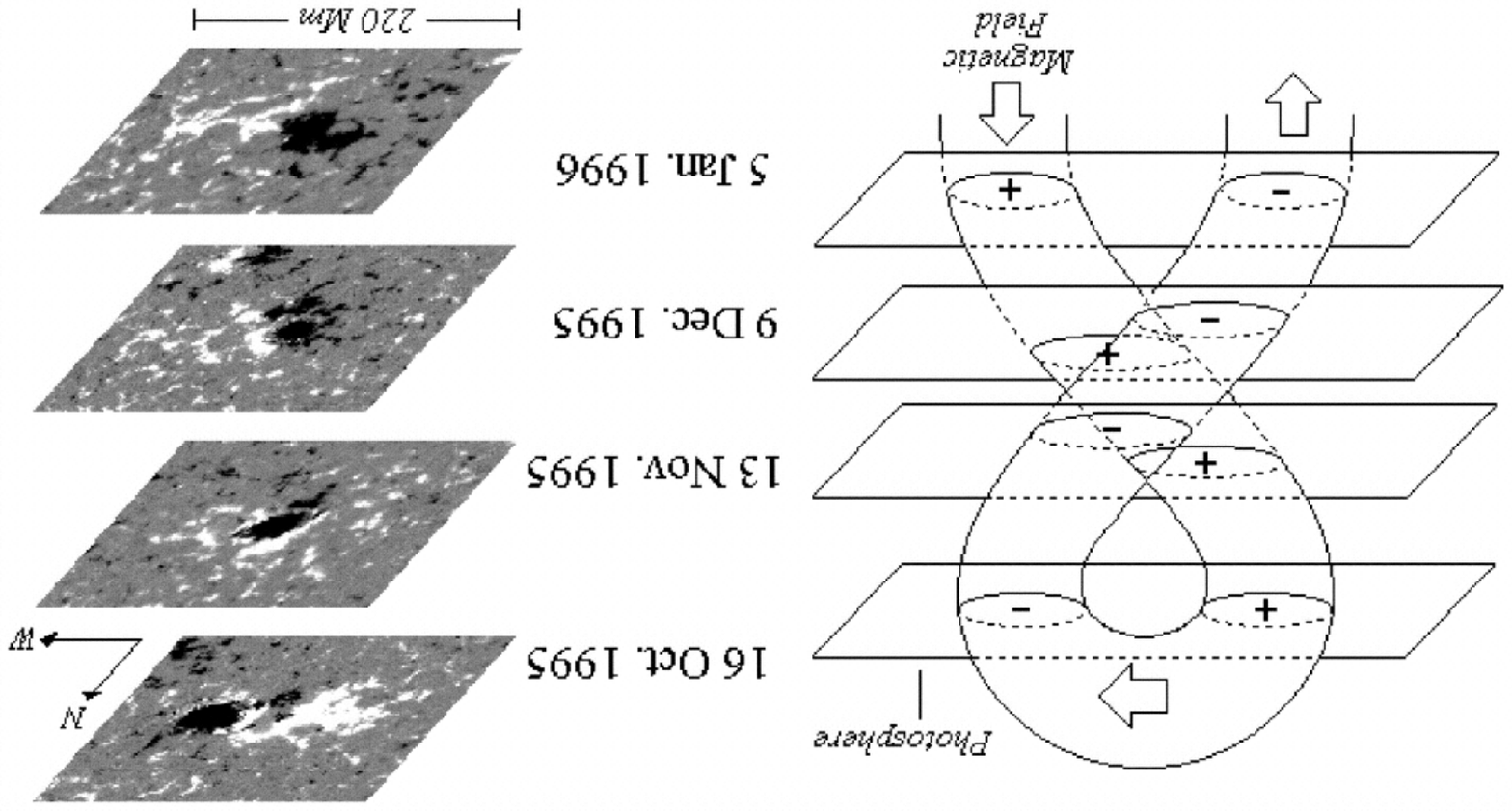}
    \caption{Observation of the long-term (three-month) evolution of the line of sight magnetic 
    field measurements of NOAA AR 7912. This shows how the magnetic fields of the region rotate by
    $\sim 180^{\circ}$ over this time frame. The cartoon to the left shows an inferred geometry
    for the sunspot region which could reproduce this behavior. Figure reproduced from \citet{LopezFuentes2000}.
    \label{fig:LopezFuentes2000}}
\end{figure}

\begin{figure}[ht]
    \centering
    \includegraphics[width=0.59\textwidth]{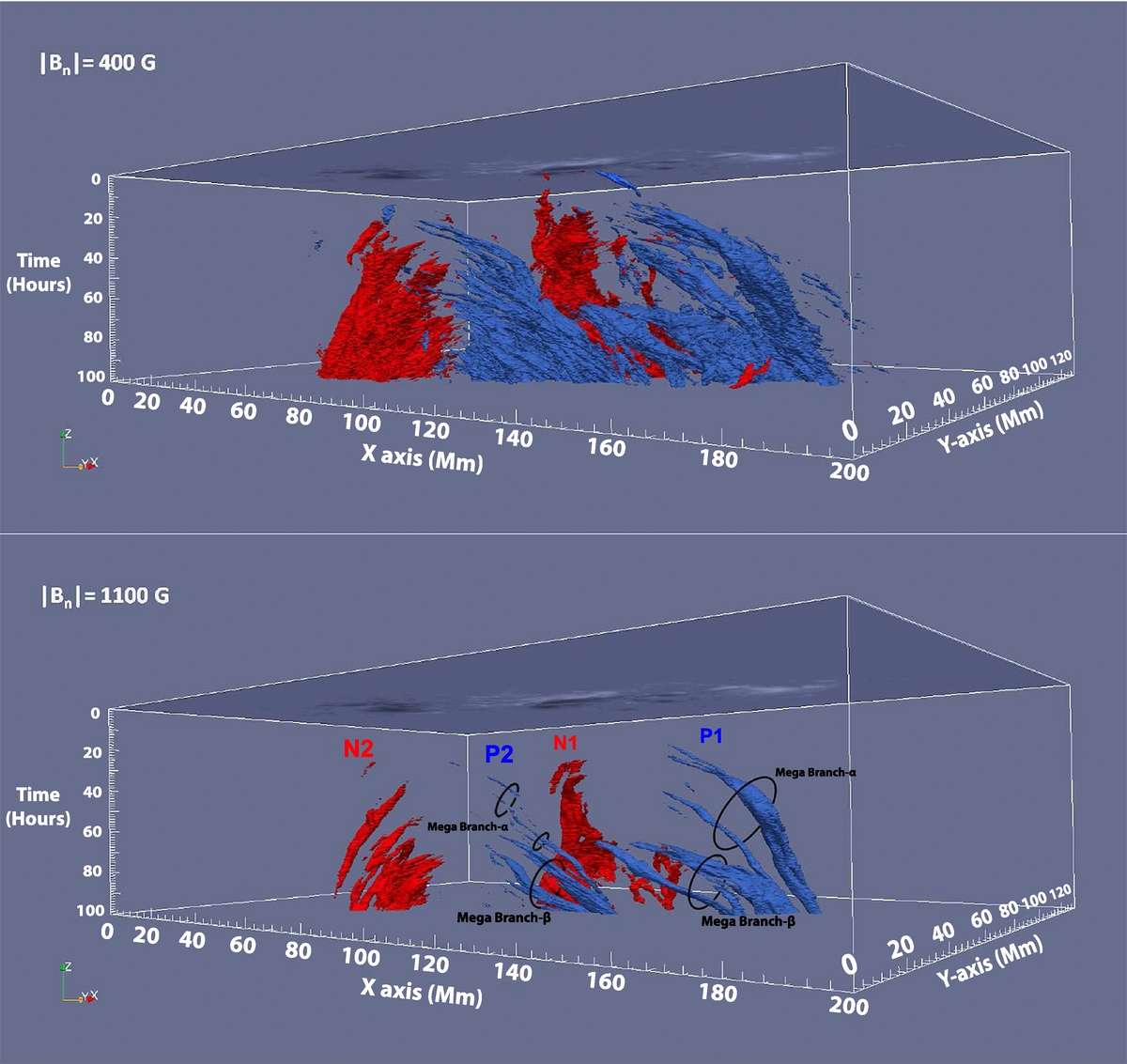}
    \caption{Height-time reconstruction of the line of sight component of magnetic field in NOAA AR 11158
    over the course of 100 hours during its emergence. The top panel shows isosurfaces at 400 G and the bottom panel shows isosurfaces at 1100G, with positive field in blue and negative field in red. This gives a three dimensional rendering of the time history of the photospheric field, and gives one an idea of the geometry
    of the structure which rises through the photosphere during this flux emergence. Figure reproduced from \citet{Chintzoglou2013}.
    \label{fig:Chintzoglou2013}}
\end{figure}

\subsubsection{Twist, rotation of sunspot fields:}
\citet{Kurokawa1991} analyzed observations of sunspot emergence with the Domeless Solar Telescope 
at the Hida Observatory of Kyoto University. By studying the time evolution of structures visible
in the chromosphere, 1000 km or so above the photosphere, they found evidence that magnetic
fields emerge as a set of twisted and sheared loop structures. This argues that emerging
structures can be in the form of so-called flux ropes, i.e., that they consist of a set
of fieldlines wrapping around a single axis, forming a tight helix. 
\citet{Pevtsov1994} analyzed vector magnetic field measurements taken
with the Stokes Polarimeter at the Mees Solar Observatory to measure the effective twist
of active regions. They measured the ratio of the vertical current to the vertical magnetic field 
at the photosphere. This gives a measure, called $\alpha$, which is related to the inverse
wavelength with which fieldlines twist about each other. They found that locally this
twist measure can reach $5\times10^{7}/m$, indicating a twist wavelength of the order of 
10,000km. \citet{Pevtsov1995} and \citet{Longcope1998} then
surveyed $\alpha$ based on the coronal geometry of fieldlines for each active region.
They found that the average $\alpha$ over each active region lies in the
range $\pm 2\times10^{-8}/m$, and increases moderately with latitude.
\citet{Luoni2011} analyzed the vertical magnetic field structure of active regions,
and found that the tongue-like structure often seen in emerging active regions,
which distorts the bipolar structure into more of a yin-yang shape, can be well
reproduced by the emergence of a twisted, arched structure through the photosphere.
If the twist in such a torus goes to zero, the tongues disappear, and the emerging
region returns to the classic, symmetric bipolar structure.

\citet{Leka1996} mapped out the rotation of pair of emerging sunspot groups about
their center of mass.  They studied an emerging, flaring sunspot region with the 
Imaging Vector Magnetograph at the Mees Solar Observatory. By following the location of 
footpoints of the two primary concentrations of magnetic polarity in the photosphere
as a function of time, they were able to generate a series of 
time versus photospheric location maps of these footpoints. These maps indicated that
the observed sunspot regions were emerging with a large scale helical rotation, or writhe
(a rotation of the axis of a tube about itself).

\citet{LopezFuentes2000, LopezFuentes2003} found significant rotation of some sunspot regions 
over longer timescales as well.  Following sunspot regions over the course of several months 
(i.e., including several passes around the back side of the Sun), they found net rotations of 
these regions about their center of masses by up to 180 degrees, as shown in Figure \ref{fig:LopezFuentes2000}. 
They found that the majority of the sunspot regions they observed
had the same sign of twist (rotation of the magnetic field of a flux tube about its axis) 
as of rotation, indicating that the rotation was likely
due to convective motions writhing the flux ropes and adding twist to them as they rose 
to the surface. This is consistent with the $\Sigma$ effect predicted by 
\citet[][see discussion below in \S \ref{section:TFT}]{Longcope1997}. 

\subsubsection{Emergence rate, separation distance:} 
\citet{Chintzoglou2013} extended the distance-time reconstruction of \citet{Leka1996}
to generate 3D maps of entire active regions, such as the one shown in Figure
\ref{fig:Chintzoglou2013} for
NOAA AR 11158 (SDO/HMI). This map shows 100 hours of photospheric magnetic field
measurements, with time plotted on the vertical axis and two-dimensional measurements
of the line of sight magnetic field on the solar surface plotted on the horizontal axes.
Isosurfaces of magnetic field strength are drawn for 400G in the top panel, and for
1100G in the bottom panel. These isosurfaces are surfaces drawn such
that the volume enclosed within the surface has field strength larger than the isosurface
magnetic field strength and the volume outside that surface has a field smaller than that 
level.  The resulting map illustrates both the separation of positive from the negative
emerging regions with time, and the internal coalescence of each of these two structures with time, particularly for the 400G isosurface level. Note that this presents a useful way to visualize the variation in horizontal location and extent of the emerging field with time, 
and helps to infer the magnetic topology that is transferred through the photosphere, but
the result should not be taken as a reconstruction of the three-dimensional shape 
of the resulting coronal field. 
This is in part because vertical emergence flows do not factor into this reconstruction, 
so the apparent vertical extent of the structure may vary from one location to the next (thus the vertical axis here is labeled with time, rather than height), and in part because the expansion of the emerging field into the corona once it passes through the photosphere and
chromosphere will likely reshape and reconfigure the field.

\citet{McClintock2016} investigated sunspot separation as a function of sunspot area.
They found that the separation of the opposite polarity spots is about 40,000 km for
small regions when their spot area is at its maximum, the separation is about 45,000
to 50,000 km for mid-size regions at that stage in their evolution, and large regions
reach about 60,000 km at that stage. All regions then proceeded to move further apart
in the 2-3 days following peak spot area, with large regions reaching 70,000 to 80,000 km
in separation. Here small areas are defined as regions with maximum umbral areas 
less than 45 MSH (millions of a solar hemisphere,
MSH = 3 million square kilometers), mid size regions are between 45 MSH and 90 MSH, 
and large regions are larger than 90 MSH.
Interestingly, all regions have a separation distance of $\sim$ 40,000 km early
in their evolution. \citet{McClintock2016} hypothesize that this may be due
to the influence of so-called supergranular convection cells, which have a similar 
diameter,
as originally proposed by \citet{Frazier1972}. This proposal had been countered by
\citet{HarveyMartin1973}, who found out that the emerging fields they studied
did not collect at supergranular boundaries, as would be expected if the emergence was driven
by these flows. However, these recent results do suggest that, even if supergranules
are not directly responsible for emerging flux, they may play a role in setting
the scale. For example, the convective dynamo simulations by \citet{Nelson2013} find that 
large scale convective rolls cause fields to detach from toroidal wreaths
of magnetic field on those convection scales and
then rise to the surface. Determining what sets this scale is a critical remaining 
question of dynamo and flux emergence theory, and poses a promising task for comparisons
of observation and modeling.

\begin{figure}[ht]
    \centering
    \includegraphics[width=0.49\textwidth]{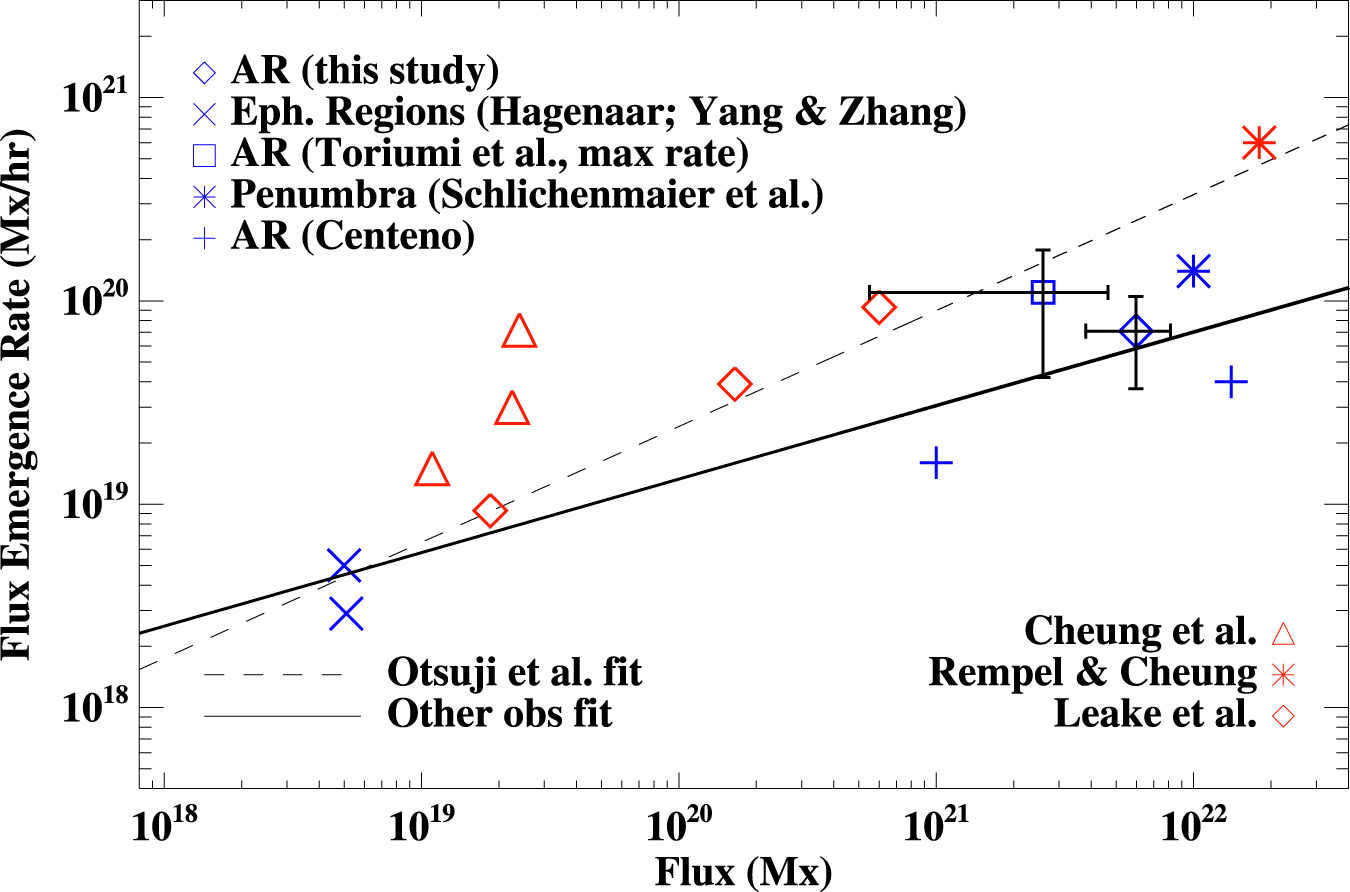}
    \caption{Magnetic flux emergence rate versus peak magnetic flux from a number of studies.
    Observations are represented by blue symbols: the \citet{Norton2017} study of ten emerging regions 
    is indicated by the diamond, the $x$ symbols show data from \citet{Hagenaar2001} and \citet{Yang2014},
    the square shows data measured by \citet{Toriumi2014}, and the plus symbol shows data measured by 
    \citet{Centeno2012}. Simulations are represented by red symbols: the triangle show simulations from 
    \citet{Cheung2007, Cheung2008}, the asterisk shows simulation results from
    \citet{RempelCheung2014}, and the diamonds show
    simulation data based on \citet{Leake2017}. The dashed line shows a fit to 
    observations made by \citet{Otsuji2011}, while
    the solid line shows a fit to all of the observed data shown on this plot, made by \citet{Norton2017}.
    This figure is reproduced, with permission, from \citet{Norton2017}.
    \label{fig:Norton2017}}
\end{figure}

\citet{Norton2017} then investigated the rate at which flux emerges in small
to mid-sized bipolar active regions, studying in-depth the evolution of 10 active 
regions whose emergence was fully captured on the visible solar disc. This
study found that flux emerged at a mean rate of $\sim 7 \times 10^{19}$ Mx/hr to 
generate a mean peak active region flux of $\sim 6 \times 10^{21}$ Mx. They present
a comprehensive review of other, complementary flux emergence rate studies which
preceded theirs, and summarize these various results it in Figure \ref{fig:Norton2017}.
This figure shows the flux emergence rate versus peak flux for all of these studies,
as well as the same measure for a number of flux emergence simulations (see the discussion
of flux emergence simulations in \S \ref{section:FE-theory_modeling}).
Together, these various studies cover three orders in magnitude of peak flux. 
Both this composite of various studies and the study of \citet{Otsuji2011}
find that flux emerges at a rate that depends on the peak flux to a power: 
$d\Phi/dt \propto \Phi_{max}^B$. \citet{Norton2017} found a power law index of B=0.36, 
while \citet{Otsuji2011} found a power law index of B=0.57. This again raises 
interesting questions about the flux emergence process. Namely, what sets
these emergence rates, and why are they well described by a power law function
of total emerging flux?

Critical to the rate of emergence, and possibly to the form that emergence takes,
is the speed at which active region magnetic flux rises to the surface prior to
emergence. This is difficult to measure, as photon signatures are not detectable
below the surface, and so helioseismology methods must be used to infer rise speed.
\citet{Ilonidis2011} reported a helioseismological detection of active region flux 
at about 65 Mm below the surface one to two days prior to the eventual surface detection
of the associated flux. This implied a rise speed for that last 65 Mm of $\sim 500$ m/s.
This is consistent with the rise speeds reported by earlier helioseismology studies \citep[see, e.g., review by][]{Kosovichev2000}, 
but conflicts with the later analysis of \citet{Birch2016}
which found a maximum rise speed of 150 km/s at 20 Mm below the photosphere. 
Further study is needed to more accurately
determine the magnitude and range of magnetic flux rise speed prior to active region
formation, both to elucidate the dynamics occurring in the convection zone, and to
impose constraints on the rise speeds which should be explored in numerical models (see
\S \ref{section:FE-theory_modeling}).

\newpage

\subsection{Theory and Modeling of Flux Emergence \label{section:FE-theory_modeling}}

\begin{figure}[ht]
    \centering
    \includegraphics[width=0.99\textwidth]{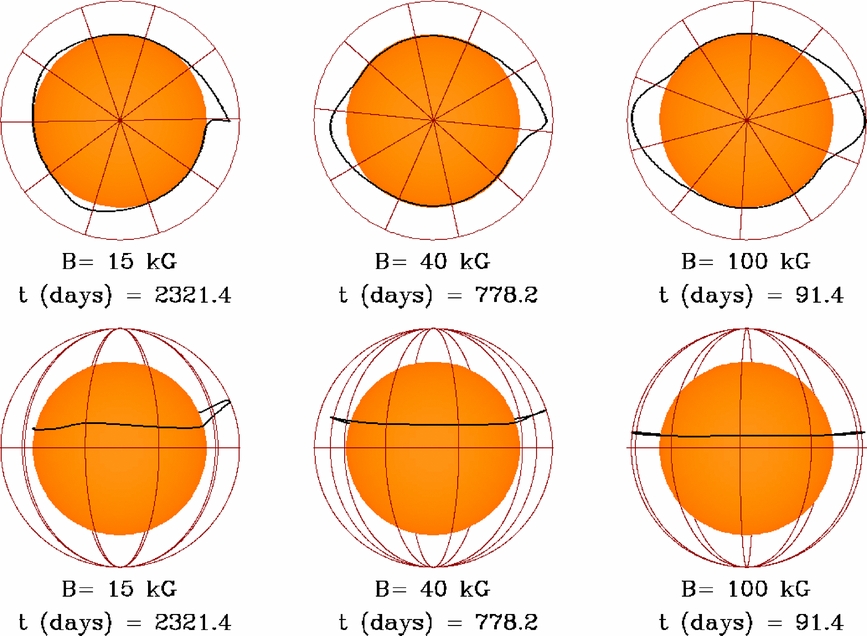}
    \caption{Figures showing the shape of a rising thin flux tube in a three dimensional
    convection zone, without convection, from simulations by \citet{Weber2011}. The solid black 
    line shows the thin flux tube, at a time when its apex nears the top of the convection zone, 
    for three simulations: an initial magnetic field strength of 15 kG (left panels), 
    40 kG (center panels), and 100 kG (right panels). The orange sphere in each image
    shows a sphere just above the base of the convection zone. In the top three panels, the viewpoint is from the solar North pole, with the Sun rotating counterclockwise. In the bottom
    three panels, the viewpoint is from the equator. 
    Figure reproduced from \citet{Weber2011}.
    \label{fig:Weber2011}}
\end{figure}

\subsubsection{Thin flux tube studies:\label{section:TFT}}
The emergence of magnetic fields to form sunspots at the solar photosphere
are the most obvious manifestation of the solar magnetic field. However, the mechanisms
which generate these magnetic fields in the convection zone below and the means
by which they rise to the surface and emerge are critically important to understanding
these surface manifestations. For a more extensive review of the theory of magnetic field
evolution in the convection zone and of magnetic flux emergence,
see \citet{Fan2009LRSP} and \citet{CheungIsobe2014LRSP}. The predominant theories explaining 
the sources of these magnetic fields are described in the review of solar dynamo theories
in \S \ref{section:dynamo}.
As discussed there, most dynamo models predict that the source of sunspot fields
lies at the base of the convection zone in a toroidal band of field stored in 
the shear layer at that depth. Therefore, significant research efforts have
been invested in modeling how these magnetic fields rise to the surface to form 
sunspots. As discussed above, observations indicate that many sunspot pairs
can be explained by invoking the emergence of a coherent loop of
magnetic field from the convection zone through the photosphere. The logical
question to investigate, then was how a coherent loop of magnetic field would
rise from the base of the convection zone to the surface. Until recently,
this was prohibitively expensive to model computationally for fully resolved three dimensional
(3D) flux loops. However, significant
early progress was made by invoking the so-called thin flux tube approximation.
\citet{Spruit1981} used this approximation to develop a set of one
dimensional thin flux tube equations.
This model first assumes that the radius of the flux rope being studied is significantly 
smaller than other length scales of importance to the problem, in particular
the local curvature of the flux tube axis, and the local length scale of the
stratified pressure gradients of the convection zone (i.e., the pressure scale height).
Second, it assumes that the internal structures of the flux rope, for example
how the magnetic field and the gas pressure profiles vary over the cross section of
the tube, can be represented by an average of those structures over the cross section.
This model then gives a set of one dimensional differential equation for the motion
of the flux tube axis through the stratified background convection zone, and for
the evolution of the velocity parallel to the tube axis, the magnetic
field parallel to the axis, the density, and the temperature \citep{Spruit1981}.

\begin{figure}[ht]
    \centering
    \includegraphics[width=0.50\textwidth]{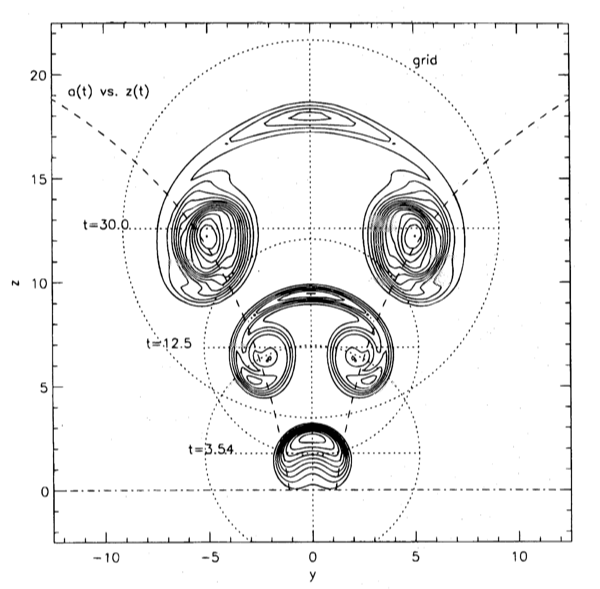}
    \caption{Two dimensional simulation of a rising magnetic flux tube. The solid lines 
    show contours of constant under-density (buoyancy), generated by the displacement
    of plasma by magnetic pressure in the flux tube. The three snapshots at t=3.54, 12.5 and 30.0
    show how the flux tube cross section is split into two counter-rotating flux tubes,
    joined by a thin section of flux tube arching over them. Figure reproduced from \citet{Longcope1996}.
    \label{fig:Longcope1996}}
\end{figure}

A critical feature of this, and of more sophisticated two dimensional and three
dimensional models of magnetic flux tube dynamics in the convection zone, is that
these flux tubes are naturally buoyant \citep{Parker1955a}. This follows from the 
fact that magnetic
fields exert forces on plasmas, and in particular, a magnetic field has an
effective pressure which is proportional to the magnetic field squared. Thus,
for a magnetic flux tube to be in pressure equilibrium with its surroundings, the
plasma plus magnetic pressure inside the flux tube must be equal to the plasma
pressure outside the flux tube. The plasma pressure inside the tube is therefore
reduced with respect to the plasma pressure outside the tube. If one assumes
that (given enough time) a flux tube settles into temperature equilibrium with
its surroundings, then the ideal gas law, where pressure is proportional to temperature 
times density, requires that the flux tube be less dense than its surroundings. A 
less dense flux tube will therefore be buoyant and will start to rise up through 
the convection zone. Assuming adiabatic expansion of the tube as it rises into 
lower pressure layers of an approximately adiabatically stratified
convection zone, the tube remains buoyant and can in theory rise all the way to 
the surface if no other forces disturb its rise.

\begin{figure}[ht]
    \centering
    \includegraphics[width=0.45\textwidth]{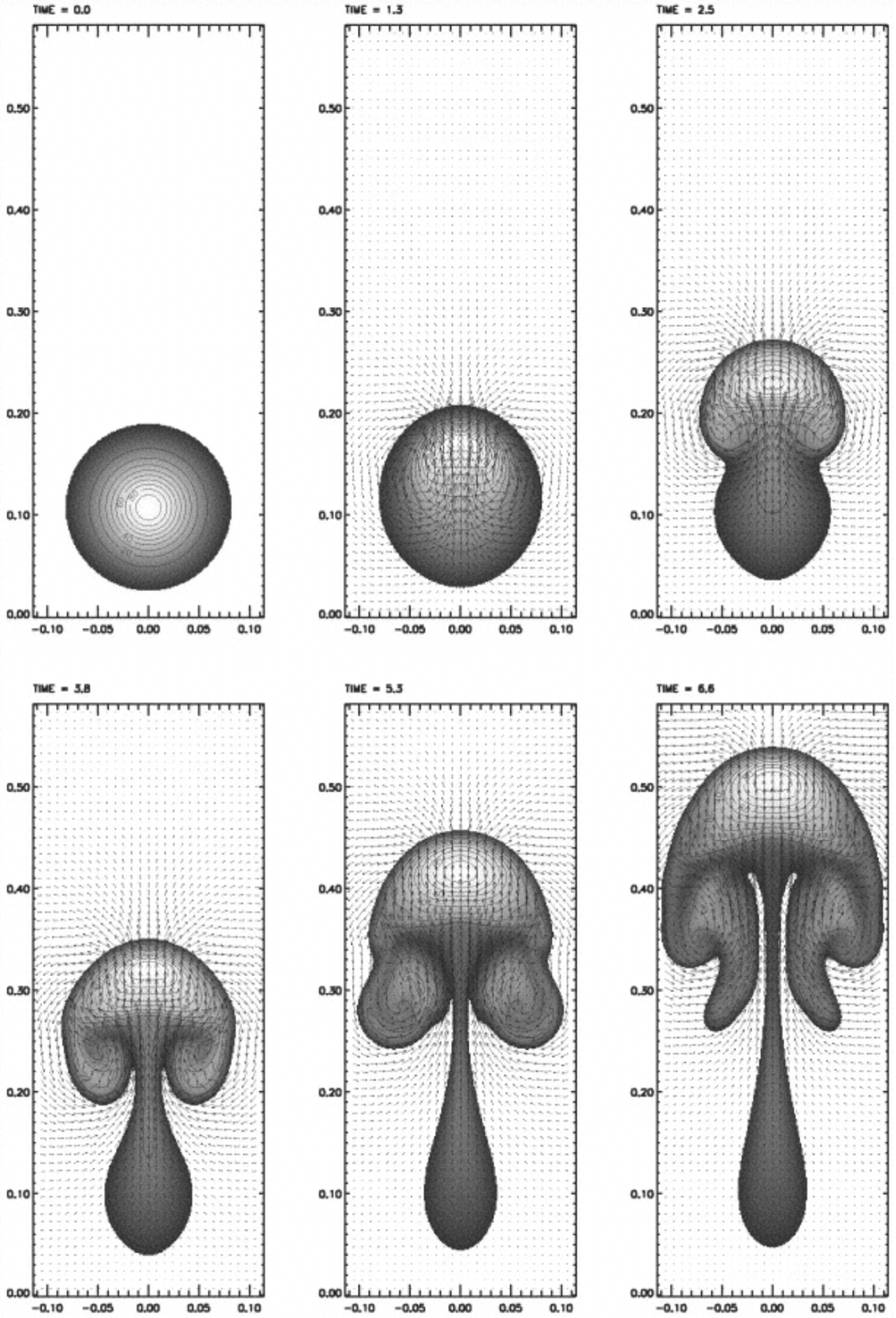}
	\caption{Two dimensional simulation of the rise of a buoyant, twisted flux tube
	through the stratified convection zone. The axial magnetic field strength is shown 
	by the greyscale and the contour lines. The vectors show the local velocity.
	The twist in the tube keeps the center of the tube, where the twist tension is strongest,
	coherent during its rise. Outer parts of the flux tube, with weaker twist tension, are still
	pulled from the flux tube by the fluid flowing past the tube. Figure reproduced from \citet{Emonet1998}.
    \label{fig:Emonet1998}}
\end{figure}

These one dimensional models had significant success in explaining many of the
large scale features of sunspot emergence described above. In particular,
by including the effect of solar rotation, they showed how the Coriolis
effect will alter the trajectory and shape of rising flux tubes. 
These simulations explored how flux tubes rising with different initial field strengths,
and therefore buoyancies, and from different latitudes can explain the observed
latitudinal ranges over which sunspots emerge \citep{Fan1993, Schussler1994}.  
\citet{D'Silva1993} and \citet{Fan1994} showed how rising flux tubes will be rotated 
out of their initial plane of symmetry by the Coriolis effect. Namely, for flux tubes initially
lying parallel to the equatorial plane, the leading side of the rising loop will rotate 
toward the equator while the following side of the rising loop will rotate
away from the equator. This introduces a tilt with respect to the equator of the tubes 
at their apex which will have a left handed sense 
in the Northern hemisphere and a right handed sense in the Southern hemisphere, 
causing the leading spots in both hemispheres to emerge closer to the equator than the
following spots. \citet{D'Silva1993} and \citet{Fan1994}
showed that the magnitude of the tilt will increase with emergence latitude, in 
agreement with observations.  A number of authors then predicted that flows will be 
set up along the rising flux tube axis, counter to the sense of solar rotation, with
the result that the magnetic field strength in the leading spot will
be more concentrated, and therefore stronger, than that in the following
spot \citep{Fan1993, Moreno-Insertis1994, Caligari1995}.
This may explain observed asymmetries in leading versus following spots,
where the leading spot is often more concentrated and coherent than the following spot.
An example of several of these thin flux tube simulations are shown in Figure
\ref{fig:Weber2011}, from \citet{Weber2011}, which shows snapshots of rising 
tubes encircling the Sun from simulations starting with 15 kG, 40 kG and 100 kG field 
strengths. The top three panels in this figure are shown looking down from solar North,
so that solar rotation is counterclockwise from this perspective. This perspective shows
how the leading side of the loop slopes gradually up to the surface, while 
the following side of the loop has a steeper slope up to the surface, as the counter-rotational
flow pushes the loop in toward this following side. 

Subsequent studies then focused on the effects that convection and turbulence have on such
thin flux tubes as they pass through
the convection zone. \citet{Longcope1997} derived a set of equations to describe
how the internal twist of a flux tube, i.e., the rotation of flux tube's fieldlines
about its axis, would be modified if the flux tube itself were `writhed,' i.e., distorted
into helical shape. \citet{Longcope1998} then investigated how thin flux tubes
would be modified by this effect, called the `$\Sigma$ effect,' when exposed to the
turbulent spectrum present in the convection zone. They found the $\Sigma$ effect
can explain the magnitude and spread in observed twist measured in sunspots at the
photosphere, as well as the dependence of the average twist on latitude
\citep{Pevtsov1995}. \citet{Weber2011, Weber2013SP} then took the analysis
further by simulating the evolution of thin flux tubes, modeled by the equations derived in 
\citet{Spruit1981}, as they move through a three dimensional realization of the dynamic
convection zone, simulated with the ASH code \citep{Miesch2006}.
With this, \citet{Weber2011, Weber2013SP} were able to study how the emergence latitude
and the tilt of the emerging regions depended on the initial field strength.
A key conclusion from this study was that flux tubes with initial field 
strengths of 40-50 kG  were best able to reproduce Joy's law and observed asymmetries 
between leading and following spots.

\begin{figure}[ht]
    \centering
    \includegraphics[width=0.40\textwidth]{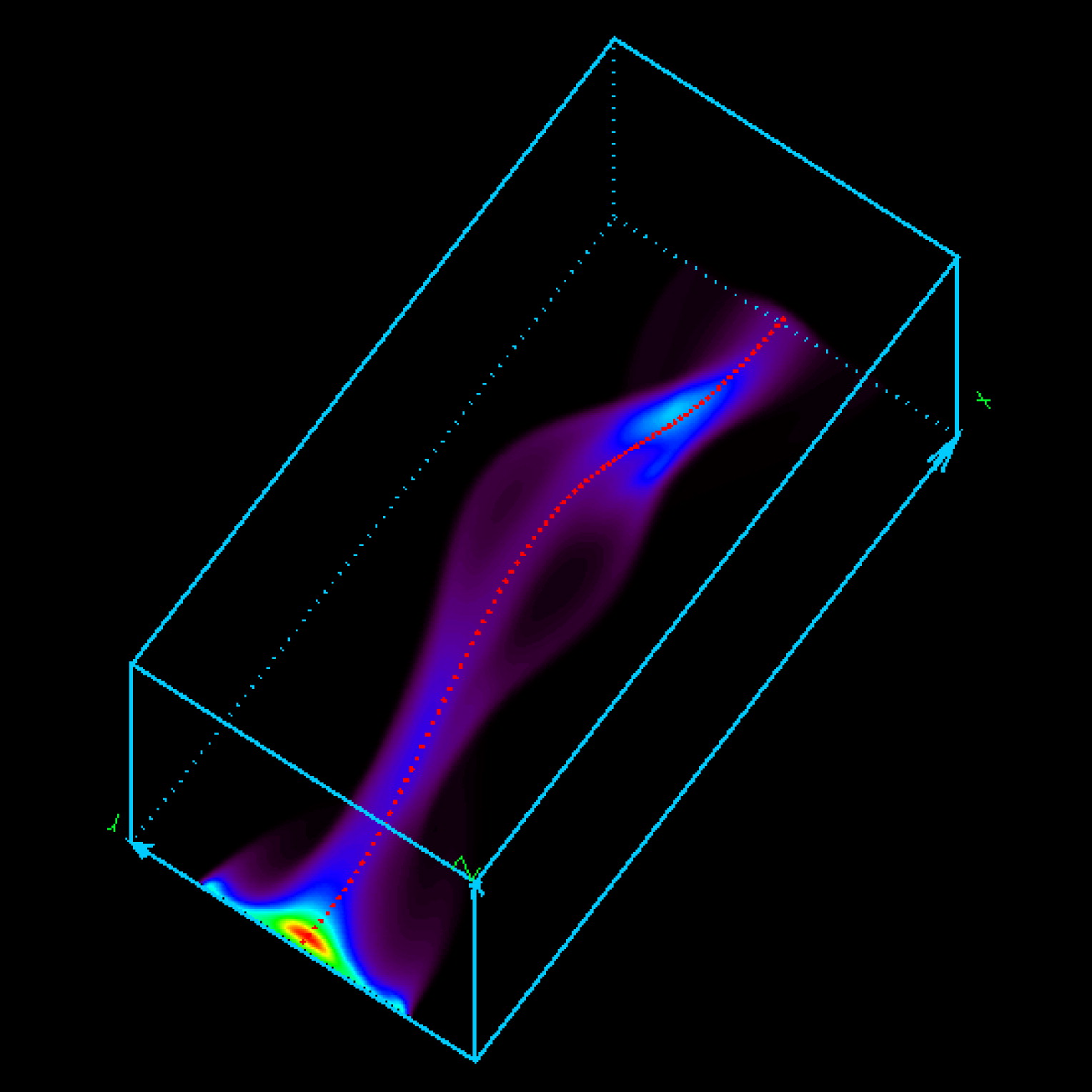}
    \includegraphics[width=0.40\textwidth]{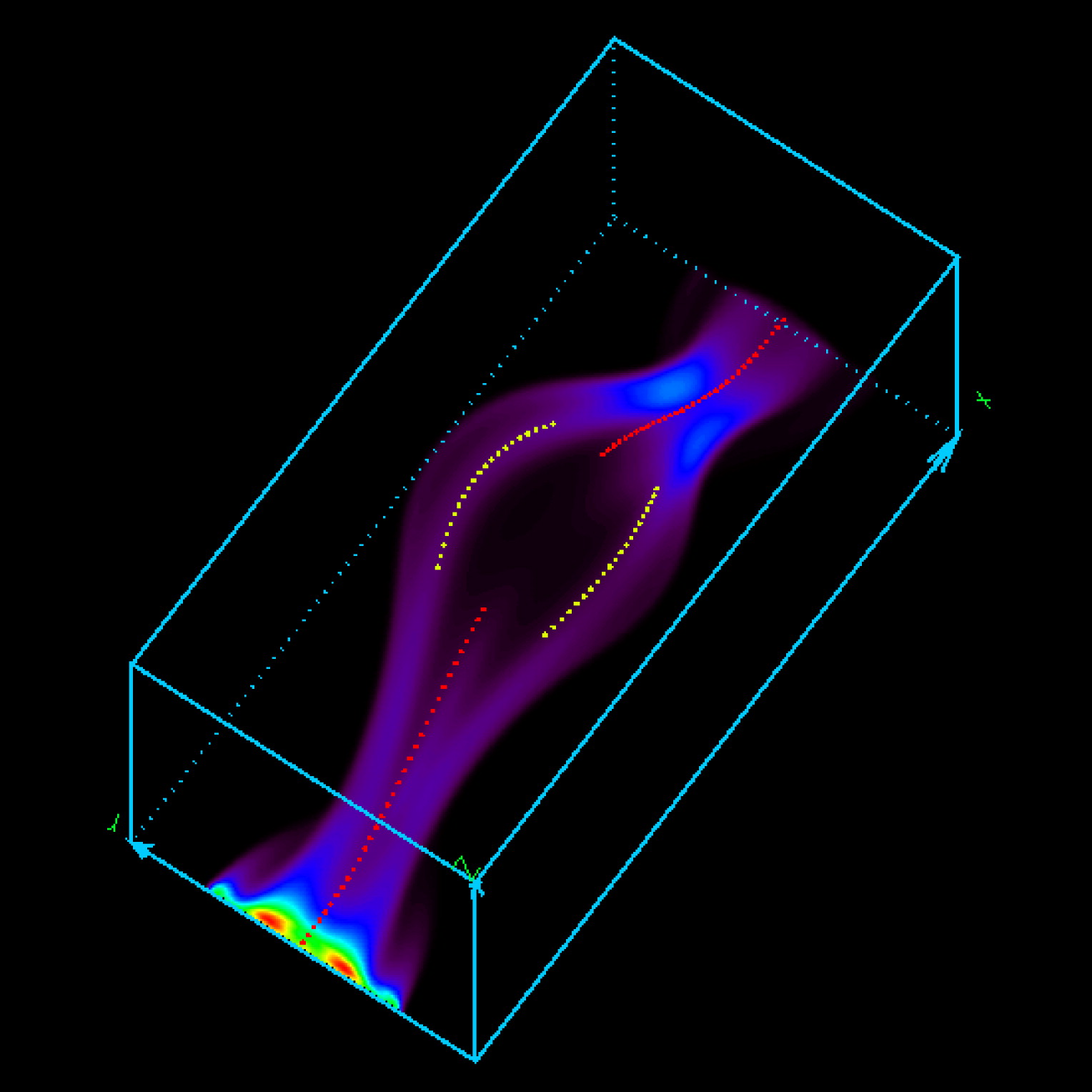}
	\caption{Snapshots from two simulations of three dimensional buoyant flux tubes 
	rising through a stratified convection zone. The left panel shows a flux
	tube with a finite twist which stays coherent as it rises. The right panel
	shows a tube with no twist which breaks into two as it rises. Reproduced
	from \citet{Abbett2000}
    \label{fig:Abbett2000}}
\end{figure}

\subsubsection{Two dimensional and three dimensional buoyancy studies:\label{section:2Dbuoyancy}}
In addition to the behavior of the large scale features of buoyant magnetic flux tubes,
the internal structure and dynamics of these flux tubes are also important, 
in part because they can modify the large scale dynamics. \citet{Schussler1979} explored
the effect of hydrodynamic forces on two dimensional flux tubes. In addition to applying 
a drag on the buoyant rise of these flux tubes, as described in \citet{Parker1979II},
the counter-rotating eddies generated by the flow of fluid around a tube as it rises through
the convection zone can break it up into a pair of tubes. \citet{Longcope1996}
followed up on this study with a series of two dimensional simulations, one of which is
shown in Figure \ref{fig:Longcope1996}, exploring how quickly such
a flux tube will break up, and showing that the previously buoyant flux tube will
lose its buoyancy once it breaks up, as the pair of counter-rotating flux tubes
generated by this splitting exert a downward drag on each other.

\citet{Emonet1998} then explored how magnetic twist in a buoyant 
two dimensional flux tube could maintain the tube's coherence against these destructive
vortex flows. Such twisted flux tubes are often referred to as flux ropes.
Here the twist field, wrapping around the flux rope axis in a helical 
manner, acts as a set of hoops holding the flux rope together
via the inward directed tension force of these twisted fields.  Their simulations
showed that the transition from coherent to incoherent tubes occurs 
when the Alfv\'en speed of the twist field exceeds the terminal rise speed of the 
flux tube, i.e., the speed at which the buoyancy force equals the drag force. Figure
\ref{fig:Emonet1998} shows one of these simulations where a significant
section of the flux tube maintains its coherence against the destructive eddies:
here, the outer parts of the flux rope with weaker field strength and therefore
smaller Alfv\'en speed are sloughed off by the flow, but the inner, stronger
parts of the flux rope maintain their coherence and continue to rise.

\begin{figure}[ht]
    \centering
    \includegraphics[width=0.3\textwidth]{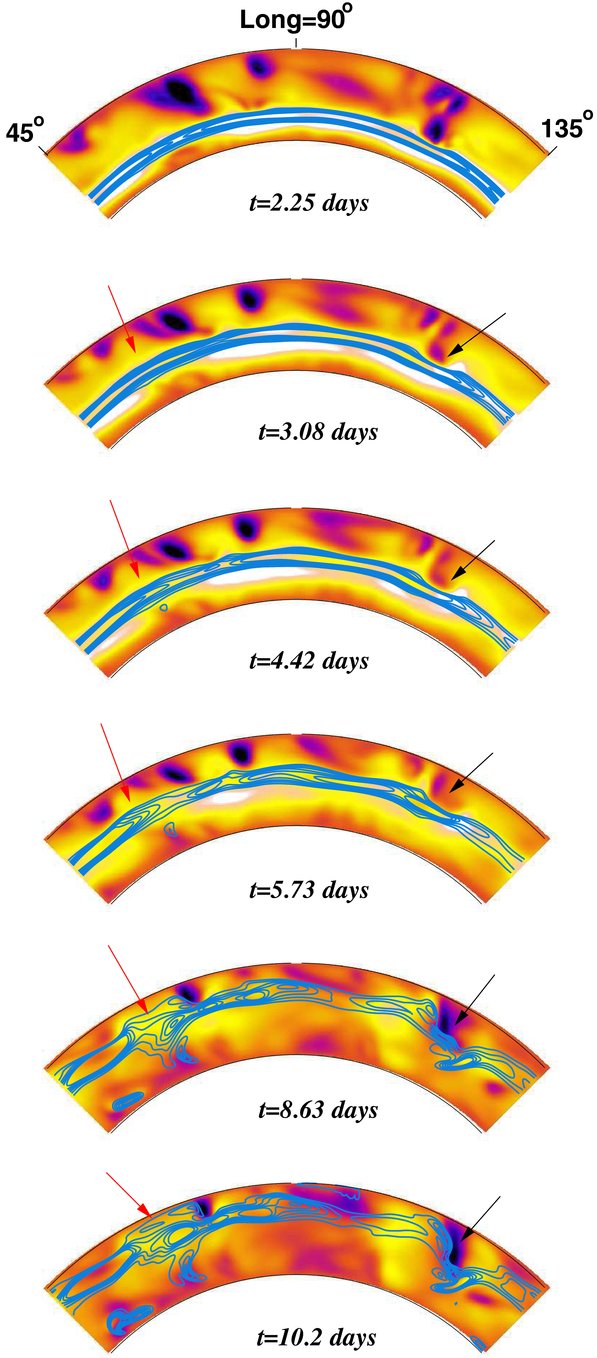}
	\caption{Simulation of a flux tube rising through a rotating convection zone
	over the course of ten days. The color scale shows radial flow: blue for
	downflow and yellow for upflow. The blue contour lines show contours of magnetic
	energy, indicating the location of the magnetic flux tube. The black arrow
	points out a downflow, while the red arrow points out an upflow.
	Figure reproduced from \citet{Jouve2009}.
    \label{fig:Jouve2009}}
\end{figure}

\citet{Abbett2000, Abbett2001} then explored several
three dimensional effects which can also help buoyant convection zone flux tubes
to maintain their coherence against these hydrodynamic forces. \citet{Abbett2000} 
simulated the rise of an initially horizontal flux tube,
where only a segment near the center of the length of the flux tube was made buoyant.
This flux tube, as shown in Figure \ref{fig:Abbett2000} then rises with
an $\Omega$ shape. They found that this tube will remain coherent if it has
sufficient twist (left panel) but that it will break into two tubes if there
is no twist in the tube (right panel), as it did in the two dimensional 
simulations of \citet{Schussler1979} and \citet{Longcope1996}.
Interestingly, they found that the tubes remain more coherent in all cases
in three dimensional $\Omega$-shaped tubes relative to equivalent two dimensional
rising tubes, and that the coherence increases as the radius of curvature of the buoyant
$\Omega$-loop decreases.
The authors conclude that this coherence for short buoyant sections was 
due to the twisting up of tubes during breakup, which effectively
adds the stabilizing magnetic twist component explored by \citet{Emonet1998}.
\citet{Abbett2001} then followed up this study with an exploration
of the effects of the Coriolis force on rising flux tubes. Here, they found
that the extra flows induced in the flux tube by the Coriolis force could also
keep a rising, three dimensional flux tube coherent.

\citet{Jouve2007, Jouve2009}, \citet{Fan2008}, and \citet{Jouve2013}
carried out a series of global-Sun simulations exploring how a three-dimensional
twisted flux tubes initially circling the Sun at the base of the convection
zone would rise buoyantly toward the surface. \citet{Jouve2007} and \citet{Fan2008}
studied the rise of flux tubes in a stable (non-convecting) solar interior, and
confirmed that a minimum twist is required in these tubes for them to successfully
reach the surface. \citet{Fan2008} found that for even relatively high twist
(relative to what is observed at the surface), tubes can lose about half of their
flux as they rise to the surface. Several studies were then carried out
to explore the effects of pre-existing convection on these rising flux tubes.
Figure \ref{fig:Jouve2009}, from \citet{Jouve2009}, shows a time sequence for one of
these simulations. Here, over the course of 10 days, one can see how most
of the flux rope rises to the surface, but also how the rope is significantly
distorted by the convective up- and down-flows impinging upon it. In particular,
the segment of the flux rope lying in an upflow (red arrow) rises faster
and farther than the rest of the rope, while the segment lying below a downflow
(blue arrow) rises more slowly. These results are consistent with
similar simulations carried out by \citet{Fan2003} on the dynamics
of buoyant three dimensional flux tubes in a convecting, Cartesian atmosphere.

\subsubsection{Emergence through the photosphere:}
Even after magnetic flux makes it to the photospheric surface, it still 
must undergo a further instability to emerge into the corona. This is because
the photosphere is the transition between the unstably stratified convection
zone and the stably stratified chromosphere and corona. For a
review of these emergence instabilities and associated phenomena, see
\citet{Archontis2012}. A flux tube which
is buoyant enough to rise to the photosphere will lose its buoyancy there
and stall just below the photosphere until enough flux builds up there to 
push it through the photosphere.  \citet{Shibata1989} performed an
early investigation of this mechanism, simulating the two dimensional
evolution of a horizontal sheet of magnetic field which rises up from
the high convection zone to the photosphere, and then later emerges
to either populate a magnetic field-free corona, or to interact with
a pre-existing coronal field. \citet{Matsumoto1993, Matsumoto1998} then extended this study
to three dimensions, first exploring the interaction of multiple emerging flux bundles,
and then exploring the formation of multiple active regions from a single,
helically kinking flux tube, as shown in Figure \ref{fig:Matsumoto1998}.

\begin{figure}[ht]
    \centering
    \includegraphics[width=0.90\textwidth]{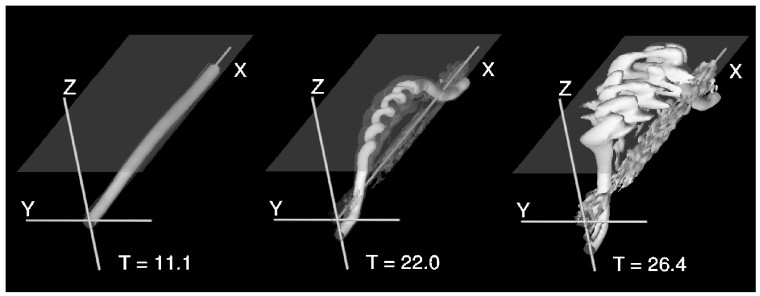}
	\caption{Simulation of the rise and emergence of a kinking magnetic flux rope.
	The grey plane in each panel is the solar surface. Two isosurfaces are
	shown in each panel, one for weak field (magnetic pressure over gas pressure is .04)
	and one for strong field (magnetic pressure over gas pressure is .36).
	This shows the tube both arching in an $\Omega$ shape and kinking as it
	rises and emerges. Figure reproduced from \citet{Matsumoto1998}.
    \label{fig:Matsumoto1998}}
\end{figure}

\citet{Magara2001} simulated the 2D emergence of the cross section of a flux tube, 
showing that the tube pinches down at the surface, with the bottom half of 
the flux tube piling up below the surface, and the top half of the flux tube 
emerging and expanding into the corona.  \citet{Fan2001} explored a similar scenario 
in 3D, simulating the emergence of a buoyant segment of a flux tube. This
study showed that even in 3D, the bottom half of a twisted flux tube is
trapped below the photosphere during emergence. \citet{Fan2001} found that this
is because dense convection zone material is trapped on the concave-up fieldlines
which lie below the axis of a twisted flux tube. The result, shown in Figure
\ref{fig:Fan2001} is that the axis of the flux rope lies below the surface,
with the two legs of the flux rope moving off to the side to form a pair of 
sunspot-like magnetic field concentrations. 
\citet{Archontis2004} pointed out that the magnetic interchange instability
\citep{Schubert1968, Acheson1979} is a useful way to diagnose when a bundle
of flux trapped below the photosphere will emerge. This instability
is similar to the Rayleigh-Taylor instability, where a low density plasma
lies underneath and supports a higher density plasma, but in this case,
the lower density plasma is a low density, high magnetic field plasma.
\citet{Toriumi2010, Toriumi2011} explored this effect in detail,
focusing on how initially buoyant magnetic fields will be halted at
the photosphere, and then in a second stage will become buoyant again
and emerge into the corona. \citet{Hood2009} and \citet{MacTaggart2009}
explored scenarios which would allow more of the flux tube cross section
to emerge.
They hypothesized that the emergence of a tube with a significantly arched 
axis could allow convection zone plasma to drain out of an emerging flux tube, 
even below the axis. The idea here was that the downward arch of the tube itself 
would be larger than the upward arch of the fieldlines within the tube, and so there
would be no concave up fieldlines to trap dense plasma during the emergence. 
To test this, they simulated the emergence of an initially toroidal twisted flux tube, with
a higher curvature than tested before. These simulations showed that
the axis of the flux tube then easily emerges, confirming their hypothesis.

\begin{figure}[ht]
    \centering
    \includegraphics[width=0.60\textwidth]{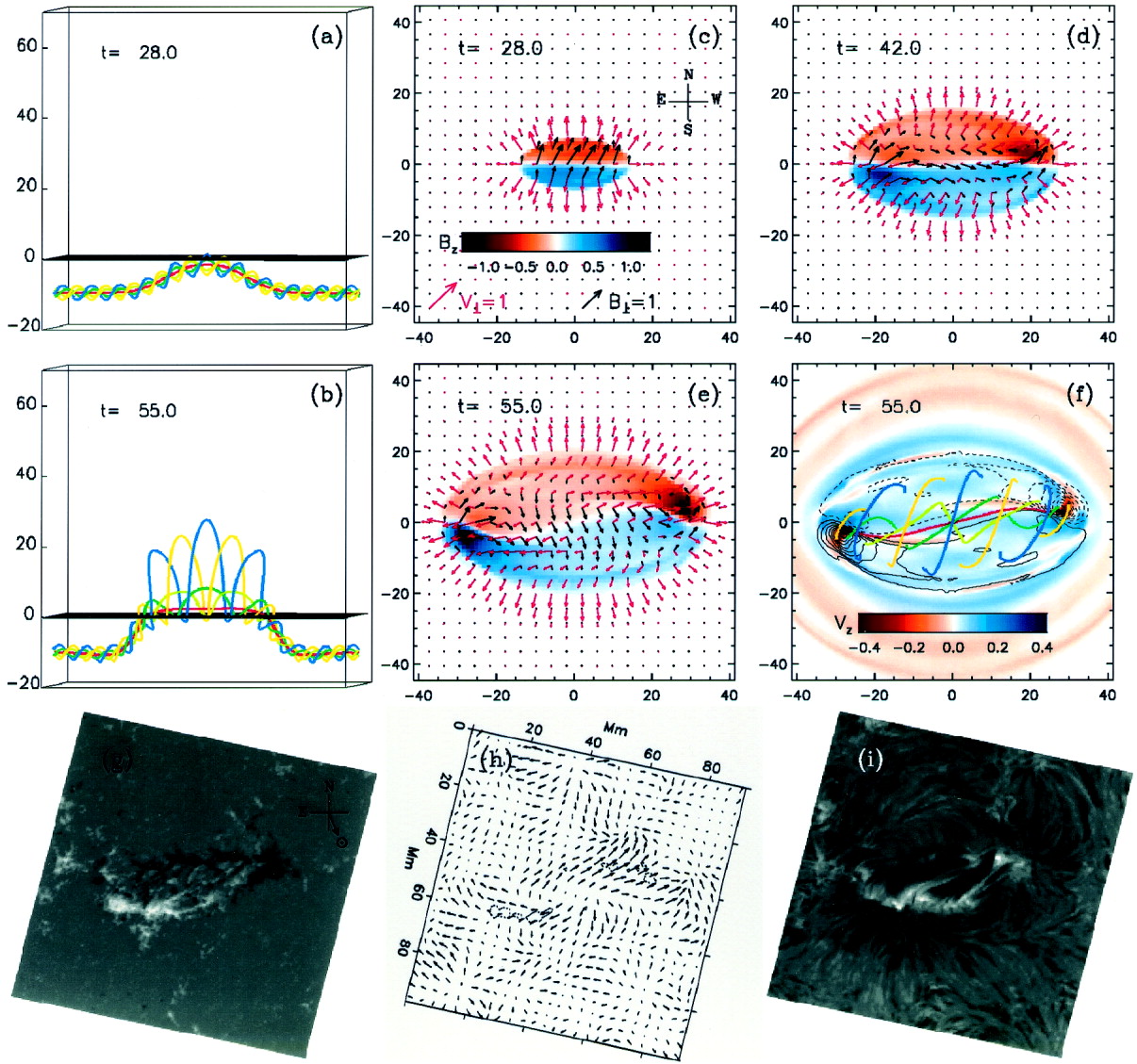}
	\caption{Simulation (top two rows) of the rise and emergence of an $\Omega$
	loop formed by a buoyant magnetic flux rope in the solar convection zone. Fieldlines
	are shown in panels (a) and (b), showing how the concave-up fieldlines below
	the flux rope axis are trapped at the photosphere (horizontal plane). Panels
	(c-e) show the vertical magnetic field at the photosphere during this emergence (color scale) as well as the horizontal magnetic field (black vectors) and the horizontal
	velocity (red vectors). Panel (f) shows the same fieldlines as panel (b) but from
	above the photosphere. Panels (g-i) show an example of the observed features of an emerged active region: the line of sight magnetic field (g), the
	horizontal velocity field (h), and an image in the Hydrogen $\alpha$ wavelength (i), from \citet{Strous1996}. Figure reproduced from \citet{Fan2001}.
    \label{fig:Fan2001}}
\end{figure}

The question of why some fields emerge and some don't is important
for understanding the process of sunspot formation, and for understanding
how much of the magnetic field populating the convection zone eventually
manifests itself in the corona.  \citet{Murray2006} and \citet{Murray2008} attacked this problem
by carrying out a parameter study investigating effect of magnetic
field strength and magnetic twist on the efficiency of emergence.
They showed that flux tubes with larger field strength are better
able to emerge, as do flux ropes with larger twist.  Lower field strength 
or twist in a buoyant flux tube results in it being trapped for a
longer time below the photosphere, and sometimes these tubes are unable to emerge.
\citet{Murray2006} postulated that this failure to emerge at low field strength was due to the
field strength being too weak to excite the interchange instability discussed above \citep{Acheson1979}. Similarly for weak twist this instability is not
excited, now because the weak twist is ineffective at keeping the magnetic
field coherent in the face of high pressure fluid and plasma flow forces, and 
so the field is dispersed and its strength ends up too weak even when it initially
was quite strong.
\citet{Archontis2013} then revisited this experiment and found
that flux tubes with significantly lower twist than the limit established
by \citet{Murray2006} could emerge, but that
they emerge on much longer timescales, and in a less coherent fashion.
Low twist emergence, shown in Figure \ref{fig:Archontis2013} proceeds
via the emergence of two side lobes of the arched tube, whereas higher twist tubes 
first emerge at the apex of the $\Omega$ loop.

\begin{figure}[ht]
    \centering
    \includegraphics[width=0.60\textwidth]{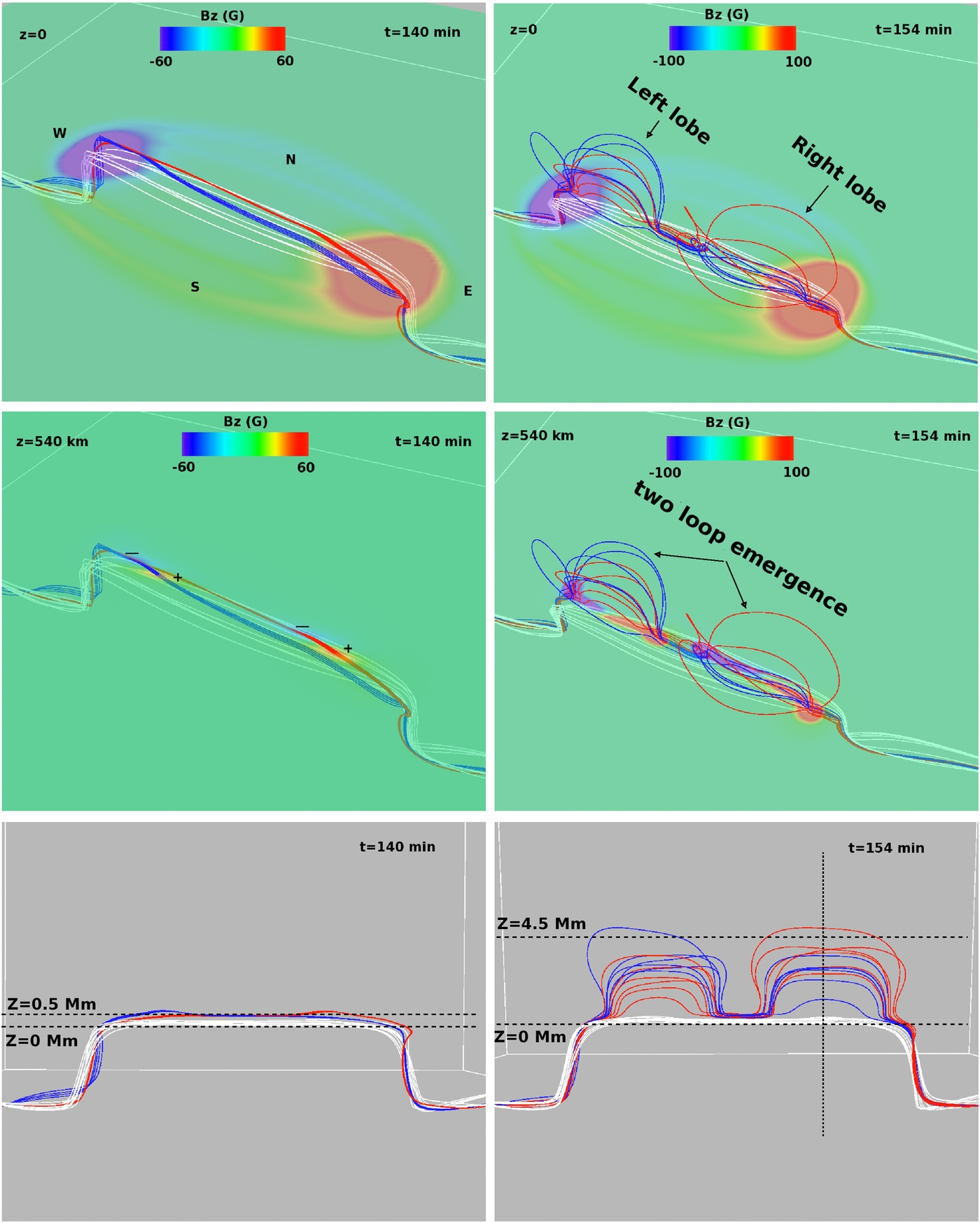}
	\caption{Simulation of the emergence of a very low twist flux rope into the corona. The left panels show the flux rope after it has reached and piled up at the photosphere (t=140min). The right panels show the flux rope after two lobes have emerged (t=154min). The vertical magnetic field is shown by the color scale: in the top panels, this is shown at the solar surface (z=0), while in the center panels, this is shown 540 km above the surface. The fieldlines are plotted as the colored lines. The bottom
	panel shows the side view. Figure reproduced from \citet{Archontis2013}.
    \label{fig:Archontis2013}}
\end{figure}

\subsubsection{Partial ionization:} While the convection zone and the overlying corona are 
hot enough that the plasma there is nearly entirely ionized, the thin layer,
about 1000km thick, of the photosphere and chromosphere between
them is cool enough that neutral plasma can be the dominant species.
Several efforts have investigated how these
neutral species affect emerging flux. The primary mechanism expected
to be in play is the extra diffusion term from the neutral species 
interacting with the ionized species. The neutral species can flow
across magnetic fieldlines while the ionized species are bound
to magnetic fieldlines, so a friction is generated between the two, resulting
in the so-called Pedersen resistivity \citep{Cowling1956} 
\citep[see, e.g.,][for a review of this resistivity, and of the related Cowling resistivity]{Leake2014SSR}. 
\citet{Leake2006} and \citet{Arber2007} included
this resistive term in a set of flux emergence simulations, and found
that the extra resistivity allows fields to emerge more quickly,
while also reducing the magnetic forces in the emerging structures.
\citet{Leake2013} later showed that in two dimensional
configurations (where the plasma cannot drain along the tube's axis),
this diffusion term provides an efficient way for excess convection
zone plasma to drain from an emerging flux rope.
\citet{Martinez-Sykora2015} then showed that the extra Pedersen
conductivity plays an important role in heating the chromosphere during
flux emergence, partially overcoming the excessive cooling seen in 
the chromosphere in fully ionized flux emergence simulations,
such as that reported in \citet{Archontis2004}.

\begin{figure}[ht]
    \centering
    \includegraphics[width=0.45\textwidth]{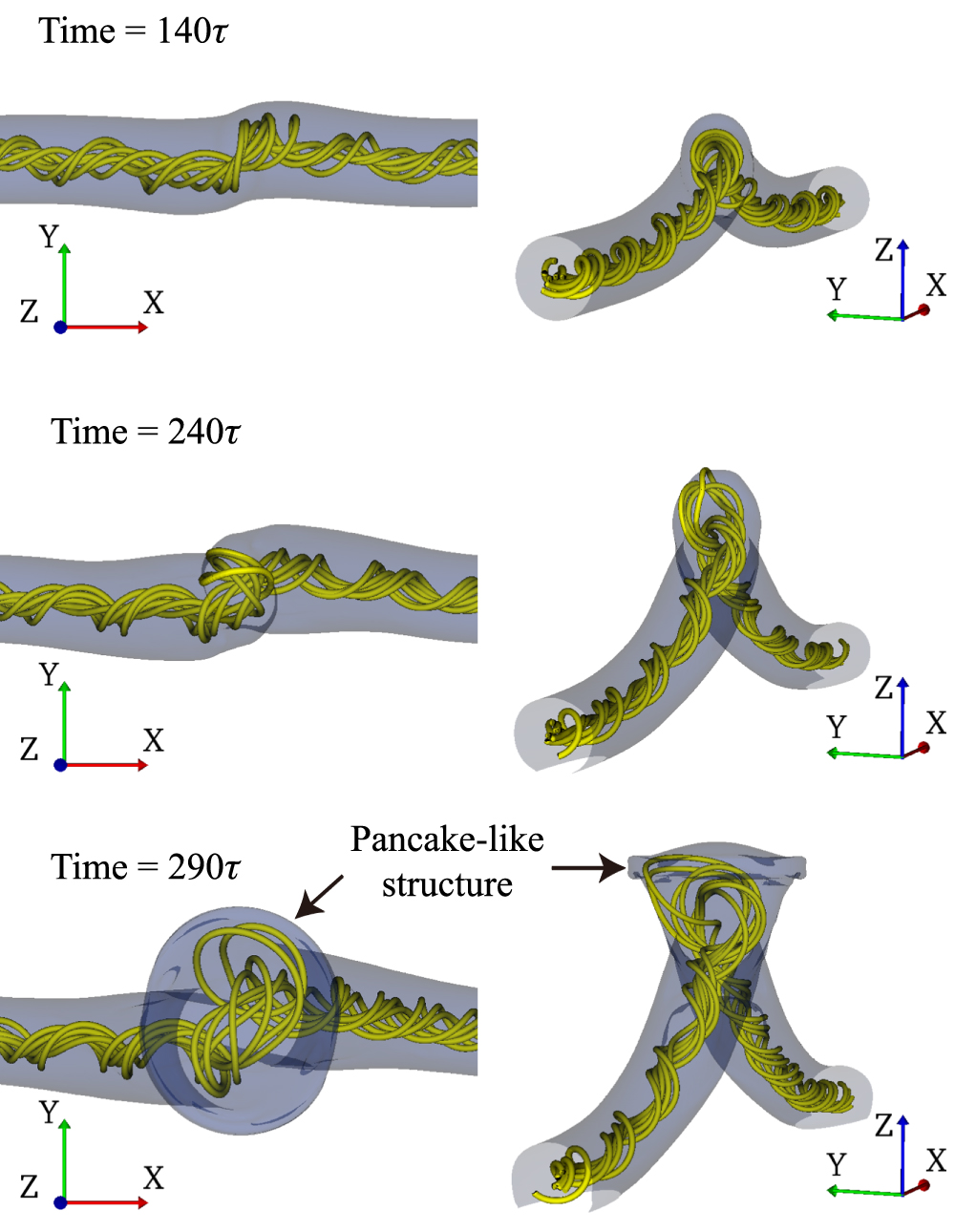}
	\caption{Simulation of the emergence of a kinking flux rope from the convection zone to form a $\delta$-spot like active region. Isosurfaces are shown in each panel at a value of half of the initial peak magnetic field strength in the tube. The yellow lines show magnetic fieldlines within the tube. Times 140$\tau$ and 240$\tau$ show how the rise of the	tube in an initial $\Omega$ shape is distorted into a helical, writhed shape by the kink instability. The bottom panels, at 290$\tau$ show the structure as it hits and
	emerges through the surface in a compact, rotated configuration, consistent with $\delta$-spot characteristics.
	Figure reproduced from \citet{Takasao2015}.
    \label{fig:Takasao2015}}
\end{figure}

\subsubsection{$\delta$-Spots:} 
As discussed above, most active regions emerge to form well separated
pairs of sunspots, aligned nearly parallel to the equator, but a
small number emerge in a compact, highly rotated formation called
a $\delta$-spot formation. Correspondingly, most flux emergence
studies aim to explain the classic bipolar, well separated regions.
But flux emergence studies also provide a promising way to explore 
the formation of $\delta$-spots. Following the hypotheses of \citet{Tanaka1991}
that these spots may be formed by the helical deformation of an
emerging flux rope, \citet{Linton1996, Linton1998, Linton1999} and 
\citet{Fan1998, Fan1999}
explored the implications of the helical kink instability for
$\delta$-spot characteristics. These studies explored the helical
deformation which acts on twisted flux tubes with a twist in excess
of the kink stability threshold \citep[see, e.g.,][]{Shafranov1957, Kruskal1958}.
\citet{Linton1999} simulated the formation of a kinked, knotted
structure in a convection zone environment, and argued that if
this structure were to emerge into the corona, it would form
a compact sunspot, due to its knotted structure, and that it would
rotate significantly with respect to the usual Hale orientation,
due to the helical writhe of the kink. \citet{Fan1999} showed
that this compact distorted structure would naturally form,
even for initially stable structures, as a flux rope rose
through the convection zone and expanded, as predicted by \citet{Linton1996}. 
\citet{Matsumoto1998}, \citet{Takasao2015}, \citet{Toriumi2017}
and \citet{Knizhnik2018} carried this study further exploring
the photospheric and coronal signatures of kinking flux tubes
when they were allowed to emerge through the photosphere. Figure
\ref{fig:Takasao2015}, from \citet{Takasao2015}, shows
how the emerging structure is indeed compact and rotated, with
a high level of twist, consistent with many $\delta$-spot structures.
\citet{Toriumi2014b}, \citet{Fang2015}, \citet{Toriumi2017}, and \citet{Jouve2018}
explored alternate scenarios for forming these complex structures wherein 
multiple flux ropes interact with each other during their rise
through the convection zone, or where multiple parts of a single flux 
rope collide with each other prior to emergence. These studies 
also found a good correspondence between key aspects of these
structures and common types of $\delta$-spots.

\begin{figure}[ht]
    \centering
    \includegraphics[width=0.60\textwidth]{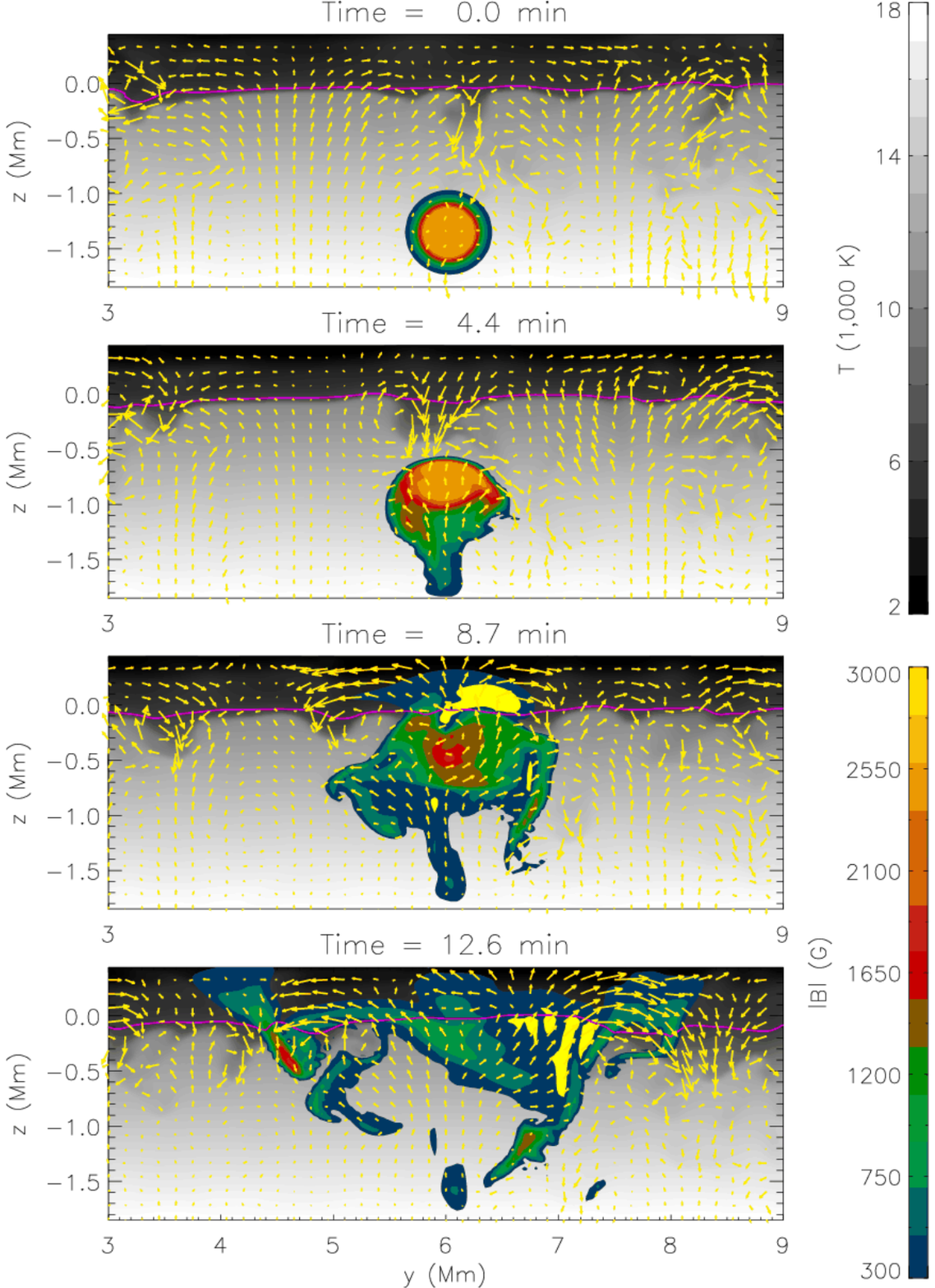}
	\caption{Two dimensional cuts from a three dimensional simulation of the rise
	of a twisted magnetic flux tube through convective flows just below
	the photosphere. Temperature is shown by the greyscale. Magnitude of
	magnetic field is shown by the color scale, flows are shown by the yellow
	vectors, and the red line shows the surface. Time advances from top to bottom,
	showing how the buoyant tube rises to the surface and emerges all the while
	being torn apart by the convective flows. Figure reproduced from \citet{Cheung2007}.
    \label{fig:Cheung2007}}
\end{figure}

\subsubsection{Emergence into Coronal Fields:}
A number of explorations have focused on how emerging fields
interact with pre-existing coronal fields. \citet{Heyvaerts1977} proposed and explored the theoretical consequences of a mechanism by which newly emerging magnetic flux would interact with pre-existing coronal fields to generate a reconnecting current sheet and drive flaring, plasma heating, and particle acceleration. \citet{Yokoyama1995}
explored such a mechanism numerically by simulating the emergence of a loop from a horizontal sheet of magnetic
flux into an inclined coronal field, finding that this initiates
reconnection and jet-like flows up into the corona. \citet{Archontis2004}
and \citet{Galsgaard2005} simulated emergence into a horizontal overlying field, 
finding that while much of the flux rope is trapped below the photosphere,
sufficient flux emerges to generate current sheets and reconnection
at the interface between the emerging flux and the coronal field. 
This interaction of emerging fields with both a field-free corona
and with overlying coronal fields has since led to numerous models of 
coronal flares and eruptions \citep[see, e.g.,][]{Manchester2004, 
Archontis2008, Leake2014}. Chapter 4 explores in
more detail these flares and eruptions resulting from the emergence
of magnetic fields into the corona.

\begin{figure}[ht]
    \centering
    \includegraphics[width=0.60\textwidth]{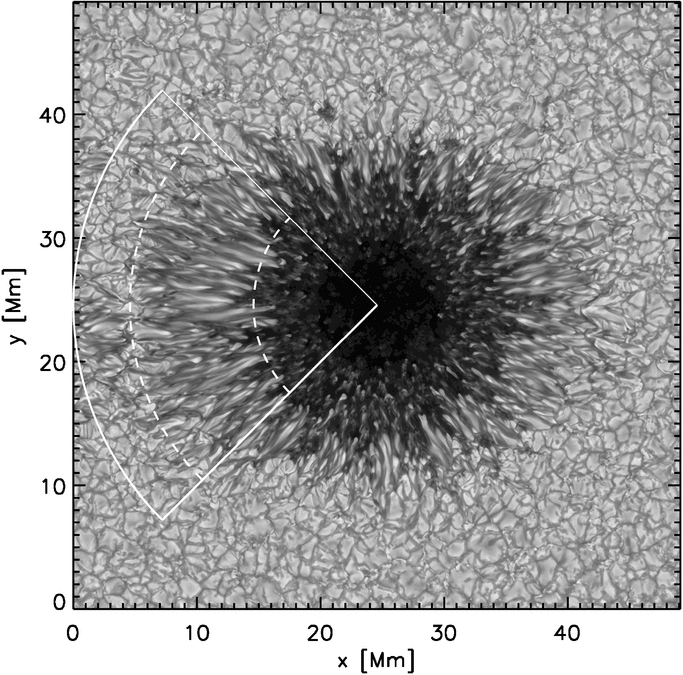}
	\caption{Simulated white light emission from modeled sunspot formation in a convecting atmosphere. These simulations by \citet{Rempel2009a, Rempel2009b, Rempel2011}, where a magnetic field is added to a convecting atmosphere to generate a sunspot, produce realistic umbral and penumbral structure and emission. Figure reproduced from \citet{Rempel2011}.
    \label{fig:Rempel2011}}
\end{figure}

\begin{figure}[ht]
    \centering
    \includegraphics[width=0.60\textwidth]{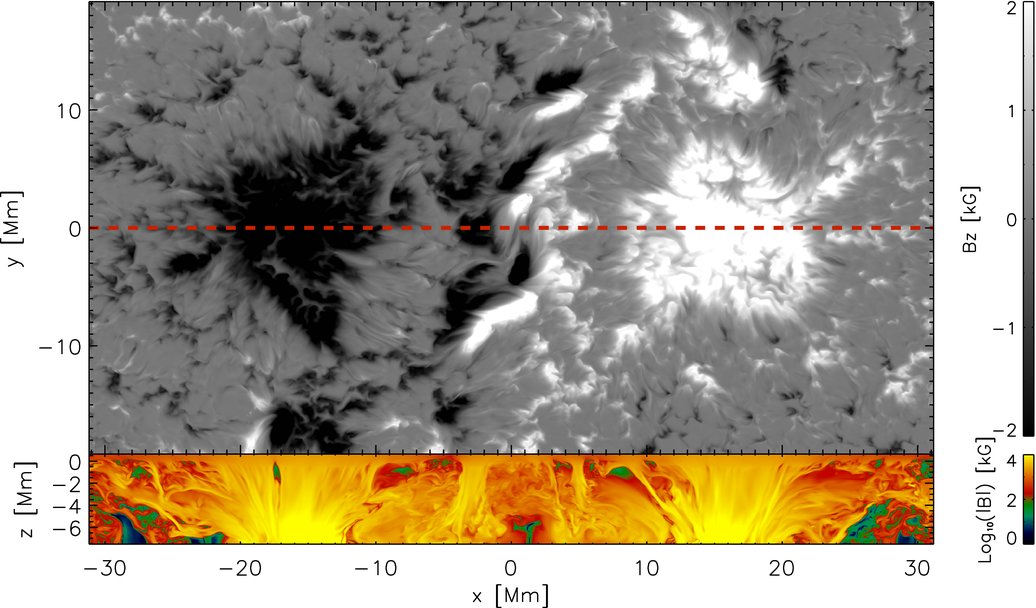}
	\caption{Synthetic magnetogram from a simulation of the formation of an active region
	sunspot pair from the emergence of magnetic flux from the convection zone. The greyscale shows the vertical magnetic field at the photosphere. The color scale shows the log
	of the magnetic field strength in a cut into the convection zone, perpendicular to the photosphere, made where the dashed red line crosses the photospheric plot. Figure reproduced from \citet{Cheung2010}.
    \label{fig:Cheung2010}}
\end{figure}

\subsubsection{Effects of radiation and convection on emergence:}
Convective motions, as discussed above, can play an important role in
the evolution of rising flux tubes.  \citet{Dorch2001} first investigated the 
interaction of buoyant, rising flux tubes with surrounding convective motions. 
They found that even for a flux rope which in theory had sufficient twist to 
remain coherent during its rise, convective flows strip off fields from the flux 
rope as it rises. \citet{Fan2003} simulated a similar scenario, and established 
that the magnetic energy density of buoyant flux tubes must be roughly three 
times the kinetic energy of convective downflows for the tubes to successfully 
rise toward the surface.

A number of studies have investigated how the transition from the convecting
atmosphere below the photosphere to the stable
chromosphere and corona affects and interacts with emerging flux. To
achieve this transition, radiative effects must be included in simulations,
so that the loss of heat from the photosphere can generate the required
transition from a convectively unstable to a convectively stable zone across
a thin layer. \citet{Cheung2007} carried out one such study, investigating
both the dynamics in the convection zone and during emergence. 
Figure \ref{fig:Cheung2007} 
shows one of these simulations where the tube is significantly distorted by the
convective motions, but still reaches the surface to
emerge. \citet{Cheung2007} and \citet{Cheung2008} then showed that these emerging 
fields can disrupt the convective flow pattern seen at the surface, consistent
with observations of emerging flux \citep[see also][]{Martinez-Sykora2008, 
Martinez-Sykora2009, Tortosa-Andreu2009, Fang2012}.
\citet{Stein2011} showed that such emerging 
flux can concentrate at the surface during its emergence to form small scale, high 
field strength regions often seen in observations, and referred to as pores.
See \citet{Stein2012LRSP} and \citet{Stein2012} for reviews of the interactions of 
convection and magnetic fields. 

These earlier convective emergence simulations focused on smaller scale fields,
which disrupted the convection but did not form sunspot-like concentrations. Larger
scale simulations then allowed for sunspot formation to be explored.
Figure \ref{fig:Rempel2011} shows simulated white light intensity emission from
a sunspot simulated by \citet{Rempel2009a, Rempel2009b} and analyzed in \citet{Rempel2011}. This sunspot structure is generated via the ad-hoc insertion of predominantly vertical magnetic field into a simulation in which convection has previously been set up. The structure and emission of the resulting sunspot umbra and penumbra display a remarkable resemblance to sunspot observations, such as that shown in Figure \ref{fig:Tiwari2015_white_light}. 
Figure \ref{fig:Cheung2010} shows a simulated surface magnetogram and a vertical cut of magnetic field strength below that magnetogram for a more self-consistently formed
sunspot pair which is generated by the emergence into the corona of a toroidal flux tube injected at
8,000 km below the photosphere, at the bottom boundary of a convecting simulation.
\citet{RempelCheung2014} then extended this type of simulation to investigate both
the formation of a sunspot via convective flux emergence and the decay of the sunspot via convective turbulence. Critically, in this case the torus injected at the bottom
boundary had zero twist but its field still rose to the surface to form a sunspot pair.
In this case, the field was caught up in strong updrafting convection flows and so
those flows carried the flux to the surface. This contrasts with the non-convecting
cases discussed in \S \ref{section:2Dbuoyancy} where an untwisted tube will be torn apart as it
rises through a largely static atmosphere and with the convecting examples discussed 
above, where convection which flows counter to the rise of flux trying to emerge
tears it apart rather then enabling emergence. Clearly convection can both enable
and inhibit flux emergence, and determining when and how often it will play either
of these roles is a promising avenue for further research.

A statistical study of how magnetic fields are distributed across the photosphere
in emerging active regions by \citet{Dacie2016, Dacie2017} found that convective
simulations such as these accurately reproduce the observed spectral distributions 
of magnetic fields
during their evolution, while simulations of flux emergence without convection are
only able to match magnetic field distributions once the emergence is mature.
These convective simulations are remarkable in that they can reproduce statistical
measures of sunspot characters such as this, and that they can produce 
magnetic field and white light emission signatures that, to the eye, are largely 
indistinguishable from observed sunspot signatures.

These various explorations, from one dimensional models focusing on the large scale
evolution of flux tubes, to two and three dimensional models focusing on the
interaction of flux tubes with the plasma surrounding them and the internal
reaction of flux tubes to their own twist, to highly sophisticated models
investigating the effects of convection and radiation loss on emerging
flux tubes have all contributed critical knowledge to our understanding
of how magnetic fields emerge into the solar corona and form sunspots
and active regions. Many important questions remain, and arguably all
of these tools will have something to contribute to advancing our
understanding to the next stage.

\vskip 0.08in
\acknowledgements  Funding for M. Dikpati was provided by the
National Center for Atmospheric Research, which is sponsored by the 
National Science Foundation. R. Howe acknowledges computing support from the National Solar Observatory and the University of Birmingham. 
Funding for M. G. Linton was provided by
the NASA Living with a Star and Heliophysics Supporting Research programs 
and by the Chief of Naval Research. SDO data was provided courtesy of SDO (NASA) 
and the 
HMI and AIA consortium. WSO data was provided via the web site http://wso.stanford.edu at courtesy of J.T. Hoeksema. The Wilcox Solar Observatory is currently supported by NASA.  This work utilizes data obtained by the Global Oscillation Network Group (GONG) program, managed by the National Solar Observatory, which is operated by AURA, Inc. under a cooperative agreement with the National Science Foundation. The data were acquired by instruments operated by the Big Bear Solar Observatory, High Altitude Observatory, Learmonth Solar Observatory, Udaipur Solar Observatory, Instituto de Astrof\'{\i}sica de Canarias, and Cerro Tololo Interamerican Observatory. BiSON is funded by the UK Science and Technology Facilities Council (STFC). This review has made use of NASA's Astrophysics Data System.

M.G. Linton thanks Aimee Norton for useful feedback on \S \ref{section:FEobservations}.

\bibliography{bib-dynamo,bib-helioseismology,bib-emergence}
%
%%%End Solar Interior chapter

%\include{ch03-09} 
\backmatter

\latexprintindex

\end{document}